\documentclass[twocolumn]{aastex62}

\usepackage{multirow}
\usepackage{color}
\received{}
\revised{}
\accepted{}
\submitjournal{ApJ}

\shorttitle{Grain growth and rim curvature}
\shortauthors{Davies et al.}

\begin{document}

\title{Simultaneous spectral energy distribution and near-infrared interferometry modeling of HD~142666}

\correspondingauthor{Claire L.\ Davies}
\email{cdavies@astro.ex.ac.uk}

\author[0000-0001-9764-2357]{Claire L. Davies}
\affiliation{Astrophysics Group, School of Physics, University of Exeter, Stocker Road, Exeter, EX4 4QL, UK}

\author[0000-0001-6017-8773]{Stefan Kraus}   
\affiliation{Astrophysics Group, School of Physics, University of Exeter, Stocker Road, Exeter, EX4 4QL, UK}

\author[0000-0001-8228-9503]{Tim J.\ Harries}   
\affiliation{Astrophysics Group, School of Physics, University of Exeter, Stocker Road, Exeter, EX4 4QL, UK}

\author[0000-0002-0911-9505]{Alexander Kreplin}  
\affiliation{Astrophysics Group, School of Physics, University of Exeter, Stocker Road, Exeter, EX4 4QL, UK}

\author[0000-0002-3380-3307]{John D.\ Monnier}   
\affiliation{Department of Astronomy, University of Michigan, Ann Arbor, MI 48109, USA}

\author[0000-0001-8837-7045]{Aaron Labdon}   
\affiliation{Astrophysics Group, School of Physics, University of Exeter, Stocker Road, Exeter, EX4 4QL, UK}

\author[0000-0003-0350-5453]{Brian Kloppenborg}    
\affiliation{Department of Physics and Astronomy, Georgia State University, Atlanta, GA, USA}



\author[0000-0002-4881-7584]{David M.\ Acreman}      
\affiliation{Astrophysics Group, School of Physics, University of Exeter, Stocker Road, Exeter, EX4 4QL, UK}

\author[0000-0002-8376-8941]{Fabien Baron}     
\affiliation{Department of Physics and Astronomy, Georgia State University, Atlanta, GA, USA}


\author[0000-0003-0447-5866]{Rafael Millan-Gabet}   
\affiliation{Infrared Processing and Analysis Center, California Institute of Technology, Pasadena, CA, 91125, USA}


\author{Judit Sturmann}   
\affiliation{The CHARA Array of Georgia State University, Mount Wilson Observatory, Mount Wilson, CA 91203, USA}

\author{Laszlo Sturmann}    
\affiliation{The CHARA Array of Georgia State University, Mount Wilson Observatory, Mount Wilson, CA 91203, USA}


\author[0000-0002-0114-7915]{Theo A.\ ten Brummelaar}   
\affiliation{The CHARA Array of Georgia State University, Mount Wilson Observatory, Mount Wilson, CA 91203, USA}







\begin{abstract}
We present comprehensive models of Herbig Ae star, HD\,142666, which aim to simultaneously explain its spectral energy distribution (SED) and near-infrared (NIR) interferometry. 
Our new sub-milliarcsecond resolution CHARA (CLASSIC and CLIMB) interferometric observations, supplemented with archival shorter baseline data from VLTI/PIONIER and the Keck Interferometer, are modeled using centro-symmetric geometric models and an axisymmetric radiative transfer code. 
CHARA's $330\,$m baselines enable us to place strong constraints on the viewing geometry, revealing a disk inclined at $58^{\circ}$ from face-on with a $160^{\circ}$ major axis position angle. 
Disk models imposing vertical hydrostatic equilibrium provide poor fits to the SED. 
Models accounting for disk scale height inflation, possibly induced by turbulence associated with magneto-rotational instabilities, and invoking grain growth to $\gtrsim1\,\mu$m size in the disk rim are required to simultaneously reproduce the SED and measured visibility profile. 
However, visibility residuals for our best model fits to the SED indicate the presence of unexplained NIR emission, particularly along the apparent disk minor axis, while closure phase residuals indicate a more centro-symmetric emitting region. 
In addition, our inferred $58^{\circ}$ disk inclination is inconsistent with a disk-based origin for the UX~Ori-type variability exhibited by HD~142666. 
Additional complexity, unaccounted for in our models, is clearly present in the NIR-emitting region. 
We propose the disk is likely inclined toward a more edge-on orientation and/or an optically thick outflow component also contributes to the NIR circumstellar flux.
\end{abstract}




\section{Introduction}
Circumstellar disks are ubiquitous across all masses of star formation \citep[e.g.][]{Andrews13, Ricci14, Ilee16, Lazareff16, Kraus17}: a consequence of the conservation of angular momentum during gravitational collapse. These disks provide the building materials and the natal environment for planets to form and evolve in. The reprocessing of starlight by dust in the innermost regions of protoplanetary disks produces strong near-infrared (NIR) continuum emission, in excess of that expected from a stellar photosphere. Developments in the field of NIR interferometry during the late 1990s enabled the first spatially-resolved observations of the circumstellar structure of Herbig Ae/Be stars -- the precursors to intermediate-mass stars \citep{Herbig60, Strom72} -- to be obtained. The milliarcsecond (mas) resolution offered by the Infrared Optical Telescope Array (IOTA) and Palomar Testbed Interferometer (PTI) showed that the inner disk regions did not extend down to the stellar surface \citep[e.g][]{Millan99, Akeson00}, in agreement with prior spectral energy distribution (SED) modeling \citep{Hillenbrand92}. As the number of Herbig Ae/Be stars observed with NIR interferometry increased, a relationship between the host star luminosity and the characteristic size of the NIR-emitting region emerged \citep{Monnier02}. The slope of this size-luminosity relationship suggests the NIR-emitting region arises from a dust sublimation rim at a temperature of $\sim1800\,$K \citep{Lazareff16}.

Early disk models incorporating a dust sublimation rim used a vertical-wall approximation \citep{Dullemond01, Natta01}. However, the strong viewing angle-dependency to the NIR emission associated with such a model is in conflict with the similar levels of NIR excess observed among Herbig Ae/Be stars over a wide range of disk inclination angles \citep{Natta01, Dominik03}. In addition, the significant closure phase ($\phi_{\rm{CP}}$) signals associated with the strongly asymmetric NIR brightness distribution in vertical rim models was not observed \citep{Monnier06,Kraus09}. Instead, curvature of the inner rim is understood to arise due to the dependence of the dust sublimation temperature and grain cooling efficiency on, for example, the gas density, the size distribution of dust grains, grain growth-induced vertical settling, and the relative abundance of different grain compositions \citep{Isella05, Tannirkulam07, Kama09, McClure13}. 

The picture was further complicated with the first \emph{sub}-mas NIR observations of Herbig Ae/Be stars, made possible with the $\sim330\,$m baselines of the Center for High Angular Resolution Astronomy (CHARA) Array. Through their observations of \object{MWC 275} and \object{AB Aur}, \citet{Tannirkulam08a} found that the ``bounce'' in the secondary visibility lobe predicted by curved rim models was not observed. Instead, to explain the relatively flat profiles of the observed second visibility lobes, an additional NIR-emitting component interior to the silicate dust sublimation front was required. Further evidence for this has been reported in studies using NIR spectro-interferometry \citep[e.g.][]{Kraus08, Eisner09}, high resolution spectroscopy \citep[e.g.][]{Ilee14}, and photometry \citep[e.g.][]{Fischer11}. The nature of this material remains unclear with plausible suggestions including a hot gas reservoir and/or more refractory grain species \citep{Tannirkulam08b, Eisner09}. 

Here, we focus on the shape, location and viewing geometry of the circumstellar disk of the Herbig Ae star, \object{HD\,142666} (spectral type A8Ve; \citealt{Meeus98}). The IR excess of \object{HD\,142666} (common aliases include V1026~Sco), first identified by \citet{Walker88}, has previously been studied using NIR and mid-IR (MIR) interferometers with operational baselines $\lesssim100\,$m. The characteristic size of the $H$- and $K$-band-emitting regions observed with the Keck Interferometer (KI), VLTI/AMBER and VLTI/PIONIER (henceforth referred to as AMBER and PIONIER, respectively) are consistent with that expected from dust sublimation ($\sim0.4\,$au at a stellar distance of $150\,$pc; \citealt{Monnier05, Lazareff16}) while the MIR emission observed with VLTI/MIDI is more extended than predicted by typically-adopted temperature gradient models suggesting a narrow, dust-free gap is present within the inner few au of the disk \citep{Schegerer13, Vural14}. However, the usual features indicative of optically thin disk regions or disk cavities are not seen in the SED of \object{HD\,142666} \citep{Dominik03} meaning the disk is not typically considered to be (pre-)transitional. Intermediate disk inclinations for HD\,142666 have been indicated via NIR and MIR interferometry ( $48.6^{\circ}\,^{+2.9}_{-3.6}$, \citealt{Vural14}; $\sim60^{\circ}$, \citealt{Lazareff16}), SED analysis ($\sim55^{\circ}$, \citealt{Dominik03}) and ALMA ($\sim60^{\circ}$, \citealt{Rubinstein18}). VLT/NACO differential imaging and ALMA indicate the disk major axis position angle\footnote{Quoted disk position angles, PA$_{\rm{major}}$, are for the disk major axis, measured east of north.}, PA$_{\rm{major}}$, is oriented along a nearly North-South direction ($\sim180^{\circ}$, \citealt{Garufi17}; $161^{\circ}$, \citealt{Rubinstein18})). 

We present new, high-resolution NIR interferometric data of \object{HD\,142666} obtained using the CLASSIC two-telescope and CLIMB three-telescope beam combiners of the CHARA Array \citep{Brummelaar13}. Section~\ref{sec:obs} details our CHARA observations and the supplementary, shorter baseline NIR interferometry retrieved from the archives. With its $\sim331\,$m maximum baseline length, our CHARA observations offer us the opportunity to distinguish between different curved rim models to understand the dominant process of rim curvature in the disk of \object{HD\,142666}. Our analysis builds upon that of \citet{Tannirkulam08b} who used the TORUS Monte Carlo radiative transfer code \citep{Harries00} to model the NIR interferometric visibilities of two other Herbig Ae stars -- \object{MWC\,275} and \object{AB\,Aur} (spectral types of A1 and A0, respectively; \citealt{Mora01, Hernandez04}) -- obtained with CHARA/CLASSIC (henceforth referred to as CLASSIC). In addition to considering a later-type Herbig~Ae star, (i) our $(u,v)$-plane coverage is much improved compared to the \citet{Tannirkulam08b} study, (ii) we probe $H$- as well as $K$-band emission, and (iii) with the addition of CHARA/CLIMB (henceforth referred to as CLIMB) data, we use $\phi_{\rm{CP}}$ information to further constrain our modeling. 

A two-fold approach is used in our analysis. First, we constrain the stellar flux contribution to the NIR flux by fitting stellar atmosphere models to optical photometry and employ centro-symmetric geometric models to constrain the viewing geometry of the NIR-emitting region. We then build on these models using the TORUS Monte Carlo radiative transfer code to explore the physical bases behind the location and shape of the observed inner disk rim. Our modeling approach is outlined in Section~\ref{sec:methodology} while the results of our geometric and radiative transfer analysis are presented in Sections~\ref{sec:geoFitResults} and \ref{sec:RTresults}, respectively. In Section~\ref{sec:discussion}, we discuss our results in the context of grain growth to micron sizes in the inner rim and comment on indirect evidence for further complexity in the NIR-emitting region.

\section{Observations and complementary archival data}\label{sec:obs}
\subsection{CHARA interferometry}
CHARA is a Y-shaped array of six $1\,$m-class telescopes located at Mount Wilson Observatory offering operational baselines between $34$ and $331\,$m \citep{Brummelaar05}. CLASSIC and CLIMB \citep{Brummelaar13} were used to obtain $K$-band observations of \object{HD\,142666} between 2009 June and 2013 June. Additional $H$-band observations of \object{HD\,142666} were obtained with CLIMB in 2014 May and June.

\begin{deluxetable*}{lclcc}
\tabletypesize{\scriptsize}
\tablecolumns{5}
\tablecaption{CHARA observation log\label{tab:obslog}}
\tablehead{
 \colhead{Date (UT)} & \colhead{Beam Combiner} & \colhead{Stations} & \colhead{Filter} & \colhead{Calibrator(s)}
}
\startdata
2009 Jun 24 & CLASSIC & S2 E2 & K & 1, 2\\
2010 Jun 15 & CLASSIC & S2 W1 & K & 3, 4\\
2011 Jun 15 & CLIMB & E1 W1 W2 & K & 3\\
2011 Jun 20 & CLIMB & (E1)\tablenotemark{a} W1 W2 & K & 3\\
2011 Jun 23 & CLIMB & S1 W1 W2 & K & 3, 5\\
2011 Jun 25 & CLIMB & S1 E2 W2 & K & 6, 7\\
2011 Jun 27 & CLIMB & S1 W1 W2 & K & 6\\
2011 Jun 28 & CLIMB & S1 E2 W2 & K & 6\\
2011 Aug 03 & CLIMB & S1 E2 W1 & K & 3, 5\\
2012 Jun 29 & CLIMB & E1 E2 W1 & K & 3\\
2012 Jul 01 & CLIMB & S1 S2 W1 & K & 6\\
2013 Jun 10 & CLIMB & S1 W1 W2 & K & 8\\
2013 Jun 13 & CLIMB & S1 W1 W2 & K & 9, 10\\
2013 Jun 14 & CLIMB & S2 E2 W2 & K & 8, 9\\
2013 Jun 16 & CLIMB & E2 W1 W2 & K & 11\\
2014 May 28 & CLIMB & S2 E2 W2 & H & 3\\
2014 May 29 & CLIMB & S2 E2 W2 & H & 3, 5\\
2014 May 30 & CLIMB & S2 E2 W2 & H & 3, 5\\
2014 Jun 04 & CLIMB & S2 E2 W1 & H & 3, 5\\
2014 Jun 09 & CLIMB & E1 E2 W1 & H & 5\\
2014 Jun 10 & CLIMB & S2 W1 W2 & H & 3, 5
\enddata
\tablenotetext{a}{CLIMB operating as a two-telescope beam combiner.}
\tablecomments{Calibrators and their UD diameters: (1) HD\,141465, $0.28\pm0.02\,$mas; (2) HD\,143766, $0.311\pm0.022\,$mas; (3) HD\,140990, $0.230\pm0.016\,$mas; (4) HD\,141597, $0.24\pm0.05\,$mas; (5) HD\,143616, $0.222\pm0.016\,$mas; (6) HD\,148211, $0.250\pm0.018\,$mas; (7) HD\,152429, $0.272\pm0.019\,$mas; (8) HD\,148198, $0.255\pm0.018\,$mas; (9) HD\,144766, $0.324\pm0.023\,$mas; (10) HD\,145809, $0.417\pm0.029\,$mas; (11) HD\,141937, $0.307\pm0.022\,$mas.}
\end{deluxetable*}

\begin{figure}
  \centering
  \includegraphics[width=0.36\textwidth]{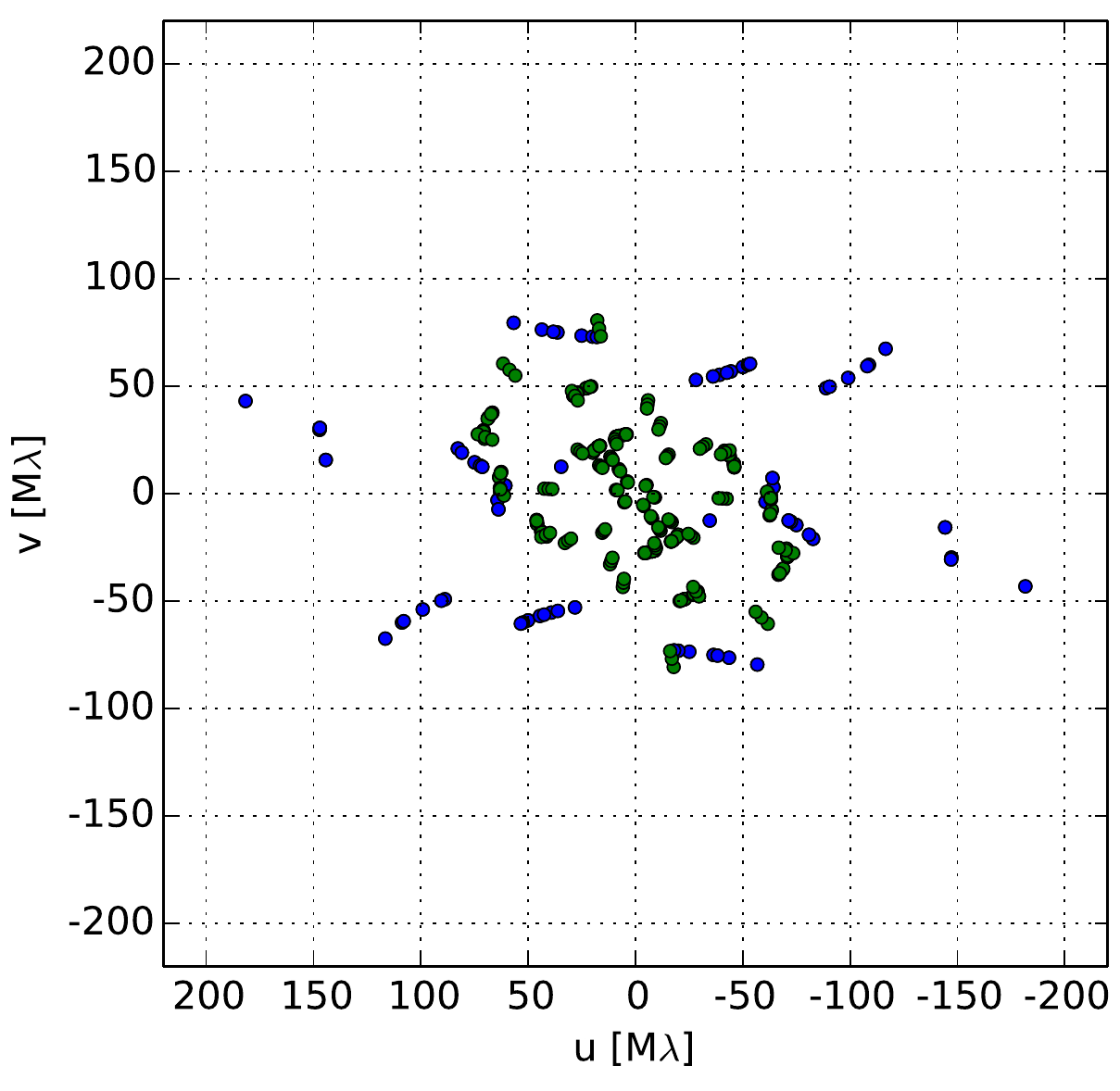}\\
  \includegraphics[width=0.36\textwidth]{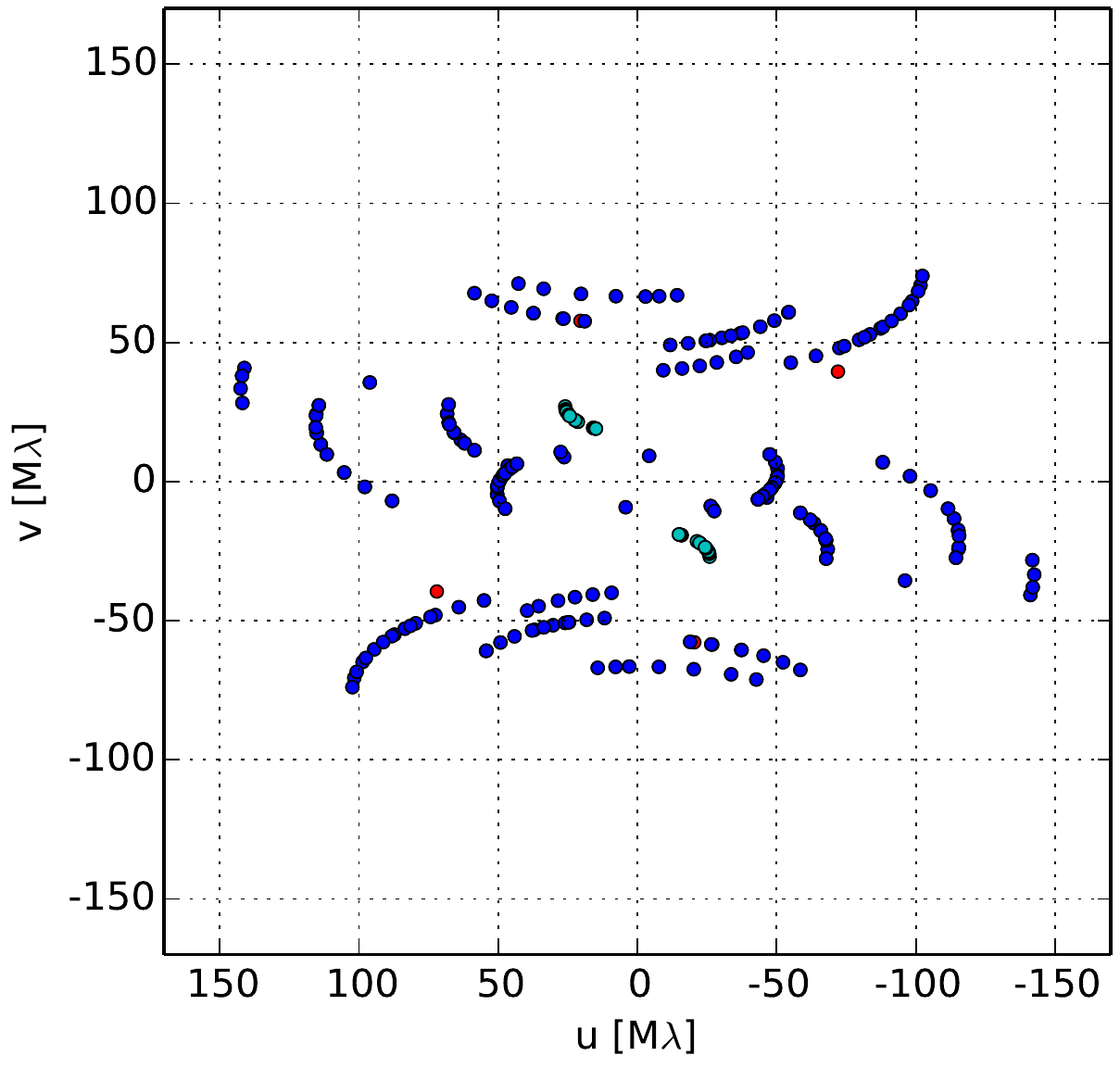}
  \caption{($u, v$)-plane coverage for \object{HD\,142666} in H- (top) and K-band (bottom). Positive values of $v$ and $u$ correspond to North and East, respectively. Our new CLASSIC and CLIMB data (see Table~\ref{tab:obslog}) are shown as red and blue points, respectively. Supplementary archival NIR interferometric data, listed in Table~\ref{tab:supldata}, are also indicated: KI (cyan); PIONIER (green).
  }
  \label{fig:uvplane}
\end{figure}

A variety of telescope configurations were used during the observing campaign with a maximum projected baseline length of $313\,$m (corresponding to an angular resolution\footnote{$\lambda/2B$ with $\lambda$ the operational wavelength and $B$ the projected baseline length.} of $0.70\,$mas). Further details regarding the individual observations are provided in Table~\ref{tab:obslog}. The resulting ($u, v$)-plane coverage is displayed in Figure~\ref{fig:uvplane}. 

The CLASSIC and CLIMB data were reduced using pipelines developed at the University of Michigan which are better suited to recovering faint fringes for low visibility data than the standard CHARA reduction pipeline of \citet{Brummelaar12}. The waterfall plot of raw data scans was first inspected for instrumental or observational effects such as drifting scans or flux drop-out on one or more telescope, for example. Any scans displaying these effects were flagged and rejected. In the majority of cases, this affected at most $5-10$ per cent of scans. Extra care was taken on the few occasions where drift or low signal-to-noise dominated the majority of scans. In these cases, the affected scans were carefully flagged while the power spectrum, averaged over the retained scans, was inspected for a signal. After this process, the foreground, background and flux recorded for each baseline pair were each inspected for flux drop-out. Finally, the power spectrum for each telescope pair and CLIMB output (P0, P1 and P2; see \citealt{Brummelaar12}) was inspected and the background level set manually. This results in one $\phi_{\rm{CP}}$ for each baseline triplet, three estimates of the square visibility, $V^{2}$, for one baseline pair and two estimates of $V^{2}$ for the remaining two CLIMB baseline pairs. 

Calibration of the $V^{2}$ and $\phi_{\rm{CP}}$ measurements were made using standard stars observed before and/or after each science observation. None of the calibrators used are known binary systems. As a further check, the $\phi_{\rm{CP}}$ signals of each of the calibrators observed more than once were inspected: no signatures of binarity were found. The uniform diameters (UDs) of each calibrator, obtained from JMMC SearchCal \citep{Bonneau06, Bonneau11}, where available, or gcWeb\footnote{http://nexsciweb.ipac.caltech.edu/gcWeb/gcWeb.jsp}, are listed in Table~\ref{tab:obslog}. The transfer function across the full observation sequence was inspected to ensure its flatness. Finally, the multiple estimates of the calibrated $V^{2}$ on each baseline pair were checked for consistency before a weighted-average value was computed. For our analysis, we thus have one estimate of $V^{2}$ for each baseline pair. The raw and calibrated data will be made available in oifits format \citep{Pauls05, Duvert17} through the CHARA archive (Jones et al.\ in prep) and the Optical interferometry Database (OiDb; \citealt{Haubois14}) of the JMMC following publication.

\subsection{Supplementary archival interferometry}
\begin{deluxetable*}{lccccc}
\tabletypesize{\scriptsize}
\tablecolumns{6}
\tablecaption{Supplementary Archival NIR Interferometric Data\label{tab:supldata}}
\tablehead{ 
 \colhead{Instrument} & \colhead{Observation Date (UT)} & \colhead{Program ID} & \colhead{Stations} & \colhead{Filter} & \colhead{Calibrator(s)}
}
\startdata
VLTI PIONIER & 2012 Mar 28 & 088.D-0185 & A1 G1 I1 K0 & H & \nodata \\
             & 2012 Mar 29 & 088.C-0763 & A1 G1 I1 K0 & H & \nodata \\
             & 2013 Jun 06 & 190.C-0963 & A1 G1 J3 K0 & H & \nodata \\
             & 2013 Jun 17 & 190.C-0963 & D0 G1 H0 I1 & H & \nodata \\
             & 2013 Jul 03 & 190.C-0963 & A1 B2 C1 D0 & H & \nodata \\
KI V2-SPR    & 2004 Mar 05 & 13         & K1K2        & K & 1\\
             & 2007 Jul 02 & 32         & K1K2        & K & 2, 3\\
             & 2009 Jul 16 & 31         & K1K2        & K & 2, 3, 4, 5\\
             & 2012 May 02 & 57         & K1K2        & K & 6\\
             & 2012 May 03 & 57         & K1K2        & K & 7
\enddata
\tablecomments{Calibrators listed in column 7 for the instances where a re-reduction of the data was required: (1) HD\,134967, $0.15\pm0.01\,$mas; (2) HD\,139364, $0.323\pm0.023\,$mas; (3) HD\,141465, $0.28\pm0.02\,$mas; (4) HD\,142301, $0.159\pm0.011\,$mas; (5) HD\,143766 $0.311\pm0.022\,$mas; (6) HD\,145809, $0.417\pm0.029\,$mas; (7) HD\,141597, $0.24\pm0.05\,$mas.}
\end{deluxetable*}

To better constrain the geometry of the disk, we supplemented our long-baseline CLASSIC and CLIMB data with shorter baseline archival NIR interferometry. Calibrated PIONIER \citep{leBouquin11} data for \object{HD\,142666}, originally published in \citet[][program IDs 190.C-0963, 088.D-0185, and 088.C-0763]{Lazareff16}, were retrieved from the OiDb. PIONIER data from UT date 2013 June 17, not available on the OiDb, were also provided by Bernard Lazareff (private communication). KI \citep{Colavita13} data were retrieved from the Keck Observatory Archive. The wide-band KI data were calibrated using the NExScI Wide-band Interferometric Visibility Calibration (wbCalib v1.4.4) tool with the flux bias correction and ratio correction options selected. Table~\ref{tab:supldata} provides further details on the collated data and, for the occasions where the data required (re-)reduction, also the names and UD diameters of the standard stars used to calibrate $V^{2}$ and $\phi_{\rm{CP}}$. 

\section{Modeling methodology}\label{sec:methodology}
We model the location and extent of the circumstellar NIR-emitting region of \object{HD\,142666} using the Monte Carlo radiative transfer code, TORUS \citep{Harries00, Harries04, Kurosawa06, Tannirkulam07}. Exploring viewing geometries using TORUS would be computationally expensive so, to allow for a more rapid exploration, we employed a series of geometric models to determine best-fit inclinations and position angles. In the subsections that follow, we outline the methodology adopted in both our analyses.

\subsection{Geometric modeling of the visibilities}\label{sec:geofit}
The visibility of the circumstellar emission is
\begin{equation}
    V_{\rm{CS}} = \frac{|V_{\rm{obs}}(F_{\star} + F_{\rm{CS}}) - V_{\star}F_{\star}|}{F_{\rm{CS}}},
\end{equation}
where $V_{\rm{obs}}$ is the observed visibility and $F_{\star}$ and $F_{\rm{CS}}$ refer to the stellar and circumstellar flux contributions, respectively. The stellar emission component of \object{HD\,142666} is expected to be unresolved as the stellar radius is much smaller than the length scales we are able to probe. As such, we set the visibility of the stellar component to unity, i.e.\ $V_{\star}=1$. 

We required an independent assessment of $F_{\star}$ at $H$- and $K$-bands to avoid degeneracies associated with fitting both the characteristic size of the emitting region and $F_{\star}$ simultaneously (see, for example, \citealt{Lazareff16}). Multi-wavelength photometry were retrieved from the literature while a post-processed, flux-calibrated \textit{Spitzer} Infrared Spectrograph \citep[IRS;][]{Houck04} spectrum \citep[AORkey 3586816]{Keller08} was retrieved from the \emph{Spitzer} Heritage Archive. The full list of collated photometry, together with the individual references, is presented in Table~\ref{tab:phot:HD} in Appendix~\ref{apen:phot}. The photospheric portion (Johnson-$B$, -$V$, and Cousins-$I_{c}$ wavebands) of the SED constructed for \object{HD\,142666} was then fit using \citet{Kurucz79} model atmospheres appropriate for the star (see Section~\ref{sec:geoFitResults}).

The measured $\phi_{\rm{CP}}$ were inspected for deviations from centro-symmetry. While no significant indication for non-zero $\phi_{\rm{CP}}$ was visible in the full $H$-band data set (CLIMB+PIONIER), a possible deviation from centro-symmetry is suggested by the full $K$-band data set (CLIMB). Assuming the NIR emission from \object{HD\,142666} emanates from the inner regions of an inclined disk, a non-zero $\phi_{\rm{CP}}$ may indicate a degree of skewness to the disk emission caused by self-shielding, for example. Alternative scenarios include, but are not restricted to, the presence of regions with enhanced brightness, possibly indicating an increased disk scale height (i.e.\ disk warp) or eluding to the presence of additional companions. If these features co-orbit with the disk, their dynamical timescales may be smaller than the four year timescale over which the $\phi_{\rm{CP}}$ were obtained. 

\begin{figure}
  \centering
  \includegraphics[width=0.47\textwidth]{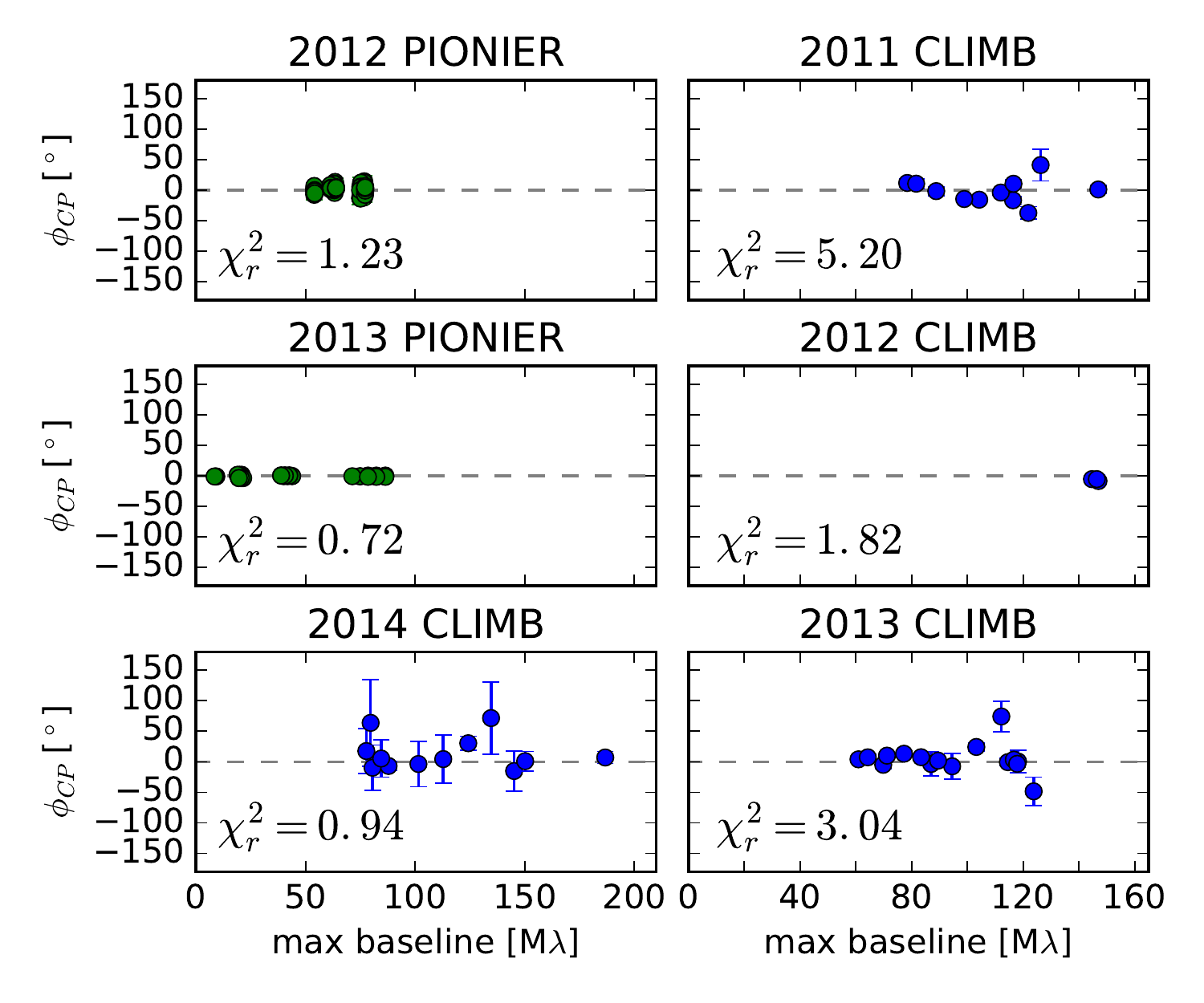}
  \caption{Closure phase as a function of the maximum spatial frequency probed by each closed triangle of baseline vectors. Left-hand panels contain the $H$-band PIONIER and CLIMB data while right-hand panels contain the $K$-band CLIMB data, each split by the year of observation (see Tables~\ref{tab:obslog} and \ref{tab:supldata}). Data points are colored as in Fig.~\ref{fig:uvplane}. The $\chi^{2}_{\rm{r}}$ for centro-symmetric models ($\phi_{\rm{CP}}=0^{\circ}$ on all spatial scales) is displayed in the lower left corner of each panel.
  }
  \label{fig:cps}
\end{figure}

In Fig.~\ref{fig:cps}, we plot the observed $\phi_{\rm{CP}}$ against the maximum spatial frequency probed by each triplet of baseline vectors, split by observational epoch and waveband. In each panel, the reduced-$\chi^{2}$ ($\chi^{2}_{\rm{r}}$) value computed for a centro-symmetric model ($\phi_{\rm{CP}}=0$ at all spatial frequencies) is displayed in the top left-hand corner. Although there may be an indication for deviation from centro-symmetry in the 2011 and 2013 $K$-band data, the $\phi_{\rm{CP}}=0^{\circ}$ model provides a good fit to all epochs. Thus, to estimate the geometry of the $H$- and $K$-band-emitting regions, we restrict our analysis to centro-symmetric models. 

In Section~\ref{sec:geoFitResults}, we consider two geometric models for the brightness distribution. Both of these use a point source component to model the stellar flux contribution and assume, for simplicity, that all non-stellar NIR emission arises from the innermost regions of a disk. In the first model, a thin ring of emission is used to emulate the disk component, corresponding to the emission expected from a centrally-illuminated vertical wall. The free parameters of this point source-plus-ring (PS+R) model are the ring radius, $R$, its inclination, $i$ (where $0^{\circ}$ corresponds to a face-on viewing geometry), and its major axis position angle, PA$_{\rm{major}}$ (measured east of north). In the second model, the disk emission is approximated as a Gaussian-smoothed ring, avoiding the sharp edges of the ring model and corresponding to a more spatially-extended NIR emitting region. The full-width at half-maximum (FWHM) of the Gaussian used in the convolution remains a free parameter in the fitting procedure. These point source-plus-smoothed ring models are henceforth referred to as PS+SR. 

During the fitting procedure, errors on the best-fit parameters (found via $\chi^{2}$ minimization) were estimated via bootstrapping. A thousand new realizations of the original visibility data sets were created and fed through the same modeling procedure as the original data. The initial values of the parameters in the fitting remained consistent between data sets in the same waveband and throughout the bootstrapping process. Histograms were created from the resulting bootstrapped model outputs and errors were estimated from $1\sigma$-Gaussian fits to each histogram. 

\subsection{Monte Carlo Radiative Transfer modeling with TORUS}\label{sec:RTmodel}
The TORUS Monte Carlo radiative transfer code uses the \citet{Lucy99} algorithm to compute radiative equilibrium on a two-dimensional, cylindrical adaptive mesh grid. Assuming that all of the circumstellar NIR emission of \object{HD\,142666} arises from a disk, we prescribe the TORUS models as follows. The initial density structure of the gas component of the disk, $\rho(r, z)$, is based on the $\alpha$-disk prescription of \citet{Shakura73}:
\begin{equation}\label{eq:gasdensity}
 \rho(r, z) = \frac{\Sigma(r)}{h(r)\sqrt{2\pi}} \exp\left\{-\frac{1}{2}\left[\frac{z}{h(r)}\right]^{2}\right\}.
\end{equation}
Here, $r$ and $z$ are the radial distance into the disk and the vertical height above the disk midplane, respectively. The parameters $h(r)$ and $\Sigma(r)$ describe the scale height, 
\begin{equation}\label{eq:scaleheight}
 h(r) = h_{\rm{0,gas}}\left(\frac{r}{100\,\rm{au}}\right)^{\beta},
\end{equation} 
and the surface density, 
\begin{equation}
 \Sigma(r) = \Sigma_{\rm{0,gas}}\left(\frac{r}{100\,\rm{au}}\right)^{-p},
\end{equation}
of the gas component of the disk, respectively. The constants $h_{\rm{0,gas}}$ and $\Sigma_{\rm{0,gas}}$ are each equated at $r=100\,$au. We keep $p=1.0$ fixed in all models.

The disk is passively heated by a single star located at the grid centre and is assumed to be in local thermodynamic equilibrium. The temperature structure of the disk and the location and shape of the dust sublimation region are established in an iterative manner using the \citet{Lucy99} algorithm. To investigate the shape of the inner rim of the disk, we use two different parameterizations of the sublimation region facilitated by TORUS. These are summarised in Table~\ref{tab:RTmodels}, discussed briefly in Section~\ref{sec:graincomp} and the interested reader is referred to \citet{Tannirkulam07} for further details. In both cases, an e-folding factor of $10\,$K to the dust sublimation temperature has been introduced to enable convergence. 

TORUS solves for radiative equilibrium with or without imposing vertical hydrostatic equilibrium. If vertical hydrostatic equilibrium is imposed, the vertical structure of the disk is modified via the equation of vertical hydrostatic equilibrium according to an adapted form of the \citet{Walker04} algorithm following each \citet{Lucy99} iteration \citep{Tannirkulam07}. The process of establishing a converged temperature and dust sublimation structure is then repeated. Typically, these models converge after the third iteration of imposing vertical hydrostatic equilibrium.

\begin{deluxetable}{lccc}
\tabletypesize{\scriptsize}
\tablecolumns{6}
\tablecaption{TORUS radiative transfer models summary showing grain fractions by mass, dust scale height ($h_{\rm{0,dust}}$) relative to gas scale height ($h_{\rm{0,gas}}$), and sublimation temperature ($T_{\rm{sub}}$) \label{tab:RTmodels}}
\tablehead{ 
 \colhead{} & \colhead{S:small} & \colhead{S:large} & \colhead{THM07} }
\startdata
$0.1\,\mu$m grain fraction ($\%$) & 100 & 0   & 90 \\
$1.2\,\mu$m grain fraction ($\%$) & 0   & 100 & 10 \\
$0.1\,\mu$m grain $h_{\rm{0,dust}}$ ($h_{\rm{0,gas}}$) & 1.0 & \nodata & 1.0 \\
$1.2\,\mu$m grain $h_{\rm{0,dust}}$ ($h_{\rm{0,gas}}$) & \nodata & 1.0 & 0.6 \\
$T_{\rm{sub}}$ (K) & $G\rho^{\gamma}(r, z)$ & $G\rho^{\gamma}(r, z)$ & 1400
\enddata
\end{deluxetable}

Following convergence, a separate Monte Carlo algorithm is used to compute model SEDs and $H$- and $K$-band images based on the optical properties of the dust species used in the particular model (\citealt{Harries00}; see Section~\ref{sec:graincomp} for details of the grain prescriptions used in our models). All model outputs were computed at a distance of $150\,$pc \citep{Lindegren16} based on the distance inferred from the Gaia DR1 parallax and consistent with that inferred from the more recent Gaia DR2 parallax ($148\pm1\,$pc; \citealt{Gaia16, Gaia18, Bailer18}), and at the best-fit disk inclinations found through our geometric modeling (see Section~\ref{sec:geoFitResults}). 

Visibility amplitudes and phases were extracted from the model images at PA$_{\rm{base}}$ and baseline lengths corresponding to the ($u,v$)-plane positions of our interferometric data (see Fig.~\ref{fig:uvplane}). Model $\phi_{\rm{CP}}$ were then computed from the sum of the visibility Fourier phases over each closed triangle of baseline vectors. Due to the combined effects of the model image resolution (which introduces errors when the image is rotated and via the interpolation between pixels to the correct baseline length) and numerical estimation of the complex visibilities, our procedure for estimating model $\phi_{\rm{CP}}$ introduces an uncertainty of $\sim1^{\circ}$. This is within our CLIMB measurement uncertainties. 

\subsubsection{Dust grain prescription and implementation of rim curvature}\label{sec:graincomp}
The location of the disk inner rim is controlled by the dust species with the highest sublimation temperature, $T_{\rm{sub}}$, and greatest cooling efficiency \citep{Isella05, Kama09}. We limit our analysis to astronomical silicate grains which sublimate at temperatures consistent with those inferred from the NIR size-luminosity relation \citep{Pollack94}. The grains are modeled as homogeneous spheres with a mass density of $3.3\,\rm{g\,cm^{-3}}$ \citep{Kim94} and optical constants prescribed by \citet{Draine03} which differ from those of \citet{Draine84} only in the details.

In one set of TORUS models, inner disk rim curvature arises due to the dependence of $T_{\rm{sub}}$ on the local gas density \citep{Pollack94, Isella05}:
\begin{equation}
    T_{\rm{sub}} = G\rho^{\gamma}(r, z).
\end{equation}
Here, $\gamma=1.95\times10^{-2}$ and the constant, $G=2000\,$K for silicate grains. Grains larger than $\sim1.3\,\mu$m in size do not contribute sufficiently to the disk opacity and thus do not play a role in determining the location of the dust rim \citep{Isella05}. As such, we adopt two different grain size prescriptions for these models: one set with small grains ($0.1\,\mu$m in size) and the other with large grains ($1.2\,\mu$m in size), consistent with the original \citet{Isella05} study. As these models use a single grain size, they are henceforth referred to as the S:small and S:large models, respectively.

\begin{deluxetable*}{cccccccccccc}
 \tabletypesize{\scriptsize}
  \tablecaption{Adopted stellar parameters\label{tab:starParam}}
   \tablewidth{0pc}
    \tablehead{
     \colhead{SpT} & \colhead{$T_{\rm{eff}}$ (K)} & \colhead{$\log g$} & \colhead{$\log Z$} & \colhead{$d$ (pc)} & \colhead{$L_{\star}/L_{\odot}$} & \colhead{$A_{\rm{V}}$} & \colhead{$R_{\star}/R_{\odot}$} & \colhead{$M_{\star}/M_{\odot}$}  & \colhead{$F_{\star ,\rm{H}}$} & \colhead{$F_{\star ,\rm{K}}$} & \colhead{$R_{\rm{out}}$ (au)} 
    }
    \startdata
     A8 & $7500$ & $4.3$ & $0.2$ & $150$ & $19.3$ & $1.63$ & $2.42$ & $1.97$ & $0.61$ & $0.35$ & $60$ 
    \enddata
 \tablecomments{References for $T_{\rm{eff}}$, $\log g$, $\log Z$, $d$ and $R_{\rm{out}}$: \citet{Dent05}, \citet{Guimaraes06}, \citet{Lindegren16}, \citet{Garufi17}, \citet{McDonald17}.}
\end{deluxetable*}

In our other set of TORUS models, rim curvature arises due to the dependence of dust sublimation on the grain size-dependent cooling efficiency \citep{Kamp04, Kama09} and settling, as originally prescribed in \citet{Tannirkulam07}. These are henceforth referred to as THM07 models. We adopt a thermally coupled mixture of $0.1\,\mu$m and $1.2\,\mu$m grains in a ratio of $9$:$1$ by mass in favour of the small grains. $T_{\rm{sub}}=1400\,$K is used for both grain sizes for consistency with the original \citet{Tannirkulam08b} study of Herbig Ae stars \object{AB Aur} and \object{MWC 275}. We also limit the disk scale height of the $1.2\,\mu$m grains to 60\,\% that of the gas, $h_{\rm{0,gas}}$, while the $0.1\,\mu$m grains inhabit the full $h_{\rm{0,gas}}$ range.

\subsubsection{Adopted parameters for the outer disk}\label{sec:RdMd}
The flux across (sub-)millimeter wavelengths provides an indication of the mass contained within the dust component of protoplanetary disks (e.g.\ \citealt{Hildebrand83, Beckwith90}) as the emission is optically thin. In preliminary modeling with TORUS, we found that the (sub-)millimeter portion of the SED was reasonably well fit using a total disk mass of $0.20\,\rm{M_{\odot}}$, assuming a radially invariant gas-to-dust ratio of $100$:$1$. As the dust grain prescription we adopt in our TORUS modeling is rather simplistic (see previous subsection), we acknowledge that this assessment of the disk mass is likely unrealistic. For instance, the adoption of different grain size distribution or inclusion of another grain species with a different mass density and optical properties would affect the inferred value (c.f.\ \citealt{Wood02}). 

The radial extent of the disk around \object{HD\,142666} has been constrained from VLT/NACO imaging \citep{Garufi17} and through analysis of rotationally-broadened emission lines of gaseous species in the outer disk regions \citep{Dent05}. In both cases, a value of $\sim60\,$au was indicated, assuming a distance to \object{HD\,142666} of $150\,$pc. This value is also consistent with the $65\,$au found recently by \citet{Rubinstein18} from ALMA band 6 continuum observations. Our preliminary TORUS models showed that an outer disk radius, $R_{\rm{out}}=60\,$au provided a good fit to the long wavelength portion of the SED. This value was adopted in all our TORUS models.

\section{Results from geometric modeling}\label{sec:geoFitResults}
As outlined in Section~\ref{sec:geofit}, an independently-assessed $F_{\star}$ estimate was used to avoid degeneracies associated with using geometric models to simultaneously fit $F_{\star}$ and the characteristic size of the NIR-emitting region. To estimate $F_{\star}$ at $1.67\,\mu$m ($H$-band) and $2.13\,\mu$m ($K$-band), estimates of the stellar effective temperature, $T_{\rm{eff}}$, surface gravity, $\log g$, and metallicity, $\log Z$, of \object{HD\,142666} were retrieved from the literature (see Table~\ref{tab:starParam}). Using the python package {\sc pysynphot} \citep{pysynphot} and the ``minimize'' function of the python {\sc lmfit} library \citep{Newville14}, the corresponding \citet{Kurucz79} model atmosphere was compared to Johnson-Cousins $BVI_{\rm{C}}$ photometry to assess the $V$-band extinction, $A_{\rm{V}}$, and the stellar radius\footnote{The radius enters the fit through the scaling factor, ($d/R_{\star})^{2}$ which arises from the \citet{Kurucz79} model being in units of surface flux.}, $R_{\star}$. As \object{HD\,142666} displays flux variations at optical and NIR wavelengths \citep[e.g.][c.f. Section~\ref{sec:disc:geom}]{Makarov94}, we ensured that the $BVI_{\rm{C}}$ photometry were obtained contemporaneously. The reddening law of \citet{Cardelli89} with a total-to-selective extinction, $R_{\rm{V}}=5.0$ was adopted based on previous analyses of Herbig Ae/Be stars by \citet{Hernandez04} and \citet{Manoj06}. The fitting procedure uses the differential evolution method which is less susceptible to regions of local minima than e.g.\ the Levenberg-Marquardt method. The values of $R_{\star}$ and $A_{\rm{V}}$ found in the fitting process were then combined with the values of $T_{\rm{eff}}$, $\log g$, $d$, and $\log Z$ to estimate $F_{\star,H}$ and $F_{\star,K}$. The best-fit values are presented in Table~\ref{tab:starParam}. 

These results were used to determine self-consistent estimates for the remaining stellar parameters required as inputs for the TORUS radiative transfer models. The stellar luminosity, $L_{\star}$, was estimated from the $V$-band magnitude and extinction using $T_{\rm{eff}}$-dependent bolometric corrections taken from Table~5 of \citet{Pecaut13} and a value of $4.755\,$mag for the bolometric magnitude of the Sun \citep{Mamajek12}. The stellar mass, $M_{\star}$, was estimated by comparing $T_{\rm{eff}}$ and $L_{\star}$ to \citet{Siess00} pre-main-sequence evolutionary models with $Z=0.02$ (without convective overshooting). These values are also presented in Table~\ref{tab:starParam}.

\begin{figure}[t]
  \centering
  \includegraphics[width=0.48\textwidth]{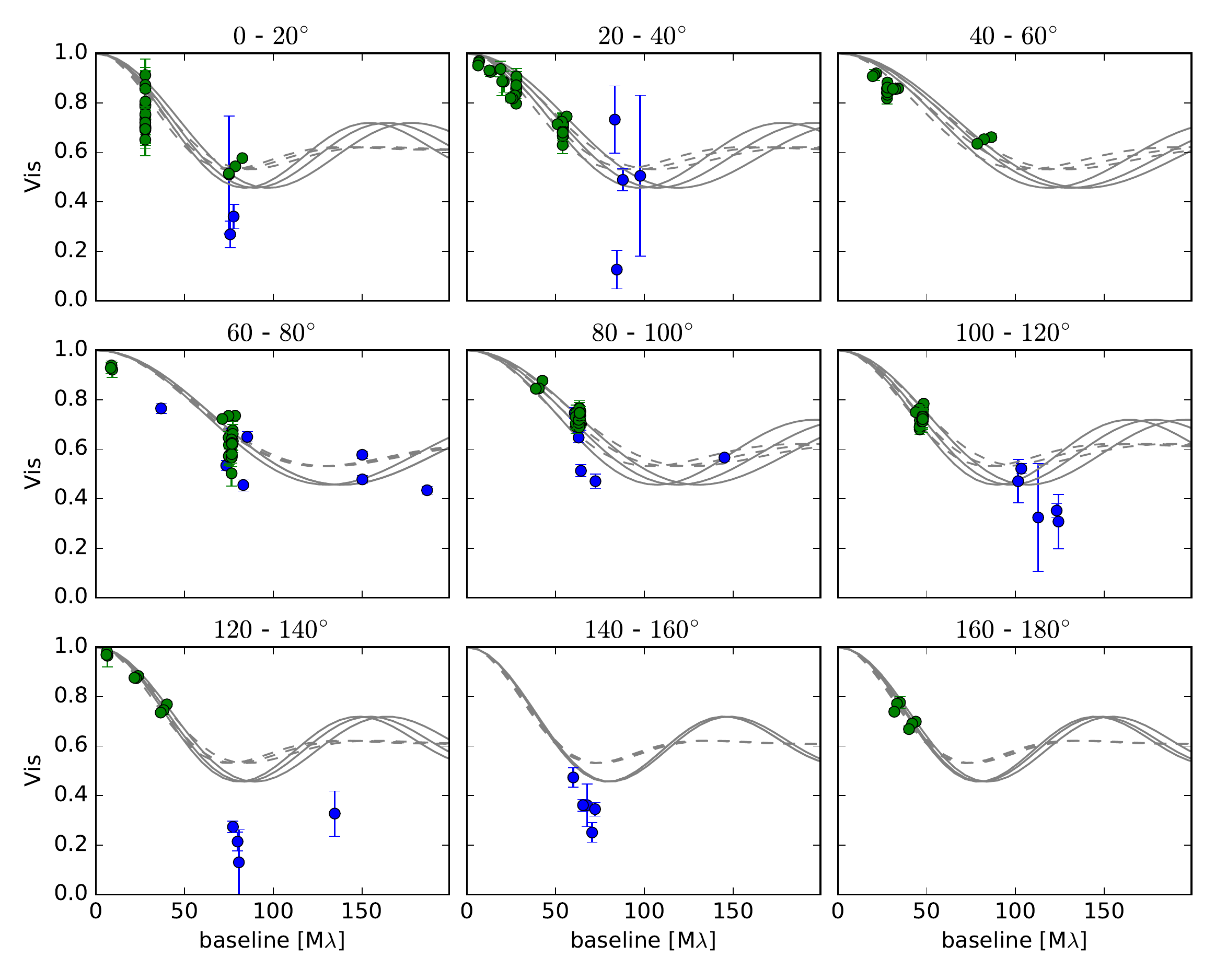}
  \caption{PA$_{\rm{base}}$-separated $H$-band visibilities plotted against spatial frequency. Grey solid lines represent the best-fit PS+R (with fixed $F_{\star,H}$) model visibility curves for the CLIMB+PIONIER data set (see Table~\ref{tab:psringHD}). Grey dashed lines represent the best-fit PS+SR model visibility curves for the same data set (see Table~\ref{tab:pssmoothringHD}). Three model curves are plotted in each panel, corresponding to steps of $10^{\circ}$ in PA$_{\rm{base}}$. The range of PA$_{\rm{base}}$ included in each figure window is listed above each window. Data points are colored as in Fig.~\ref{fig:uvplane}.}
  \label{fig:psringH}
\end{figure}

\begin{figure}[t]
  \centering
  \includegraphics[width=0.48\textwidth]{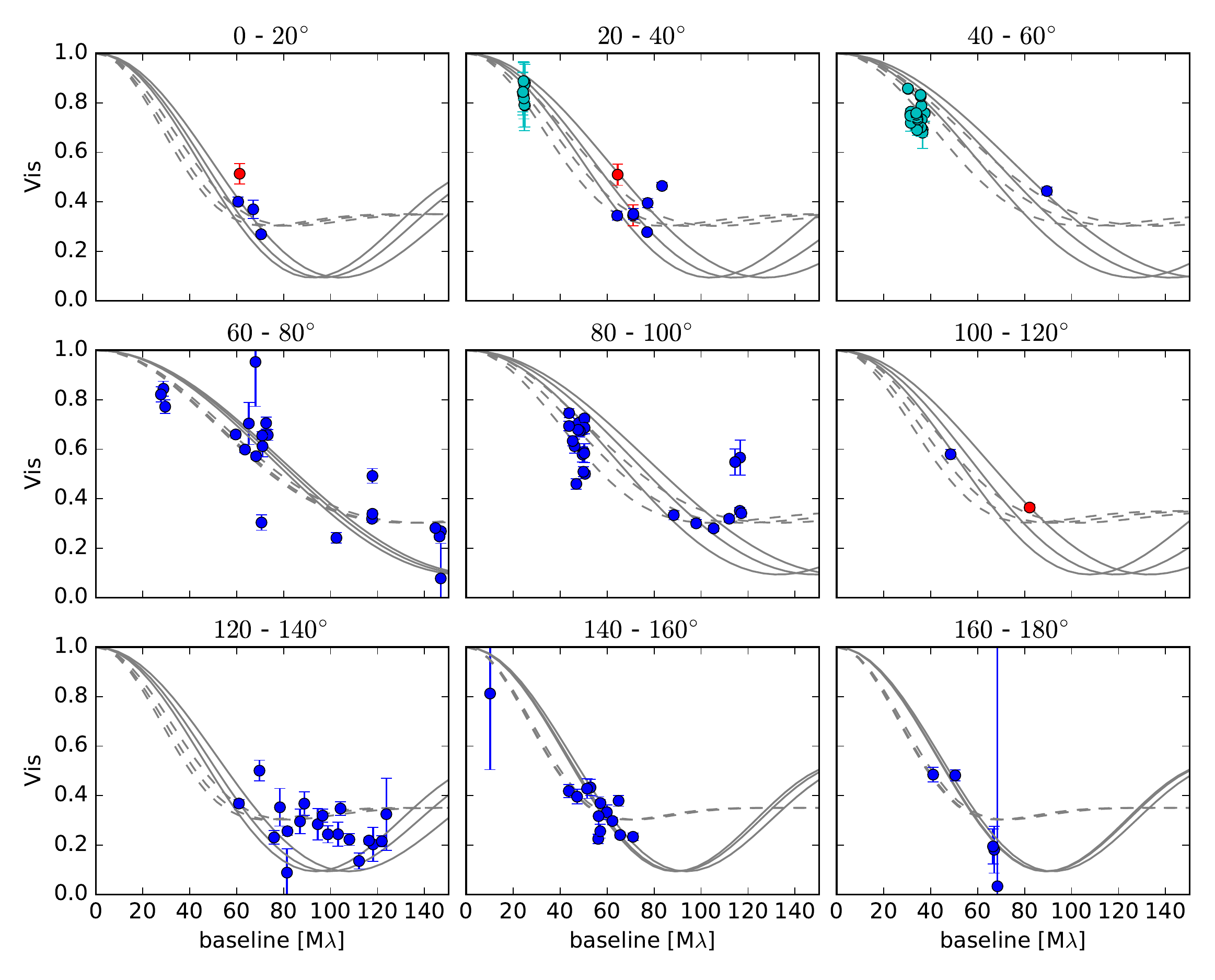}
  \caption{As Fig.~\ref{fig:psringH} but for $K$-band. Grey solid lines represent the best-fit PS+R (with fixed $F_{\star,K}$) model visibility curves for the CLIMB+CLASSIC+KI data set (see Table~\ref{tab:psringHD}) while grey dashed lines correspond to those of the best-fit PS+SR model (see Table~\ref{tab:pssmoothringHD}).}
  \label{fig:psringK}
\end{figure}

PS+R models were fit to the visibilities obtained with CLASSIC and CLIMB before repeating the process with the inclusion of the shorter baseline PIONIER ($H$-band) or KI ($K$-band) data. In each case, the $H$- and $K$-band data were fit separately. The stellar flux contribution remained fixed at the values in Table~\ref{tab:starParam}. The resulting best-fit parameters (and corresponding $\chi^{2}_{\rm{r}}$) for each model are displayed in Table~\ref{tab:psringHD}. The observed $H$- and $K$-band visibilities are compared to those of the best fit models (grey solid lines) in Figs.~\ref{fig:psringH} and \ref{fig:psringK}, respectively.

The PS+R model suggests $i\sim53-61^{\circ}$ and PA$_{\rm{major}}\sim155-166^{\circ}$, depending on the data set. Inclusion of the shorter baseline data favours slightly lower inclinations in $K$-band and reduces the uncertainty range on our estimates of $i$ and PA$_{\rm{major}}$ across both wavebands. These values for $i$ and PA$_{\rm{major}}$ agree well with previous analyses of NIR and MIR interferometric data ($30<i<60^{\circ}$ and PA$_{\rm{major}}\sim170^{\circ}$ \citealt{Vural14, Lazareff16}) together with the recent VLT/NACO imaging of \citet{Garufi17} who found a preference for an inclined disk with PA$_{\rm{major}}$ along a North-South direction.

\begin{deluxetable*}{lcclcr}
\tabletypesize{\scriptsize}
\tablecolumns{6}
\tablecaption{Results from PS+R model fits to $H$- and $K$-band visibilities \label{tab:psringHD} }
\tablehead{
 \colhead{Dataset} & \colhead{$R$ (mas)} & \colhead{$i$ ($^\circ$)} & \colhead{P.A. ($^\circ$)} & \colhead{$F_{\star}$} & \colhead{$\chi^{2}_{\rm{r}}$}
 }
\startdata
\multicolumn{6}{c}{$H$-band} \\
\tableline
CLIMB         & $1.61\pm0.08$ & $57.1\pm5.1$ & $163\pm14$ & $0.61$ (fixed) & $21.97$ \\
              & $1.61\pm0.11$ & $56.6\pm4.9$ & $166\pm18$ & $0.57\pm0.07$ & $19.84$ \\
CLIMB+PIONIER & $1.44\pm0.05$ & $55.5\pm2.1$ & $155\pm2$ & $0.61$ (fixed) & $11.41$\\
              & $1.58\pm0.09$ & $57.1\pm1.7$ & $159\pm3$ & $0.64\pm0.03$ & $10.71$\\
\tableline
\multicolumn{6}{c}{$K$-band} \\
\tableline
CLIMB+CLASSIC    & $1.29\pm0.06$ & $60.7\pm3.1$ & $157\pm3$ & $0.35$ (fixed) & $34.64$\\
                 & $1.46\pm0.06$ & $55.0\pm3.3$ & $156\pm4$ & $0.46\pm0.06$ & $21.03 $ \\
CLIMB+CLASSIC+KI & $1.28\pm0.04$ & $57.4\pm3.0$ & $162\pm4.2$ & $0.35$ (fixed) & $34.10$\\
                 & $1.50\pm0.07$ & $52.9\pm2.6$ & $164\pm5$ & $0.46\pm0.06$ & $19.98 $
\enddata
\end{deluxetable*}

\begin{deluxetable*}{lcccccr}
\tabletypesize{\scriptsize}
\tablecolumns{6}
\tablecaption{Results from PS+SR model fits to $H$- and $K$-band visibilities \label{tab:pssmoothringHD} }
\tablehead{
 \colhead{Dataset} & \colhead{$R$ (mas)} & \colhead{$i$ ($^\circ$)} & \colhead{P.A. ($^\circ$)} & \colhead{FWHM (mas)} & \colhead{$\chi^{2}_{\rm{r}}$}
 }
\startdata
\multicolumn{6}{c}{$H$-band} \\
\tableline
CLIMB         & $1.78\pm0.09$ & $57.2\pm5.0$ & $165\pm14$ & $<2\times10^{-4}$ & $22.67$\\
CLIMB+PIONIER & $1.59\pm0.04$ & $56.7\pm1.9$ & $159\pm2$ & $0.60\pm0.09$ & $9.54$\\
\tableline
\multicolumn{6}{c}{$K$-band} \\
\tableline
CLIMB+CLASSIC    & $1.51\pm0.08$ & $62.9\pm3.6$ & $153\pm3$ & $0.87\pm0.09$ & $14.24$\\
CLIMB+CLASSIC+KI & $1.50\pm0.08$ & $60.5\pm3.6$ & $159\pm3$ & $0.95\pm0.12$ & $14.36$\\ 
\enddata
\end{deluxetable*}

According to the PS+R model-fitting results, the $H$ band-emitting region has an effective radius of $\sim1.44-1.61\,$mas ($\sim0.22-0.24\,$au at $150\,$pc). Assuming this region is associated with dust sublimation, we use the \citet{Whitney04} temperature-radius relation,
\begin{equation}
    R_{\rm{sub}} = R_{\star}\left(\frac{T_{\rm{sub}}}{T_{\rm{eff}}}\right)^{-2.1},
\end{equation}
to infer a sublimation temperature, $T_{\rm{sub}}\sim1750-1820\,$K. Typically quoted values of $T_{\rm{sub}}$ for silicate grains vary between $\sim1500-1800\,$K \citep{Pollack94}, indicating that the $H$-band emission is consistent with arising from the silicate dust sublimation rim. This value of $R$ is larger than the inferred value of $1.3\,$mas from \citet{Vural14} but we note that those authors also adopt a lower value of $F_{\star, H}$ ($0.53$). The degeneracy between the stellar flux contribution and characteristic size of the emitting region found through geometric modeling is well known (see \citealt{Lazareff16} for a discussion) and is likely responsible for these differences.

Interestingly, at first glance, our PS+R model fitting suggests that the $K$-band emission from \object{HD\,142666} traces material \emph{interior} to the $H$-band-emitting region. This is counter-intuitive as longer wavelength emission traces cooler material. If stellar radiation is the dominant heating mechanism, the cooler, $K$-band emission should emerge from larger disk radii than the warmer, $H$-band emission. However, under closer inspection, this is more likely to be a result of the poorer fit provided by the PS+R model to the $K$-band visibilities compared to those at $H$-band wavelengths. We investigated alternative models in an attempt to improve the fit. Firstly, we relaxed the constraint on $F_{\star}$, allowing it to vary between $0$ and $1$ in the fitting process\footnote{The total emission remained at 1 so the circumstellar emission provided a flux contribution of $1-F_{\star}$.}. Secondly, we adopted PS+SR models with $F_{\star}$ fixed at the values in Table~\ref{tab:starParam}. The resulting best-fit values and $\chi^{2}_{\rm{r}}$ for these alternative models are presented in Tables~\ref{tab:psringHD} and \ref{tab:pssmoothringHD}, respectively. 

From Table~\ref{tab:psringHD}, we can see that the fit is improved in all cases when the constraints on $F_{\star}$ are lifted. For the $H$-band emission, the fitted $F_{\star}$ values are consistent with the values adopted in our prior fitting within their uncertainties: $0.57\pm0.07$ (CLIMB) and $0.65\pm0.04$ (CLIMB+PIONIER) compared with the value of $0.61$ found via SED fitting. For the $K$-band emission, both data sets reveal a preferred value of $0.46\pm0.06$ over the value of $0.35$ found via SED fitting. \object{HD\,142666} exhibits variability across optical and NIR wavelengths \citep{Meeus98, Zwintz09} and these discrepancies in $F_{\star}$ are consistent with the intrinsic $H$- and $K$-band variability of $0.14$ and $0.30\,$mags, respectively. At the same time, an increase (decrease) in $F_{\star}$ coincides with an increase (decrease) in $R$, highlighting the degeneracy that exists when fitting $R$ and $F_{\star}$ simultaneously. Without the constraints on $F_{\star}$, we see that the characteristic radius of the $H$- and $K$-band-emitting regions is consistent within the bootstrapped errors. 

A further reduction in $\chi^{2}_{\rm{r}}$ is provided by the PS+SR models (dashed grey lines in Fig.~\ref{fig:psringH} and \ref{fig:psringK}). The PS+R model prescribes the NIR-emitting region as a central star with circumstellar emission provided by a vertical disk rim. In comparison, the circumstellar component of the PS+SR model emulates a more rounded rim in which the emitting region is more spatially extended. The better fit provided by the PS+SR model over the PS+R model suggests that the inner rim of the disk of \object{HD\,142666} is not well-approximated by a vertical wall. This is consistent with previous studies of other Herbig Ae/Be stars (e.g.\  \citealt{Tannirkulam07, Kraus09, McClure13}). As with the PS+R models, we see that the characteristic radius of the $H$ band-emitting region found via the PS+SR model fitting is consistently larger than that of the $K$-band emission though the two are roughly consistent within their estimated uncertainties: $1.59\pm0.04\,$mas (CLIMB+PIONIER; $H$-band) compared with $1.50\pm0.08\,$mas (CLIMB+CLASSIC+KI; $K$-band). Furthermore, the FWHM of the Gaussian component used to convolve the ring in these models is larger in $K$-band than in $H$-band, suggesting that the $K$-band emission may originate over a broader range of disk annuli.

\begin{figure}
  \centering
  \includegraphics[width=0.47\textwidth]{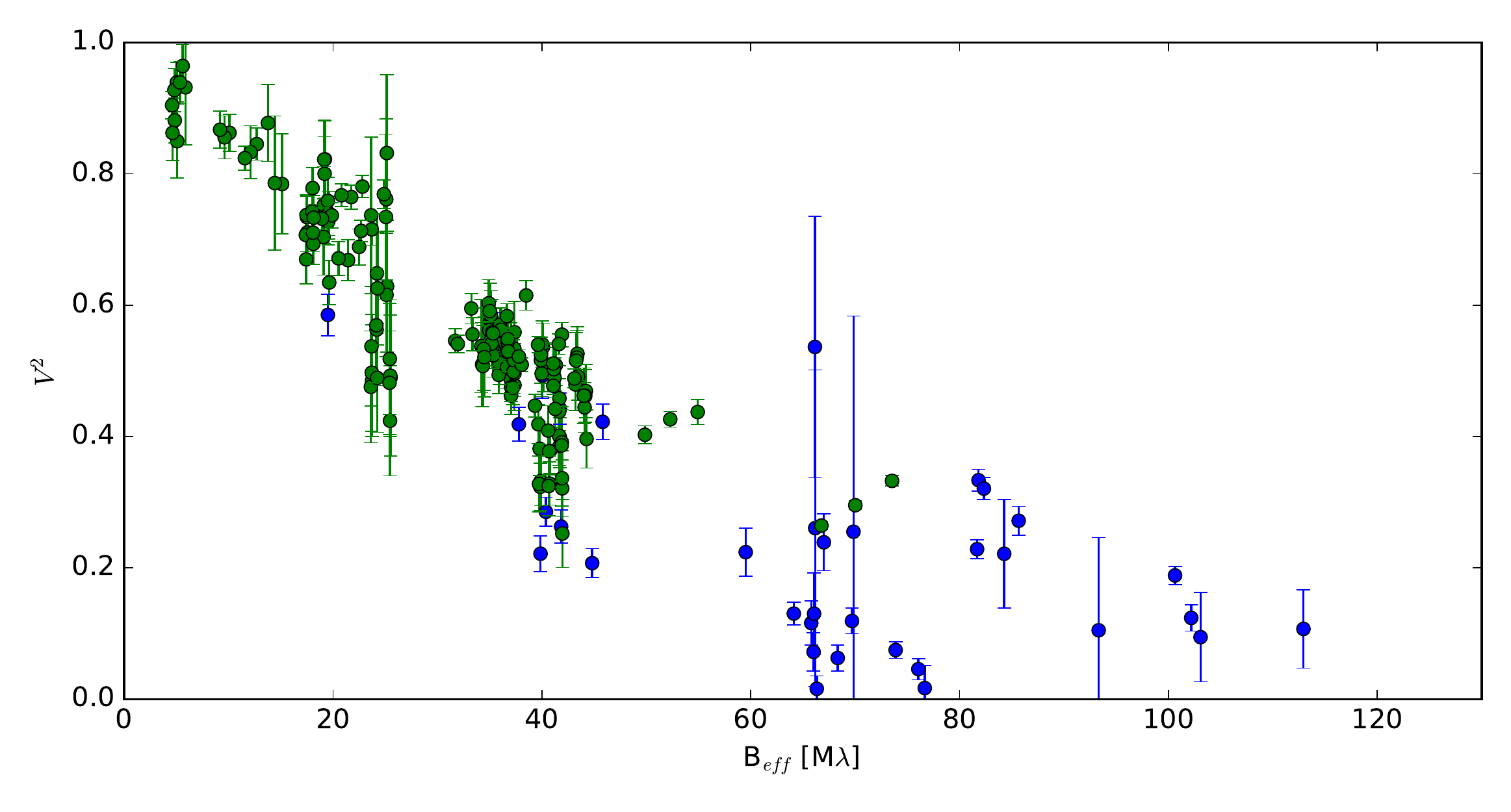}
  \includegraphics[width=0.47\textwidth]{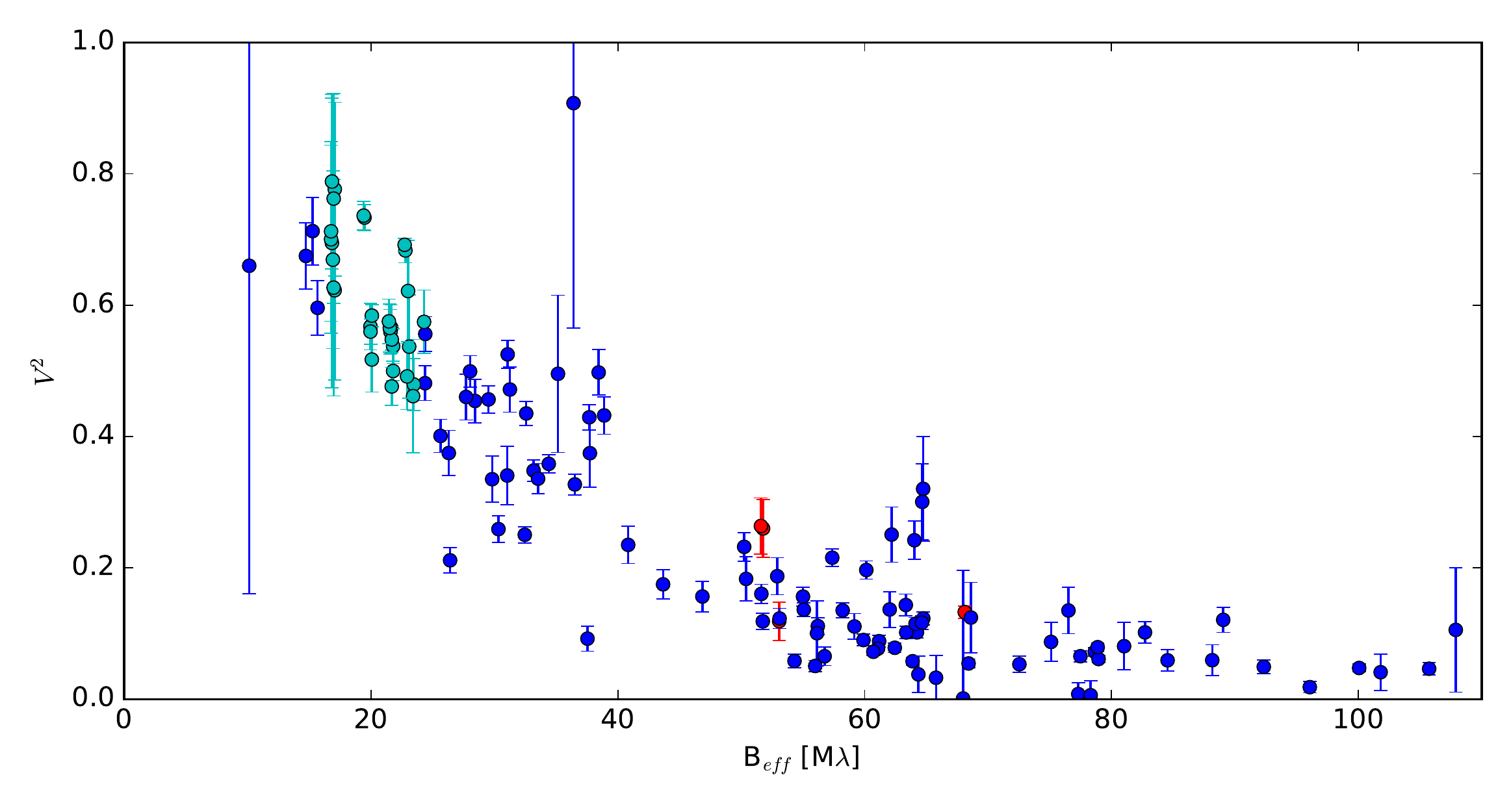}
   \caption{
   Distribution of $H$- (top) and $K$-band (bottom) squared visibilities with respect to the effective spatial frequency, computed using $i=58^{\circ}$ and PA$_{\rm{major}}=160^{\circ}$. Data points are colored as in Fig.~\ref{fig:uvplane}.
   }
  \label{fig:effbase}
\end{figure}

\section{Results from TORUS radiative transfer modeling}\label{sec:RTresults}
In light of our geometric modeling results, we adopted $i=58^{\circ}$ and PA$_{\rm{major}}=160^{\circ}$ throughout our TORUS modeling for computation of model SEDs and images. Fig.~\ref{fig:effbase} shows the distribution of squared visibilities ($V^{2}$) as a function of effective spatial frequency which accounts for the change in resolution across the uv-plane due to this inferred viewing geometry. With inclination effects accounted for, the vertical scatter in $V^{2}$ at each effective spatial frequency is assumed to arise due to the combined effects of calibration uncertainties and temporal variability (see Section~\ref{sec:disc:geom}).

When extracting $\phi_{\rm{CP}}$ from the TORUS model images, we orient the disk such that its north-eastern portion is the far side (and, thus, the brighter side) of the disk. This is based on the asymmetric brightness distribution seen in the VLT/NACO polarimetry \citep{Garufi17}.

\subsection{Models imposing vertical hydrostatic equilibrium}\label{sec:hydroresults}
\begin{figure}
  \centering
  \includegraphics[width=0.47\textwidth]{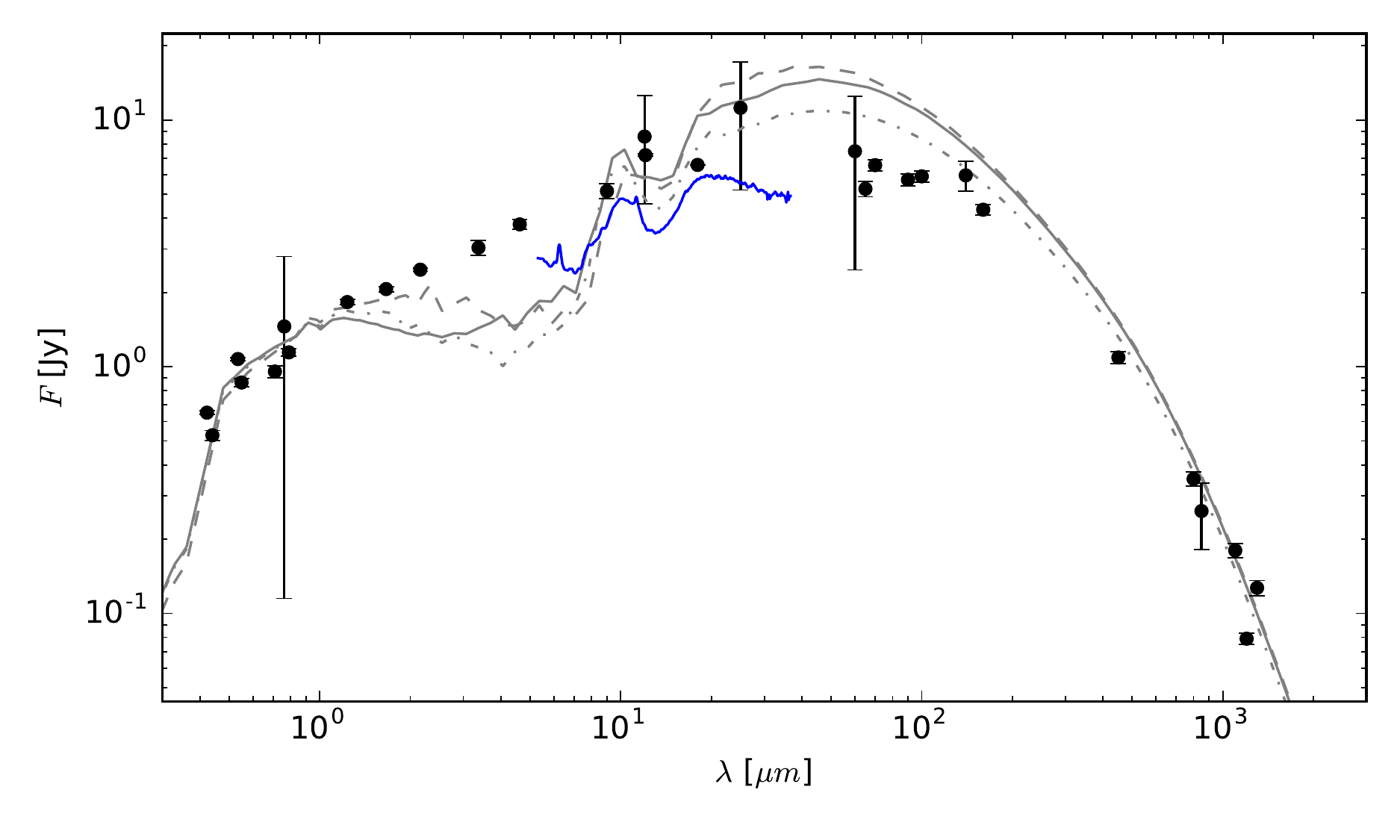}
   \caption{
   SEDs for our TORUS models with vertical hydrostatic equilibrium enforced. Photometric data are shown as black points while the \emph{Spitzer} spectrum is represented by a solid blue line. Grey lines correspond to the reddened TORUS model SEDs computed at $i=58^{\circ}$ (solid line: S:small; dashed line: S:large; dot-dashed line: THM07).
   }
  \label{fig:hydroSEDs}
\end{figure}

\begin{deluxetable}{lcccc}
\tabletypesize{\scriptsize}
\tablecolumns{5}
\tablecaption{Converged structure of TORUS models computed at disk inclinations of $58^{\circ}$ with vertical hydrostatic equilibrium imposed and their respective  stellar contribution to the total model $H$- and $K$-band fluxes \label{tab:hydro} }
\tablehead{\colhead{Model} & \colhead{$h_{\rm{0,gas}}$ (au)} & \colhead{$\beta$} & \colhead{$F_{\star,H}$} & \colhead{$F_{\star,K}$}}
\startdata
S:small & $4.6$ & $1.14$ & $0.85$ & $0.68$ \\
S:large & $5.0$ & $1.17$ & $0.67$ & $0.50$ \\
THM07   & $4.7$ & $1.18$ & $0.75$ & $0.62$
\enddata
\end{deluxetable}

We first computed S:small, S:large and THM07 models with vertical hydrostatic equilibrium established. In each case, the disk temperature and density structure converged after three iterations. We used equation~(\ref{eq:scaleheight}) to determine an approximate value of $\beta$ for the converged disk structure and present these alongside the values of $h_{\rm{0,gas}}$, $F_{\star,H}$, and $F_{\star,K}$ in Table~\ref{tab:hydro}. The SEDs computed for each model were reddened and are compared to the observed SED in Fig.~\ref{fig:hydroSEDs}. A relative dearth of NIR flux up to $\sim3\,\mu$m combined with a relative excess of flux over $\sim3-10\,\mu$m is provided by the S:small model (solid grey line) compared with the S:large and THM07 models (dashed and dot-dashed lines, respectively). Interestingly, the models including larger grains produce noticeably different SED shapes across NIR wavelengths: although the flux across $H$- and $K$-bands is underestimated in both cases, the S:large model produces NIR flux levels closest to those observed. These differences arise due to differences in the size of the disk area which directly intercepts stellar radiation in the innermost disk regions. The inclusion of larger grains extends the inner edge of the disk to smaller radii, as already discussed in, for example, \citet{Monnier02, Isella05, Tannirkulam07, Kama09} and \citet{McClure13}. The rim curvature provided by the THM07 model is also shallower and more extended than the S:large model. As such, if we consider the surface layers of the disk behind the sublimation rim, the directly illuminated disk area between disk annuli, $r$, and $r+\delta r$ will be smaller in the THM07 model than in the S:large model. As more optically thick disk material exists at hotter temperatures, this produces a larger NIR flux-emitting disk area.

In all three cases, our assumption that all the circumstellar NIR emission arises from a disk in vertical hydrostatic equilibrium leads to a poor SED fit. To better reproduce the observed SED, our models require more NIR-emission at the expense of FIR-emission. This discrepancy has previously been seen in both radiation hydrodynamic and radiation hydrostatic model fits to the SEDs of other Herbig AeBe stars \citep{Mulders12, Flock16a}. Turbulence associated with, for example, magneto-rotational instability (MRI) \citep{Turner14, Flock16} and/or the presence of magnetospheric or photo-evaporative disk winds \citep{Alexander07, Bans12} could contribute to lifting optically thick material above the disk scale heights predicted by our hydrostatic models. In addition, the presence of any optically thick gaseous material existing interior to the dust sublimation rim \citep{Tannirkulam08a, Tannirkulam08b} would affect the temperature structure of the dusty disk. As our CHARA interferometry does not reveal a bounce in the visibilities (Fig.~\ref{fig:effbase}), we are unable to comment on whether optically thick material interior to the sublimation rim contributes to the NIR flux. The extension of current optical interferometry facilities such as the CHARA Array to longer operational baselines and/or the construction of longer baseline optical interferometers equipped with NIR detectors (e.g.\ Planet Formation Imager; \citealt{Kraus14}) are essential for investigating whether NIR continuum emission also arises interior to the silicate sublimation rim in disks of later type Herbig Ae stars and their low-mass counterparts, the T~Tauri stars. The introduction of a dusty disk wind is also beyond the scope of this paper and we defer this to future study. Instead, in the subsections that follow, we focus on whether turbulence-induced scale height inflation is able to simultaneously fit the observed SED and interferometry of HD~142666. 

\begin{figure*}
  \centering
  \includegraphics[width=0.47\textwidth]{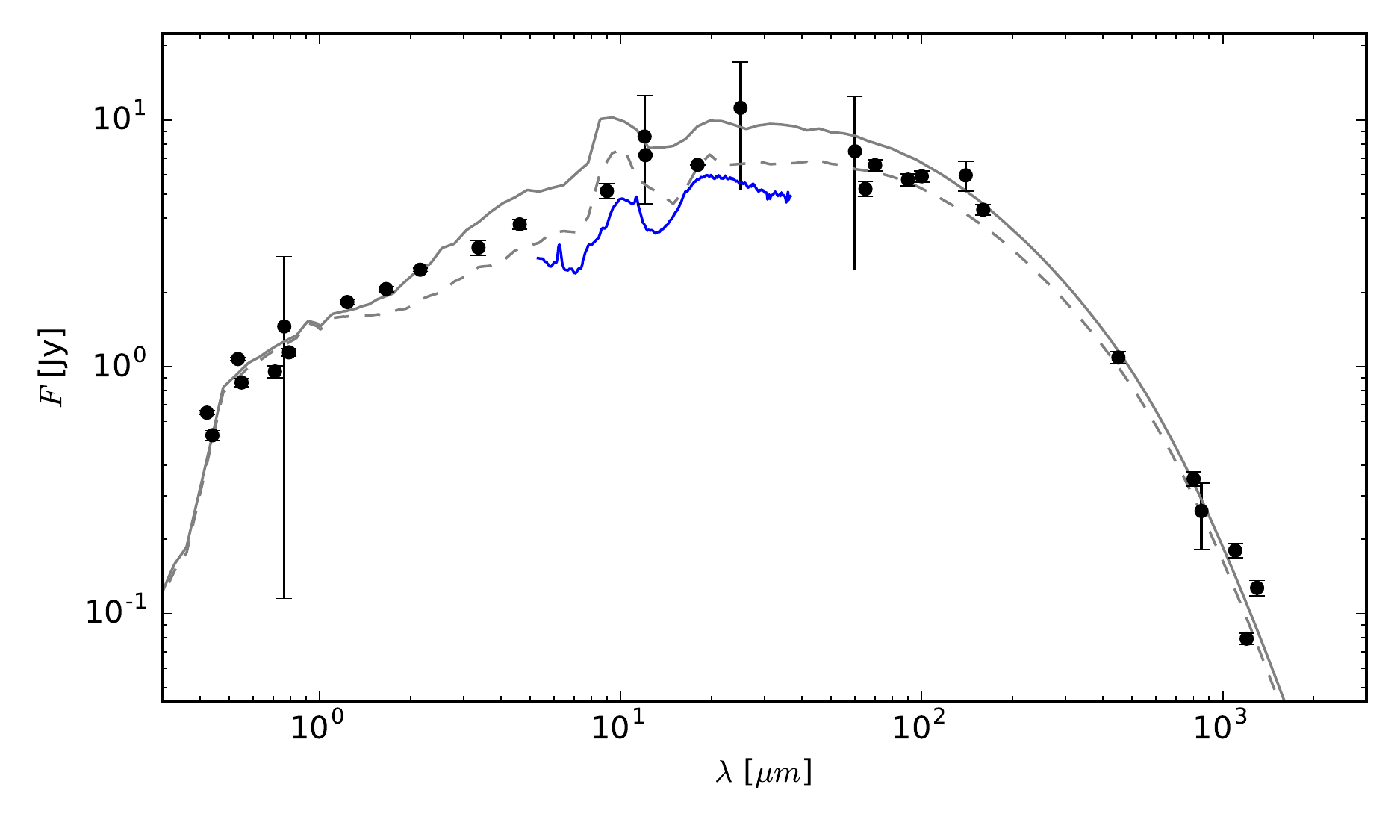}
  \includegraphics[width=0.47\textwidth]{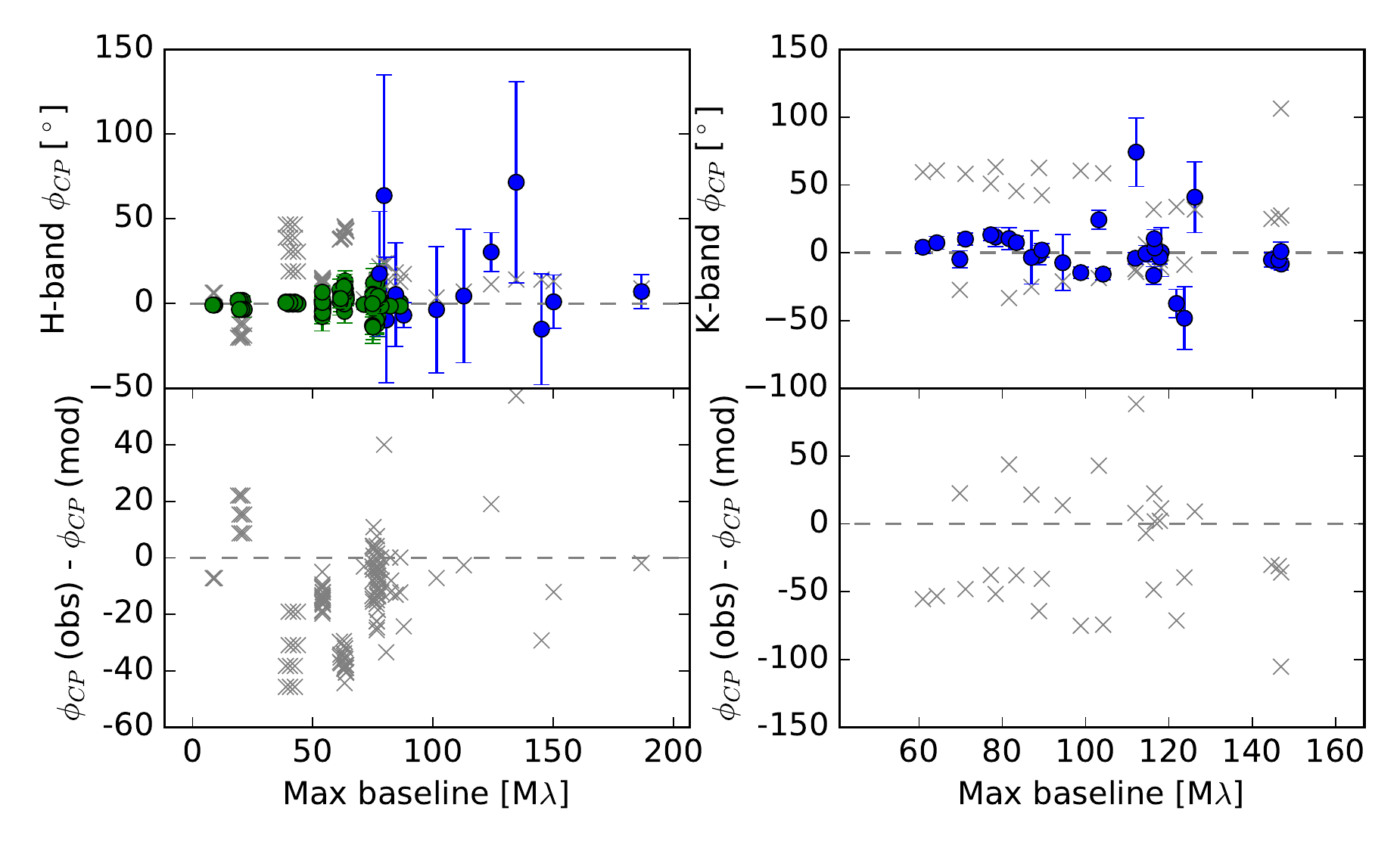}
  \includegraphics[width=0.47\textwidth]{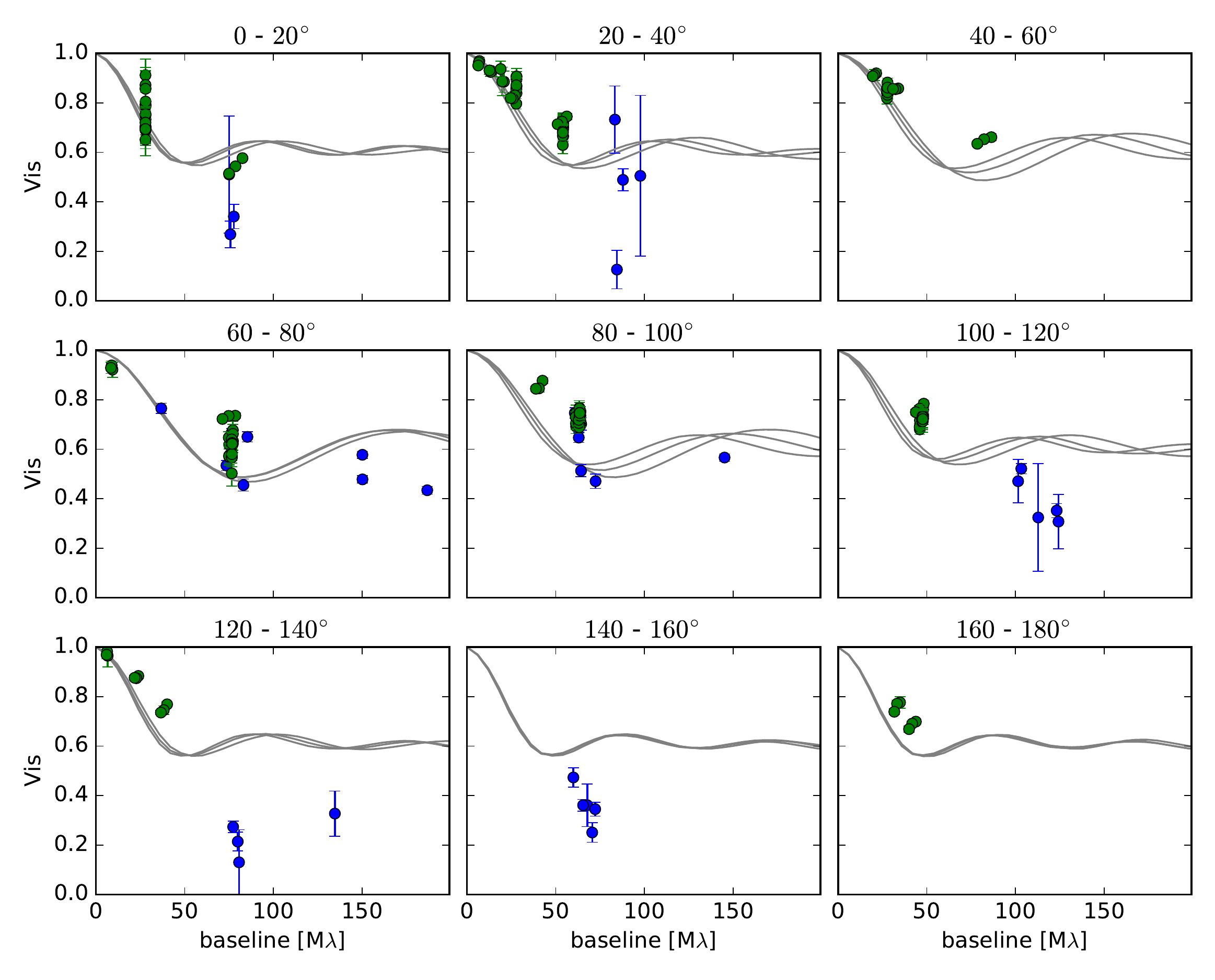}
  \includegraphics[width=0.47\textwidth]{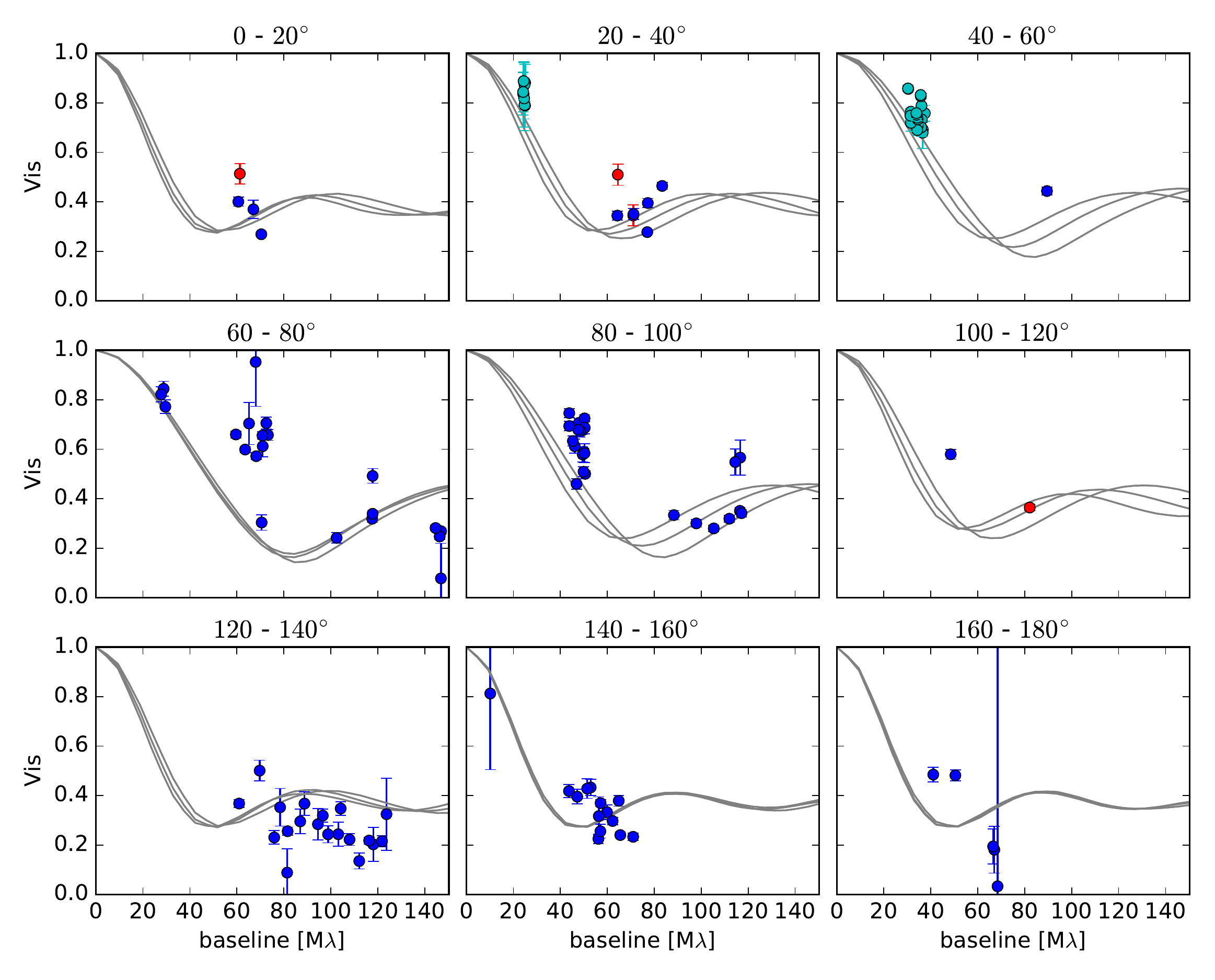}
   \caption{
   Summary plots for S:small model providing the best fit to the observed SED across the NIR ($h_{0}=7\,$au, $\beta=1.09$). Top left: As Fig.~\ref{fig:hydroSEDs} but comparing this best-fit model (solid grey line) to the S:small model with $h_{0}=10\,$au and $\beta=1.06$ (dashed grey line). Top middle and top right: $H$- and $K$-band $\phi_{\rm{CP}}$ (top) and the residuals (bottom), respectively. Both are displayed as a function of the maximum spatial frequency probed by the closed triangle of baseline vectors. Bottom rows: visibilities as a function of spatial frequency, separated by PA$_{\rm{base}}$. As in Fig.~\ref{fig:psringH}, three model visibility curves are plotted in each panel corresponding to $10^{\circ}$ steps in PA$_{\rm{base}}$. The nine panels on the left-hand side correspond to the $H$-band data while those on the right are for the $K$-band. Data in the visibility and $\phi_{\rm{CP}}$ plots are colored as in Fig.~\ref{fig:uvplane}.
   }
  \label{fig:IN05small}
\end{figure*}

To artificially emulate scale height inflation in the inner disk, we computed a series of grids of TORUS models without establishing vertical hydrostatic equilibrium. Each model grid was computed at a range of gas disk scale heights ($h_{0,\rm{gas}}=6, 7, 8, 9, 10, 11\,$au) and flaring parameters ($\beta=1.05, 1.06, 1.07, 1.08, 1.09, 1.10$) while the stellar parameters, disk mass and outer disk radius each remained fixed (see Table~\ref{tab:starParam}). 

\subsection{Small grain models}
Over the range of $h_{\rm{0,gas}}$ and $\beta$ probed, none of our S:small models were able to reproduce the observed SED across the full wavelength range. In the top left panel of Fig.~\ref{fig:IN05small}, the SED of \object{HD\,142666} is compared to the reddened SEDs of the two best-fitting S:small models. The S:small model with $h_{\rm{0,gas}}=7\,$au and $\beta=1.09$ (dashed grey line) provides the best fit to the SED over the full wavelength range but clearly provides insufficient NIR flux. To fit the NIR portion of the SED, a greater scale height was required: the $h_{0}=10\,$au and $\beta=1.06$ model (solid grey line) provides the best fit across this wavelength range while still reproducing the SED longward of $\sim100\,\mu$m. However, this latter model clearly overestimates the MIR to FIR flux.

\begin{figure*}
  \centering
  \includegraphics[width=0.47\textwidth]{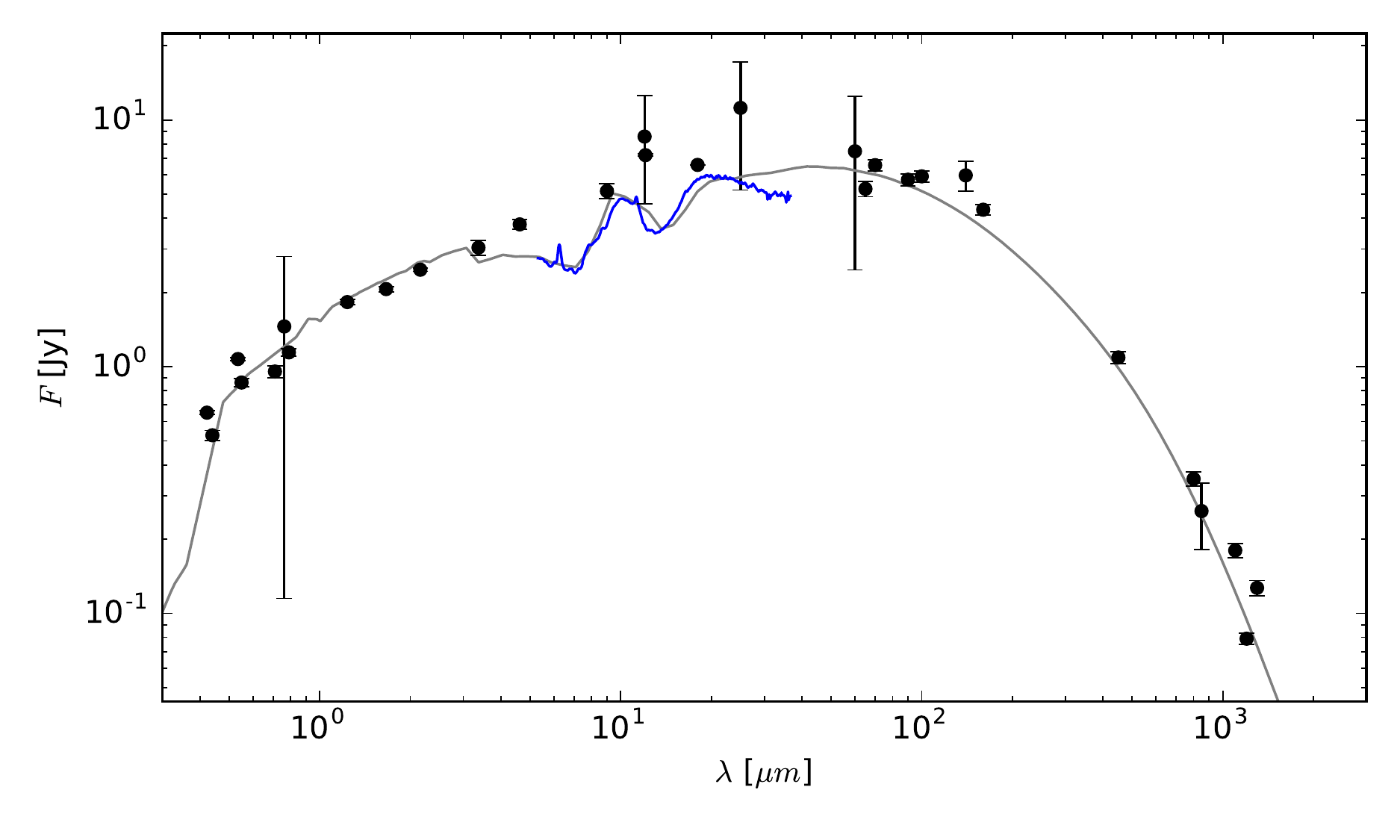}
  \includegraphics[width=0.47\textwidth]{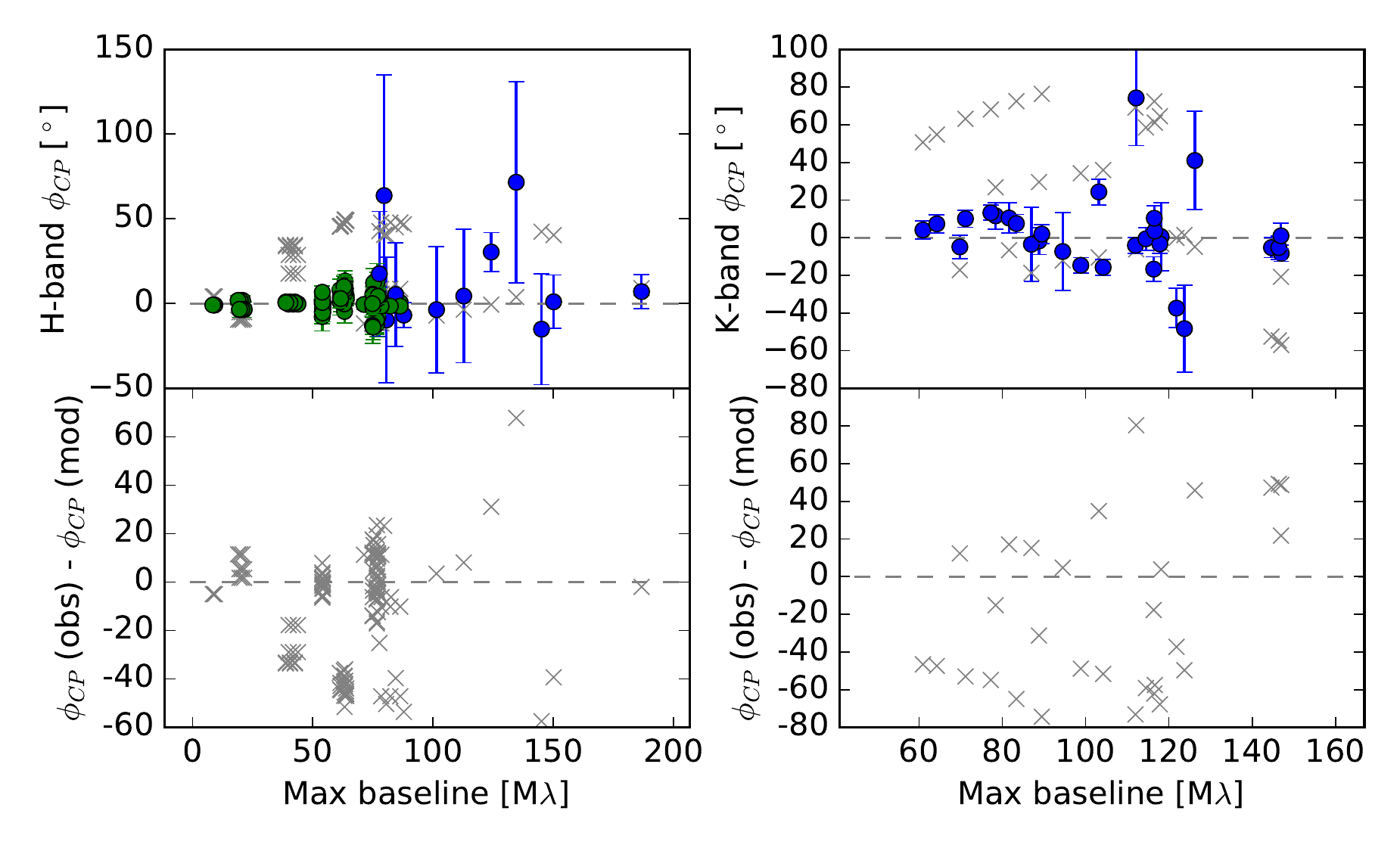}
  \includegraphics[width=0.47\textwidth]{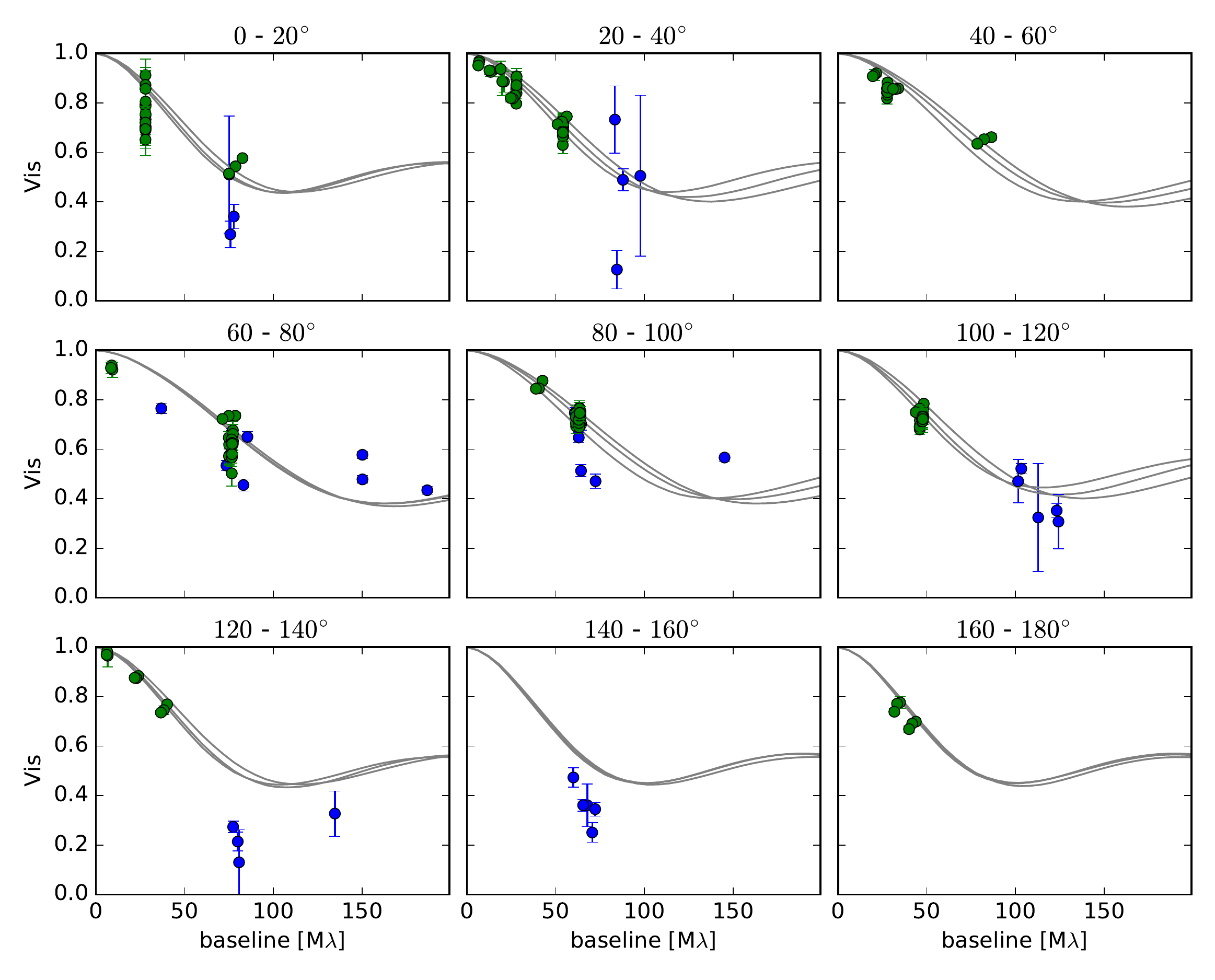}
  \includegraphics[width=0.47\textwidth]{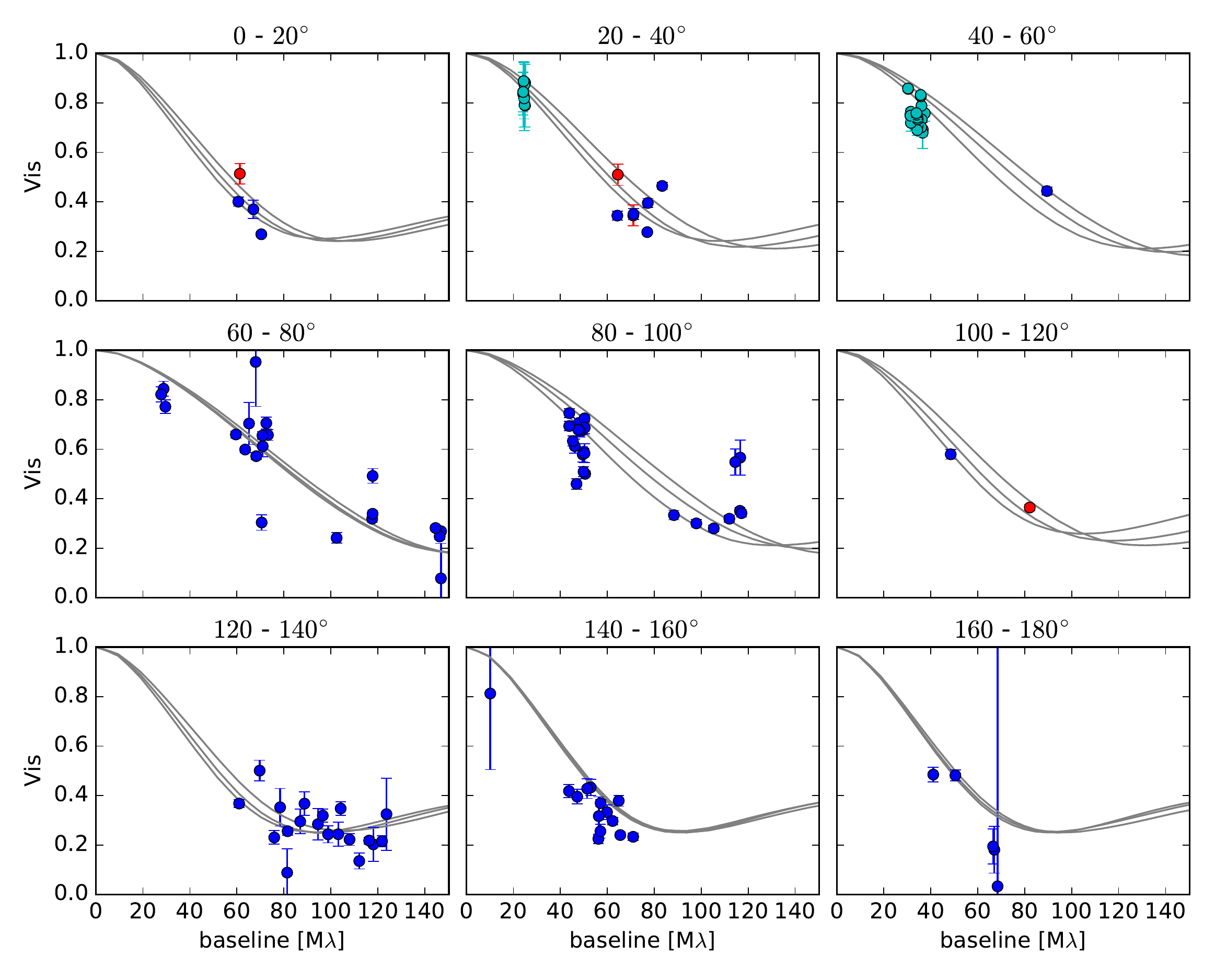}
   \caption{
   As Fig.~\ref{fig:IN05small} but for the S:large model providing the best fit to the observed SED ($h_{\rm{0,gas}}=7\,$au and $\beta=1.09$). 
   }
  \label{fig:IN05large}
\end{figure*}

As the NIR flux is well-approximated by the S:small model with $h_{0}=10\,$au and $\beta=1.06$, we examined the visibilities and $\phi_{\rm{CP}}$ for this model to inspect the rim position and shape. These are displayed in the remaining panels of Fig.~\ref{fig:IN05small}. Here, as in Figs.~\ref{fig:psringH} and \ref{fig:psringK}, the different panels show visibilities measured along different baseline position angles, PA$_{\rm{base}}$. In addition to providing too much flux across MIR-to-FIR wavelengths, we see in the bottom panels of Fig.~\ref{fig:IN05small} that the first lobe of the model visibility curves drop more rapidly at short baselines than measured by the data. As such, optically thick material is required to exist interior to the silicate dust sublimation rim location predicted by S:small models. This is consistent with results from analyses of NIR size-luminosity relations for Herbig Ae stars in which the size of the NIR-emitting region is controlled by the sublimation of larger grains ($\sim1\,\mu$m in size; e.g. \citealt{Monnier02}). 

\subsection{Models invoking grain growth}\label{sec:largegrains}
\begin{figure*}
  \centering
  \includegraphics[width=0.47\textwidth]{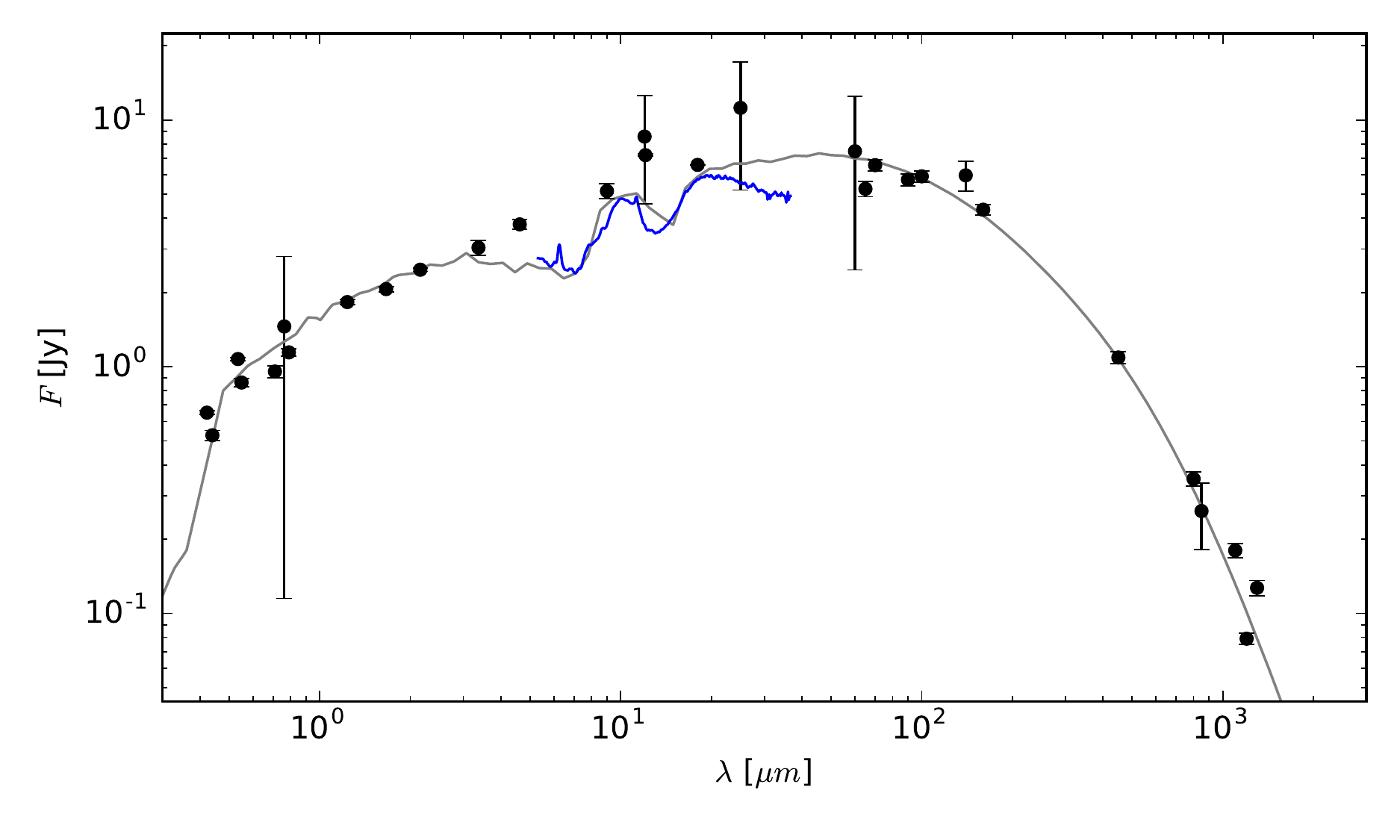}
  \includegraphics[width=0.47\textwidth]{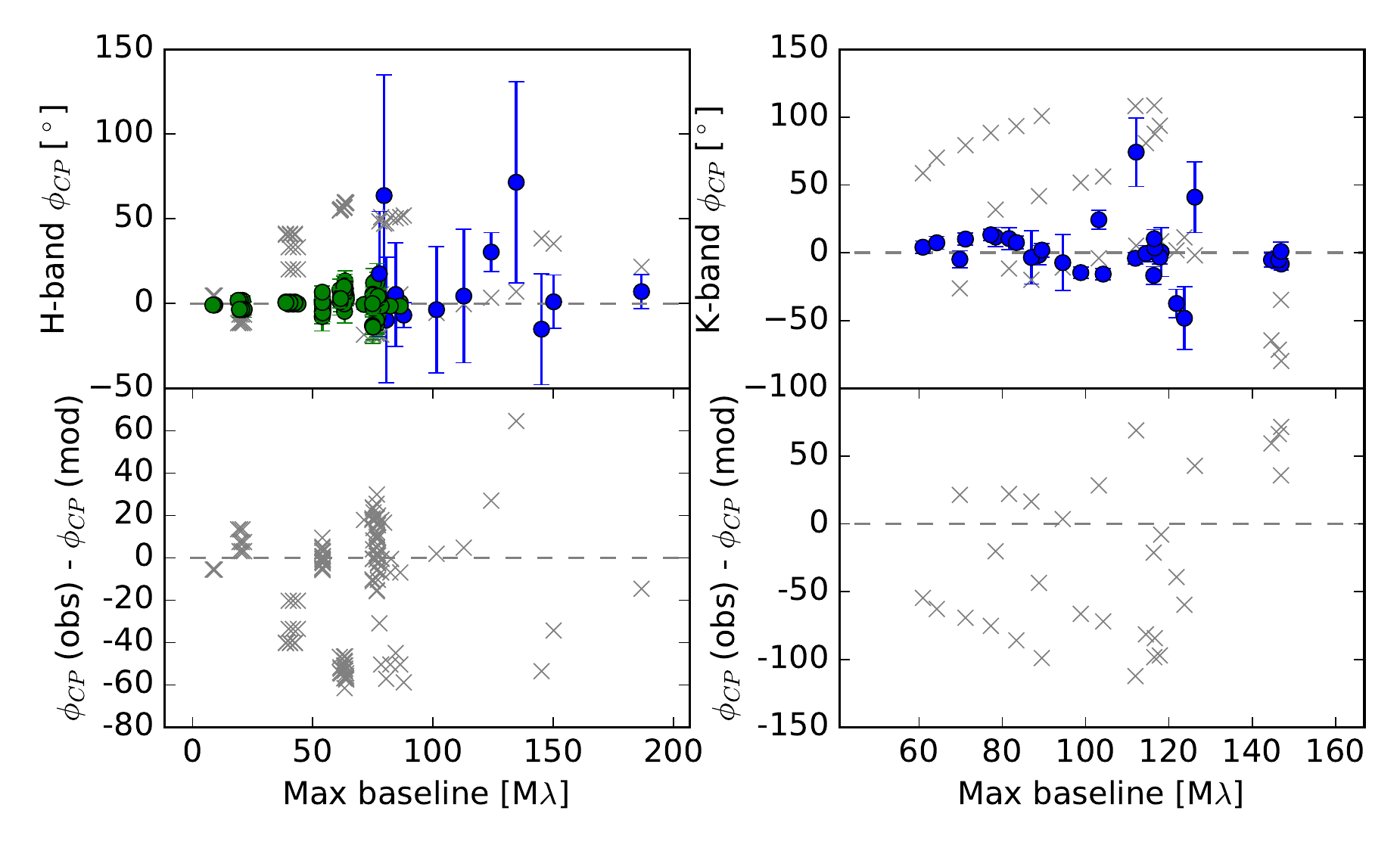}
  \includegraphics[width=0.47\textwidth]{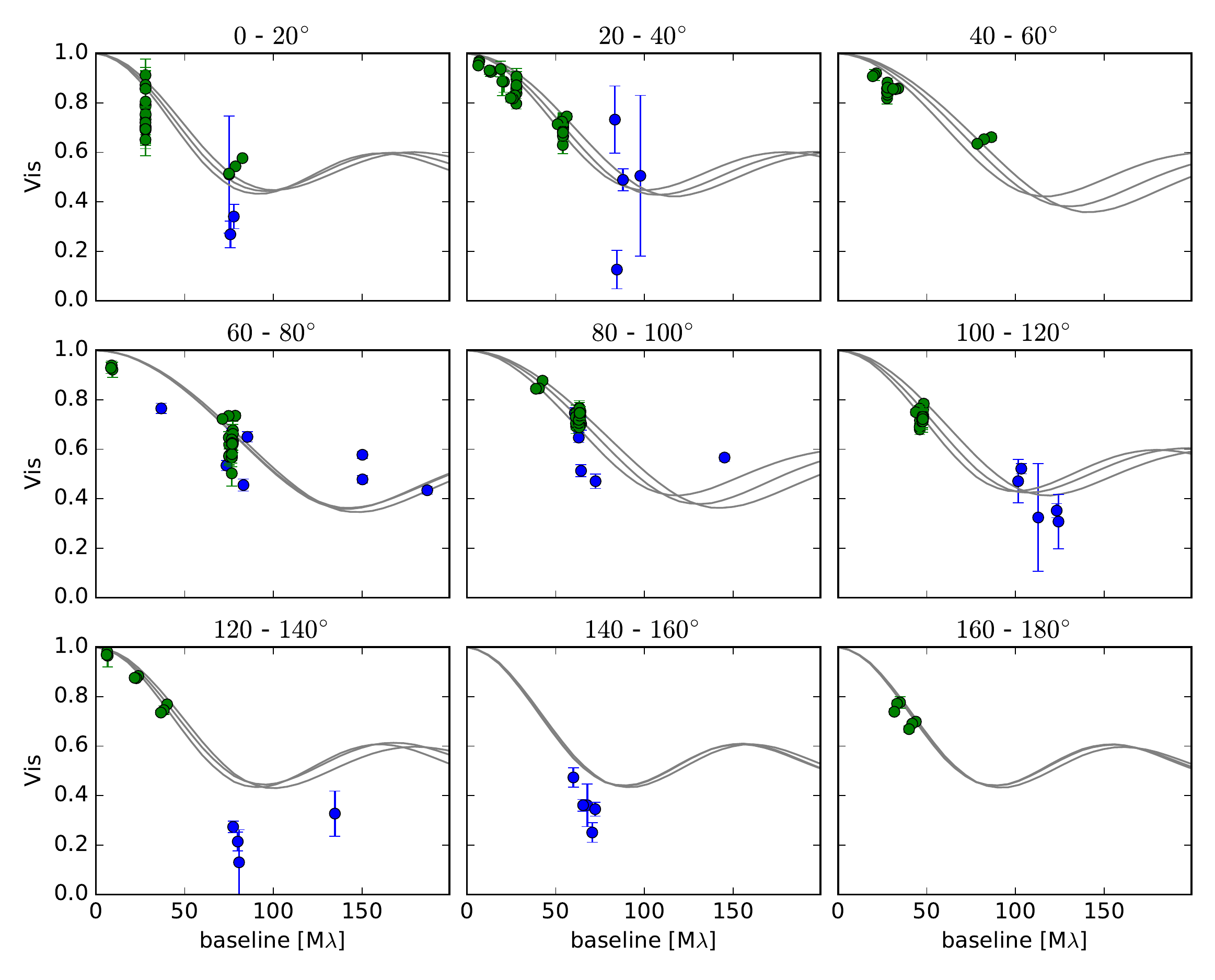}
  \includegraphics[width=0.47\textwidth]{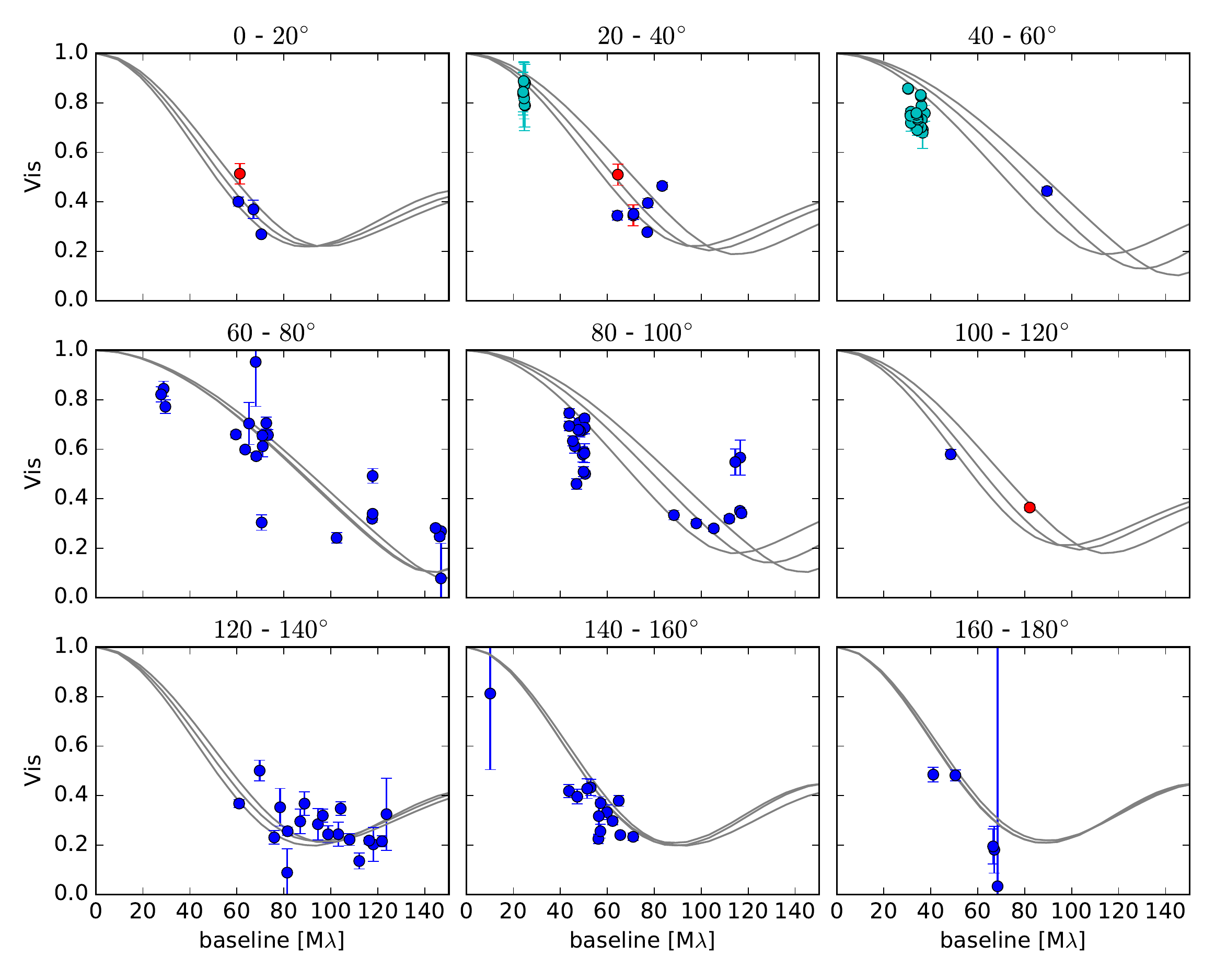}
   \caption{
   As Fig.~\ref{fig:IN05small} but for the THM07 model providing the best fit to the observed SED across NIR wavelengths ($h_{\rm{0,gas}}=8\,$au and $\beta=1.09$) . 
   }
  \label{fig:THM07}
\end{figure*}

Figs.~\ref{fig:IN05large} and \ref{fig:THM07} show the SED, visibilities and $\phi_{\rm{CP}}$ of the best-fitting S:large and THM07 models, respectively. The $\chi^{2}_{\rm{r}}$ fits to the SED ($\chi^{2}_{\rm{r,SED}}$), visibilities ($\chi^{2}_{\rm{r,vis}}$), and $\phi_{\rm{CP}}$ ($\chi^{2}_{\rm{r,CP}}$) are presented in Table~\ref{tab:chisq}. Models invoking grain growth to micron sizes clearly provide an improved fit to the SED and visibilities compared to the S:small models we explored. The best-fitting S:large and THM07 models are both able to provide a reasonable estimate of the fluxes in the SED across the full range of wavelengths probed. Though the SEDs provided by the S:large and THM07 models in Figs.~\ref{fig:IN05large} and \ref{fig:THM07} are broadly consistent with one another, the S:large model is able to reproduce the general shape of the \emph{Spitzer} spectrum out to $\sim12\,\mu$m better than the THM07 model. 

\begin{deluxetable}{lccrrr}
\tabletypesize{\scriptsize}
\tablecolumns{6}
\tablecaption{Parameters of the best-fitting TORUS models without vertical hydrostatic equilibrium imposed (adopting $i=58^{\circ}$ and PA$_{\rm{major}}=160^{\circ}$) and their $\chi^{2}_{\rm{r}}$ fits to the SED, visibilities, and closure phases ($362$, $337$, and $153$ degrees of freedom, respectively) \label{tab:chisq} }
\tablehead{\colhead{Model} & \colhead{$h_{\rm{0,gas}}$} & \colhead{$\beta$} & \colhead{$\chi^{2}_{\rm{r,SED}}$} & \colhead{$\chi^{2}_{\rm{r,vis}}$} & \colhead{$\chi^{2}_{\rm{r,CP}}$} }
\startdata
S:small & 10 & 1.06 & $109554$ & $74.3$ & $9.3$ \\
S:large &  7 & 1.09 & $1086$   & $14.4$ & $16.3$ \\
THM07   &  8 & 1.09 & $2029$   & $22.1$ & $18.2$
\enddata
\end{deluxetable}

\begin{figure*}
  \centering
  \includegraphics[width=0.40\textwidth]{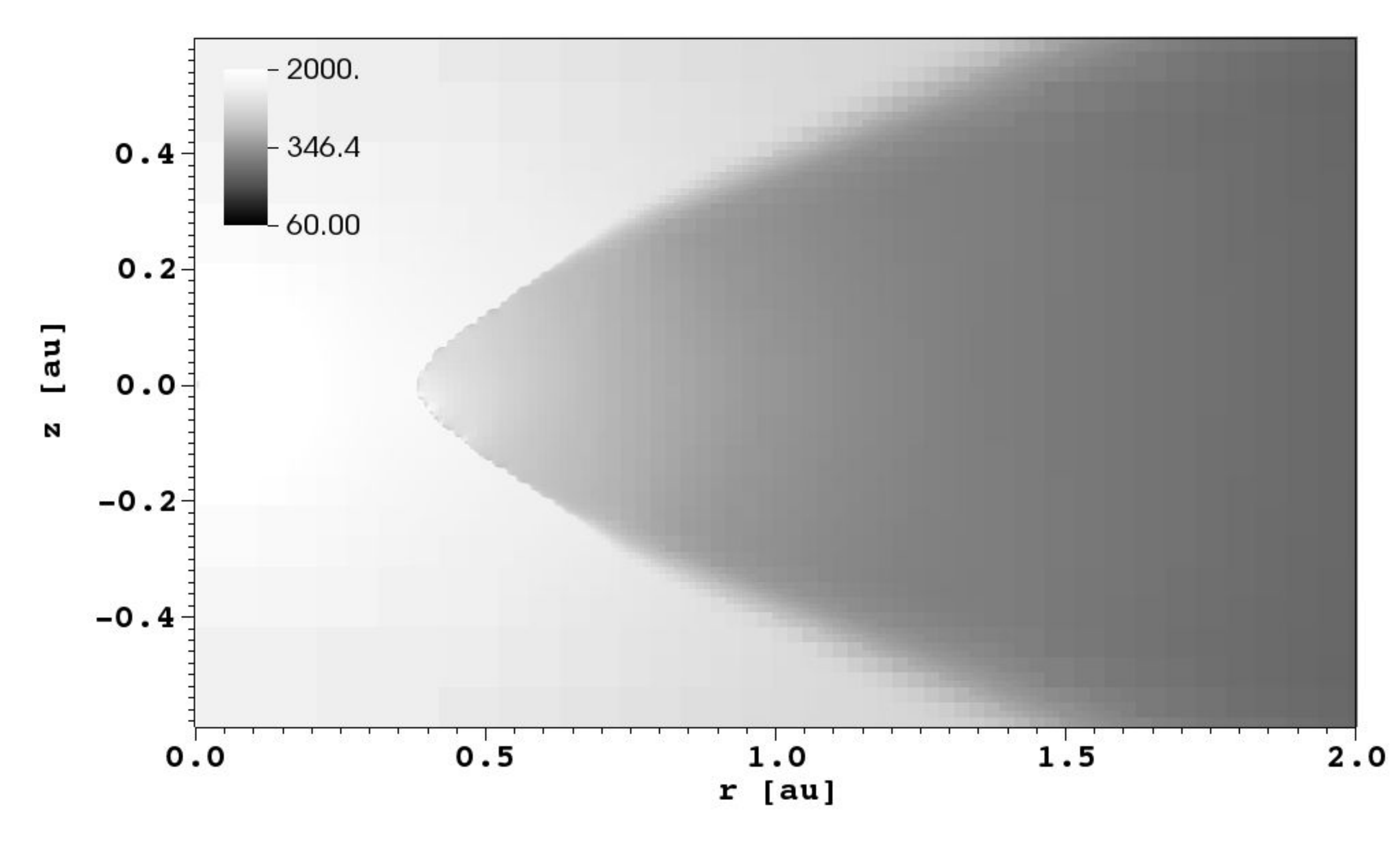}
  \includegraphics[width=0.29\textwidth]{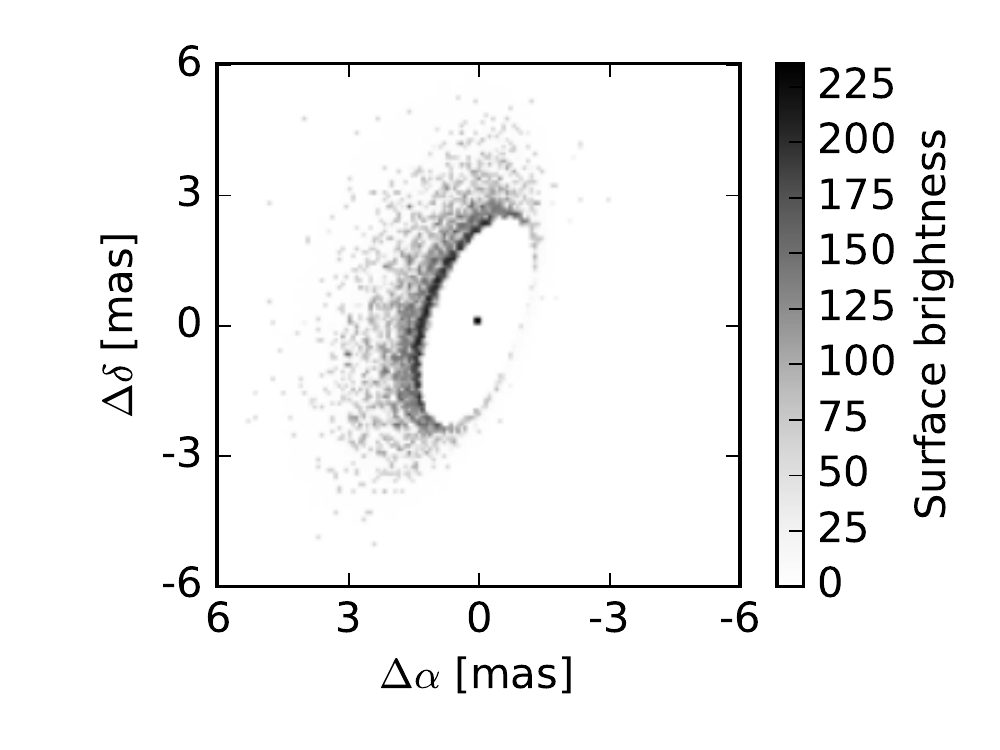}
  \includegraphics[width=0.29\textwidth]{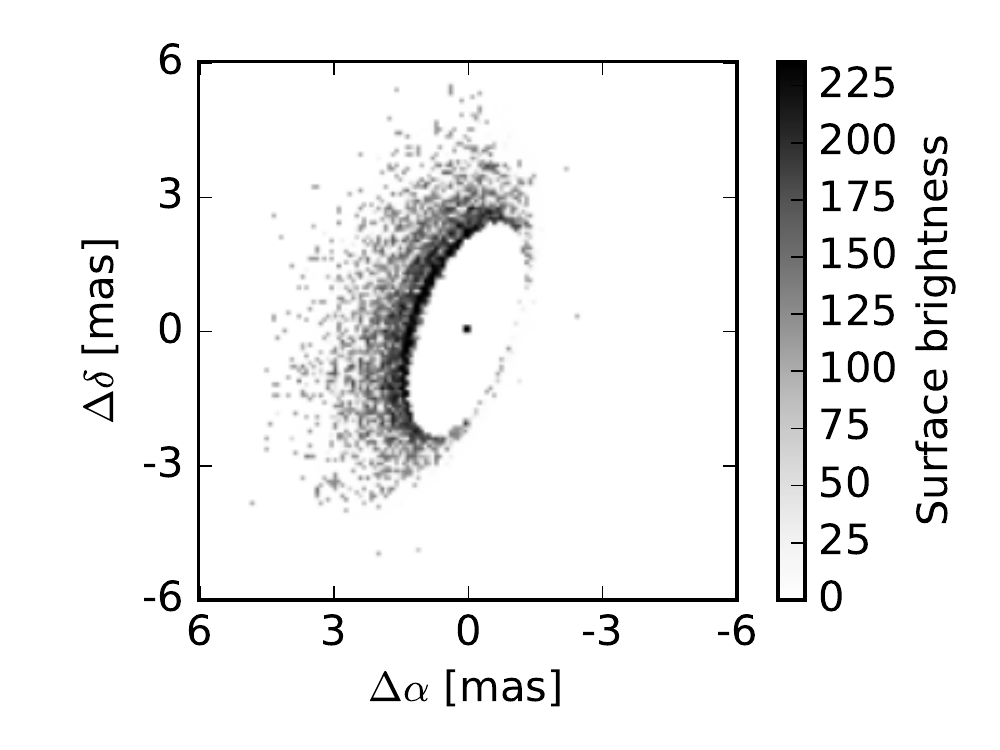}\\
  \includegraphics[width=0.40\textwidth]{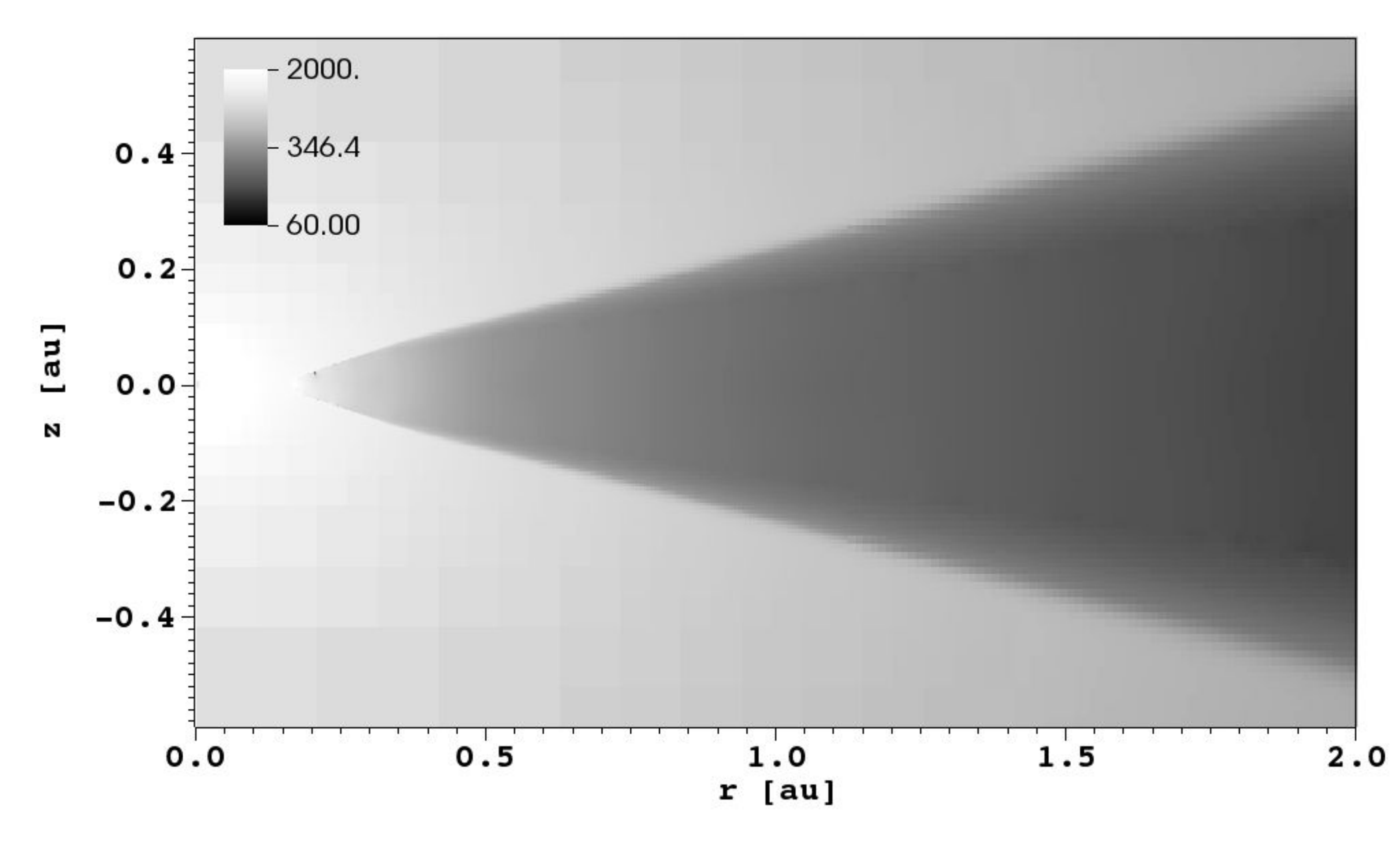}
  \includegraphics[width=0.29\textwidth]{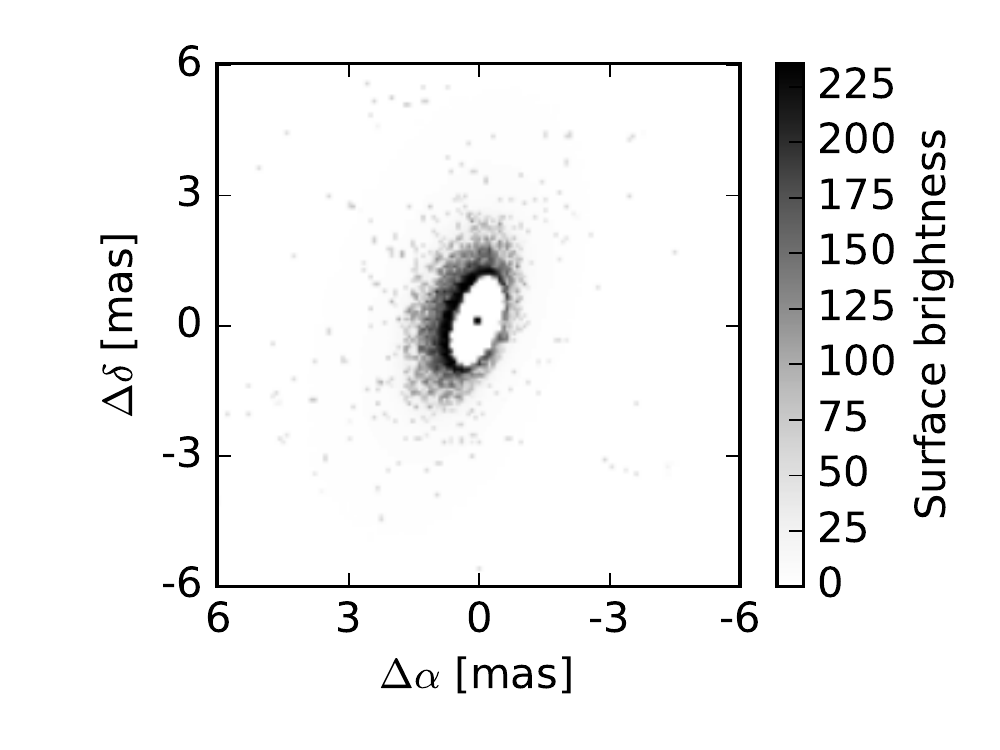}
  \includegraphics[width=0.29\textwidth]{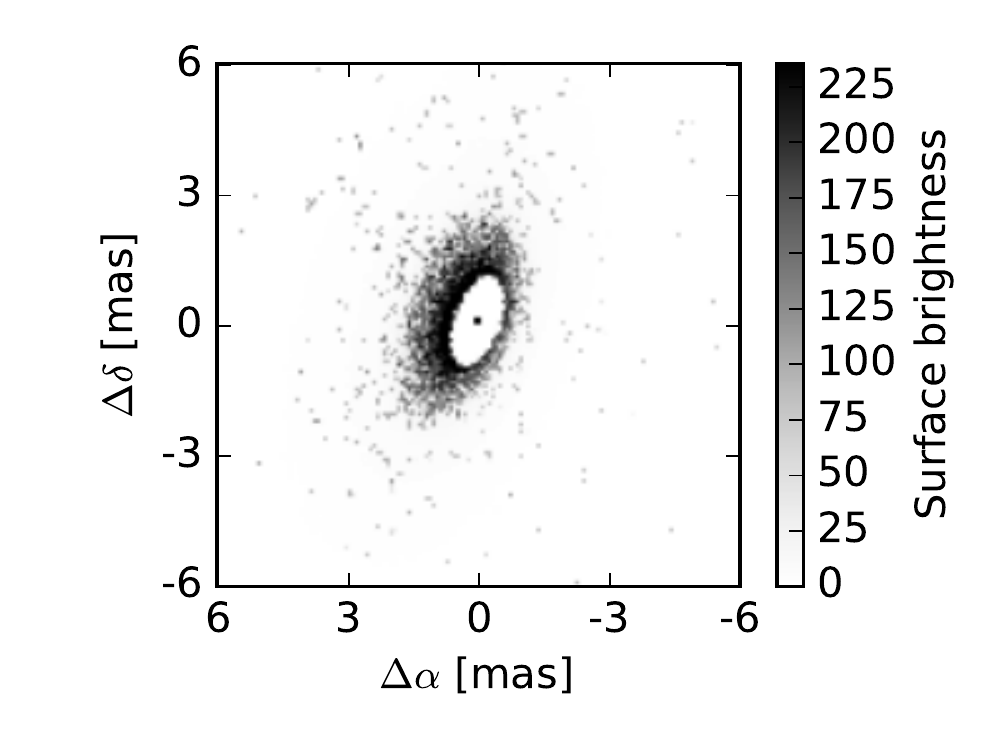}\\
  \includegraphics[width=0.40\textwidth]{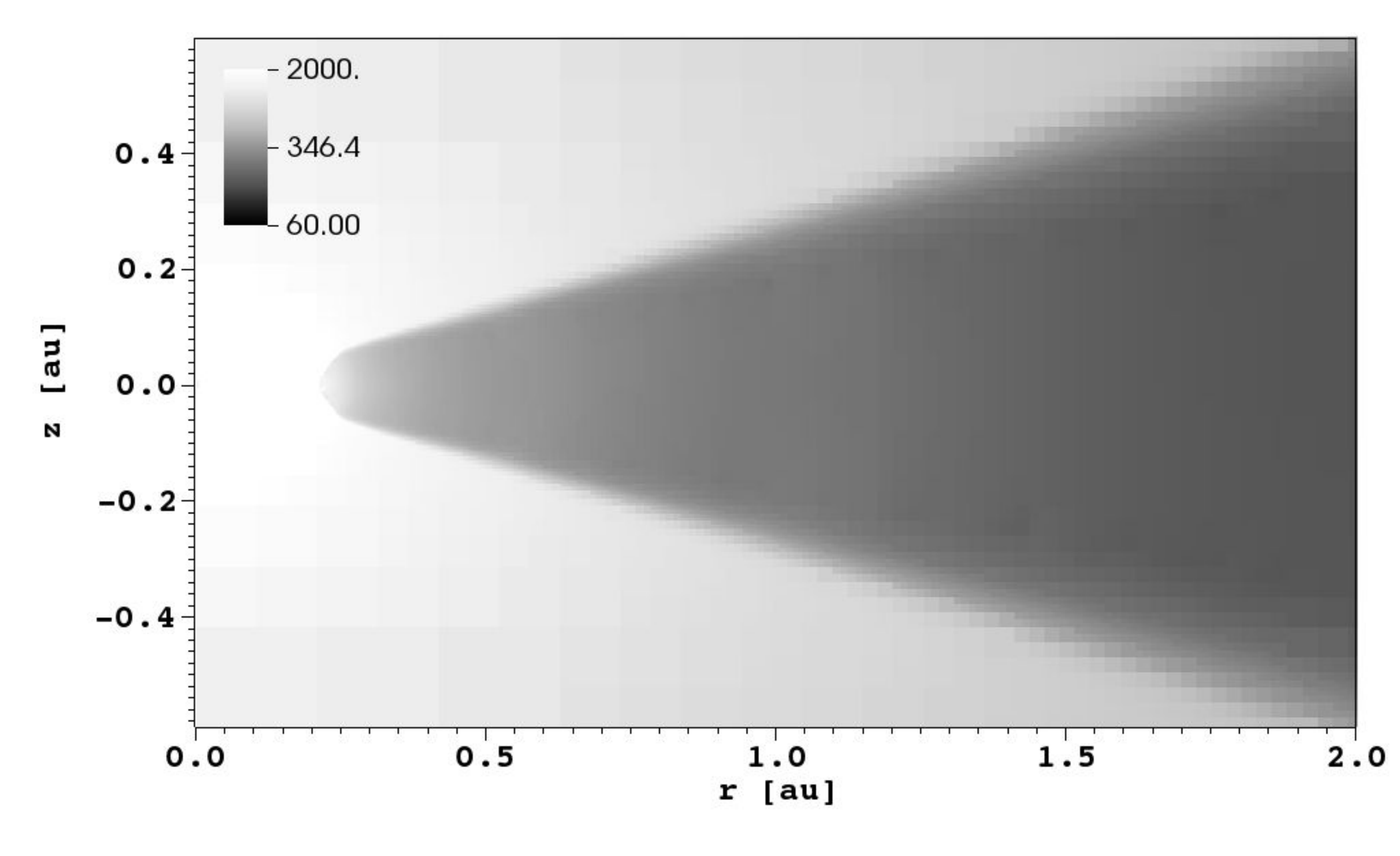}
  \includegraphics[width=0.29\textwidth]{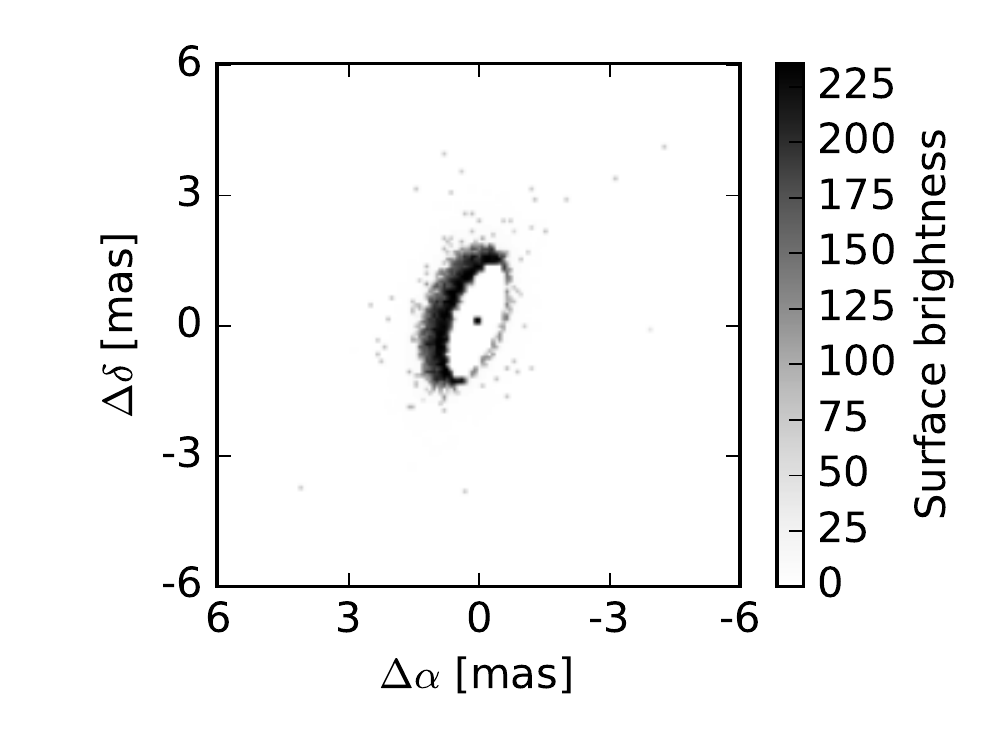}
  \includegraphics[width=0.29\textwidth]{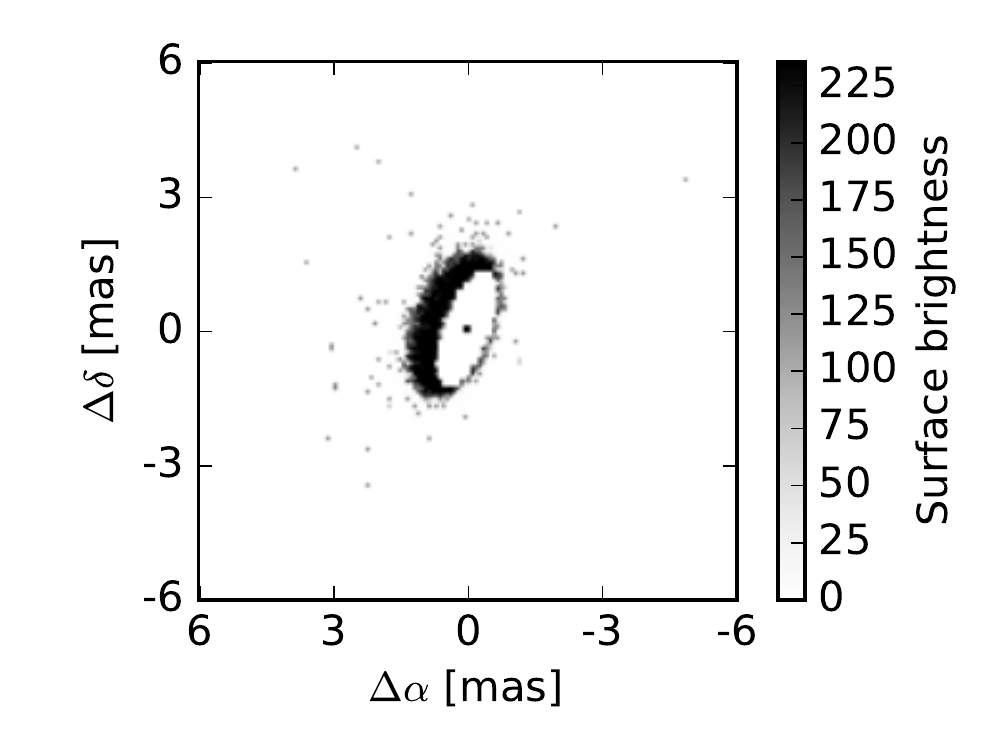}
   \caption{
   Cross-section of the disk temperature profile (in Kelvin; left hand panel) and model $H$- and $K$-band images (middle and right hand panels, respectively) output by TORUS for the S:small model with $h_{\rm{0,gas}}=10\,$au and $\beta=1.06$ (top row), the S:large model with $h_{\rm{0,gas}}=7\,$au and $\beta=1.09$ (middle row), and the THM07 model with $h_{\rm{0,gas}}=8\,$au and $\beta=1.09$ (bottom row). 
   }
  \label{fig:images}
\end{figure*}

The scale height and flaring parameters of the best-fitting models are broadly consistent:  $h_{\rm{0,gas}}=7\,$au and $\beta=1.09$ for S:large versus $h_{\rm{0,gas}}=8\,$au and $\beta=1.09$ for THM07. The S:small model which provided the best fit across the full wavelength range probed by the observed SED (while underestimating the NIR flux; see dashed grey line in the top left panel of Fig.~\ref{fig:IN05small}) also had $h_{\rm{0,gas}}=7\,$au and $\beta=1.09$. The differences in the scale heights required for the different models to produce the same NIR flux is consistent with what we saw for the models invoking vertical hydrostatic equilibrium (Section~\ref{sec:hydroresults}) whereby the different rim curvature prescriptions give rise to inner rims with varying radial extents. This is shown more clearly in Fig.~\ref{fig:images}: the rim produced by the S:small model is located further from the star than that of the S:large and THM07 models. The curvature of the rim produced by the THM07 model is also sharper than that produced by the S:large model. As a result, the NIR flux arises from a smaller range of disk radii than the S:large model. 

The best-fitting S:large and THM07 models provide similarly good fits to the observed $H$- and $K$-band visibilities with the S:large model providing a marginally better fit to the observed visibilities than the THM07 model (see Table~\ref{tab:chisq}). The first lobe of the model visibility curves is in good agreement with the data across most PA$_{\rm{base}}$, suggesting that the location of the inner disk rim of \object{HD\,142666} is consistent with the silicate sublimation region predicted by the models invoking grain growth to micron sizes. In Fig.~\ref{fig:images}, we see that the THM07 models predict a rim location which is slightly more extended than the S:large models while the S:large model is able to provide more flux in the south-west portion of the disk. 

Upon closer inspection, the visibilities and $\phi_{\rm{CP}}$ in Figs.~\ref{fig:IN05large} and \ref{fig:THM07} indicate additional complexity to the circumstellar component of the NIR emission which remains unexplained in our suite of models. In the visibility plots, the models appear under-resolved compared to the data along the apparent disk minor axis ($60-80^{\circ}$ PA$_{\rm{base}}$ panels) while the often significant ($\sim50-100^{\circ}$) $\phi_{\rm{CP}}$ signals predicted by the models are not present in the data. These discrepancies indicate the presence of additional material along the disk minor axis interior to the sky-projected location of the dust sublimation rim predicted by our models as well as a more centro-symmetric brightness distribution. We discuss this further in Section~\ref{sec:disc:geom}.

\section{Discussion}\label{sec:discussion}
\subsection{Grain growth in the disk of HD\,142666 and the inner rim location}
The results presented in Section~\ref{sec:RTresults} indicate that models in which the inner disk rim is dominated by small ($0.1\,\mu$m) grains are incompatible with the SED and NIR interferometry obtained for \object{HD\,142666}. Instead, models invoking the growth of dust grains to micron sizes provide improved fits to the observations. These results support those of \citet{vanBoekel03} who, in their analysis of the shape and strength of the silicate feature in the \emph{Spitzer} spectrum of \object{HD\,142666}, found strong evidence for growth from $0.1\,\mu$m to $2.0\,\mu$m grains with a mass ratio of 1:1.54 in favour of large grains. As MIR emission arises from the disk surface layers, and larger grains are expected to settle to lower scale heights in the disk \citep{Testi14}, the dominance of micron-sized grains in the disk midplane was anticipated to be even more pronounced. Our results support this idea as the models invoking the presence of larger, micron-sized grains (S:large model with $h_{\rm{0,gas}}=7\,$au and $\beta=1.09$ and THM07 mode with $h_{\rm{0,gas}}=8\,$au and $\beta=1.09$) are able to simultaneously reproduce the NIR portion of the SED, the shape and flux of the \emph{Spitzer} spectrum, and the observed $H$- and $K$-band visibilities.

The lower $\chi^{2}_{\rm{r,vis}}$ provided by the S:large model fit to the visibilities compared to the THM07 model (see Table~\ref{tab:chisq}) further suggests that the inner disk rim of \object{HD\,142666} is more consistent with models invoking a gas density-dependent dust sublimation temperature \citep[e.g.][]{Isella05} than those invoking constant dust sublimation temperatures where rim curvature arises due to the relative abundance of different grain sizes (in $r$ and $z$) and their relative cooling efficiencies \citep[e.g.][]{Tannirkulam07}. However, it should be noted that using (i) grains $<1.2\,\mu$m as the larger grains, (ii) a different size for the smaller grains, (iii) a different value for $h_{\rm{0,dust}}$ for the larger grains, and/or (iv) a different silicate sublimation temperature (see Table~\ref{tab:RTmodels}) would all affect the rim shape, location and temperature structure predicted by the THM07 models. The parameters we adopted in our THM07 models were chosen for their consistency with the original \citet{Tannirkulam07} study and a comprehensive evaluation of the impact of these variables is beyond the scope of this paper. However, our use of $1.2\,\mu$m-sized grains as the ``large'' grains should produce inner rim locations close to the lower limit allowed by the \citet{Tannirkulam07} and \citep{Isella05} models. This is because silicate grains larger than $\sim1.3\,\mu$m do not significantly contribute to the dust opacity and thus their inclusion would not make the rim any more compact \citep{Isella05}. 

Of the parameters explored herein, our best model (the S:large model with $h_{\rm{0,gas}}=7\,$au and $\beta=1.09$) produces a sublimation rim that remains optically thick down to within $0.17\,$au of the star (in the disk midplane). This is broadly consistent with the results of our geometric fitting (Section~\ref{sec:geoFitResults}) in which the characteristic radii of the $H$- and $K$-band-emitting regions were found to be $\sim0.22-0.24\,$au in both the PS+R and PS+SR fits. These inner radii are lower than previously published estimates by \citet{Monnier05} and \citet{Schegerer13} based on short-baseline NIR and MIR visibilities ($\sim0.38-0.39\,$au, accounting for differences in the adopted distance to \object{HD\,142666}) but consistent with those in \citet[][$0.19-0.23\,$au]{Vural14}. However, we note that the adopted stellar parameters (including the stellar flux contribution) are not consistent across these studies nor between these studies and our own. As discussed in \citet{Lazareff16}, the characteristic size of the emitting region and the circumstellar flux contribution are intrinsically linked in the visibility so it is understandable that differences in one parameter will lead to differences in the other when comparing studies. 

\subsection{Indicators of additional complexity in the NIR-emitting region}\label{sec:disc:geom}
Throughout our radiative transfer analysis (Section~\ref{sec:RTresults}), we first required our TORUS models to reproduce the observed SED before assessing the fit to the interferometry. In this way, we assume that the disk in our TORUS models accounts for all the NIR circumstellar flux. If additional NIR-emitting gaseous material exists interior to the sublimation rim \citep{Tannirkulam08a, Tannirkulam08b} and/or a dusty outflow exists \citep{Alexander07, Bans12} -- neither of which are accounted for in our models -- they will also contribute to the observed $H$- and $K$-band flux. 

Additionally, in our geometric modeling (Section~\ref{sec:geoFitResults}), we assumed all the circumstellar NIR flux could be fit using a Gaussian-smoothed ring model and, from this, estimated a disk major axis position angle and inclination of $160^{\circ}$ and $58^{\circ}$, respectively. While this viewing geometry agrees with previous assessments of the disk inclination ($40-60^{\circ}$ \citealt{Dominik03}, \citealt{Vural14}, \citealt{Lazareff16}, \citealt{Rubinstein18}) and position angle ($\sim140-180^{\circ}$ \citealt{Garufi17}, \citealt{Rubinstein18}), indirect evidence for further model complexity is suggested in the visibility and $\phi_{\rm{CP}}$ residuals. As stated in Section~\ref{sec:largegrains}, while the models invoking grain growth to micron sizes provide a good fit to the visibilities across a wide range of PA$_{\rm{base}}$, the model visibility curves appear under-resolved compared to the data along the apparent disk minor axis. In addition, the significant ($\sim50-100^{\circ}$) $\phi_{\rm{CP}}$ signals predicted by our best-fitting TORUS models are not present in the data, indicating the true brightness distribution is more centro-symmetric. 

\begin{figure}
  \centering
  \includegraphics[width=0.47\textwidth]{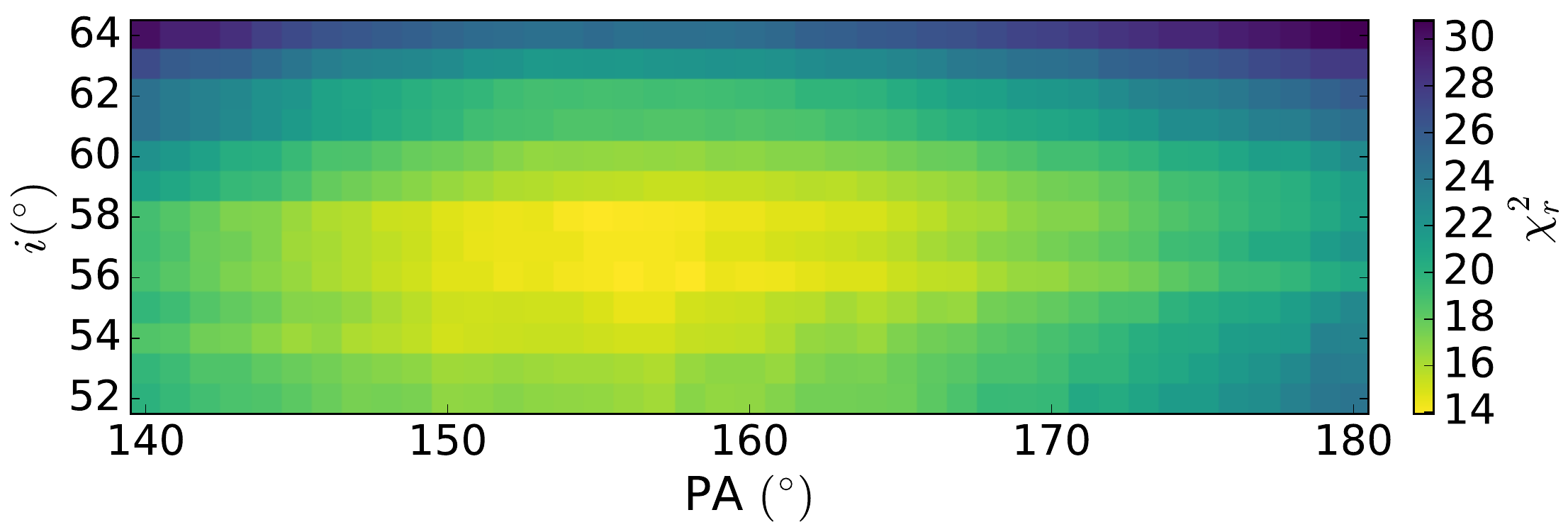}
   \caption{
   $\chi^{2}_{\rm{r}}$ map showing the comparative goodness-of-fit provided by the S:large model with $h_{\rm{0,gas}}=7\,$au and $\beta=1.09$ to the $H$- and $K$-band visibilities using disk inclinations, $52^{\circ}\leq i\leq 64^{\circ}$, and position angles, $140^{\circ}\leq$PA$_{\rm{major}}\leq 180^{\circ}$.
   }
  \label{fig:pa_inc}
\end{figure}

Further indirect evidence of additional model complexity is found when considering the UX~Ori-type phenomena displayed by \object{HD~142666} \citep{Meeus98, Zwintz09}. This type of variability is associated with line-of-sight fluctuations in opacity and is typically attributed to circumstellar disk occultation \citep{Grinin91, Natta97} although unsteady accretion \citep{Herbst99} and/or the existence of dusty outflows \citep{Vinkovic07, Tambovtseva08} have been proposed as alternative causes. As the disk- and outflow-based origins require intermediate-to-high disk inclinations for line-of-sight occultations to arise, and the inferred disk inclination of $58^{\circ}$ for \object{HD~142666} is relatively low compared to the $\sim70^{\circ}$ inferred for other UX~Ori stars (VV~Ser, KK~Oph, and UX~Ori itself; \citealt{Pontoppidan07, Kreplin13, Kreplin16}), the photometric variability observed for HD~142666 may suggest that the disk is more inclined. Alternatively, the UX~Ori variability may indicate that azimuthal and temporal variations in disk scale height exist. 

To investigate whether the residuals in the interferometry fits could be reconciled solely by changing the disk viewing geometry, we explored whether a better fit to the visibilities could be achieved if our best-fit S:large TORUS model ($h_{\rm{0,gas}}=7\,$au and $\beta=1.09$) was observed at differing viewing geometries: $52^{\circ}\leq i\leq 64^{\circ}$ and $140^{\circ}\leq$\,\,PA$_{\rm{major}}\leq180^{\circ}$. The resulting $\chi^{2}_{\rm{r}}$ map is shown in Fig.~\ref{fig:pa_inc}. At the original viewing geometry (PA$_{\rm{major}}=160^{\circ}$ and $i=58^{\circ}$), the model provides $\chi^{2}_{\rm{r}}=14.4$. Over the range of inclinations and position angles probed, the model with $i=58^{\circ}$ and PA$_{\rm{major}}=155^{\circ}$ provides the best fit to the visibilities but the improvement in $\chi^{2}_{\rm{r}}$ is small: $\chi^{2}_{\rm{r,min}}=13.9$. This revised disk viewing geometry unsurprisingly still produces model visibilities which are under-resolved along the apparent disk minor axis and model $\phi_{\rm{CP}}$ signals in excess of those observed. As such, the residuals in our TORUS model fitting cannot be explained simply be changing the viewing geometry and instead point to additional model complexity.

The disk models we have explored with TORUS assume azimuthal symmetry: we have not accounted for the possible presence of azimuthal variations of the disk scale height (i.e. disk warps). The disk of \object{HD\,142666} is not strongly flared (Section~\ref{sec:RTresults}; c.f.\ \citealt{Meeus01}) and, as such, disk regions at large distances from the star are unlikely to provide line-of-sight stellar occultations when observed at an inclination of $58^{\circ}$ (see Fig.~\ref{fig:images}). Assuming that optically thick material only exists exterior to the dust rim location predicted by the best-fit S:large TORUS model, a disk inclination of $58^{\circ}$ requires azimuthal scale height increases of around $40\%$ in the inner disk for direct line-of-sight occultation. The periods of minimum brightness observed for \object{HD\,142666} last for a maximum of $\sim2-3\,$days \citep{Zwintz09}. Comparing this to the orbital timescale at the inner disk rim ($18.2\,$days), these scale height variations would be required to extend over a maximum of $\sim10-15\%$ of the disk circumference. Furthermore, as the photometric variability is aperiodic, the scale height variations would have to rise and fall on timescales within the $\sim18.2\,$days orbital period. Taking this all into account, the $58^{\circ}$ disk inclination inferred for \object{HD~142666} appears inconsistent with a disk-based origin for the UX~Ori phenomena. In light of this, and the fact that the visibilities of the best-fit S:large TORUS model appear under-resolved along baseline position angles that probe the disk minor axis, it seems likely that either the disk is inclined at $>58^{\circ}$ or that the UX~Ori phenomena observed for HD~142666 is attributed to an outflow component of variable optical depth which is oriented perpendicular to the disk midplane. In both cases, additional NIR emitting material exterior to the flared disk we have considered here is required. 

\section{Summary}\label{sec:conclusion}
We have used geometric and radiative transfer modeling to explore the shape and structure of the inner rim of the disk of \object{HD\,142666}. Our results are summarized as follows:

\begin{itemize}

\item Fitting geometric models in which all the circumstellar emission arises from the innermost regions of a disk to the $H$- and $K$-band visibilities suggest a viewing geometry for HD~142666 of $i=58^{\circ}$ from face-on and major axis position angle of $160^{\circ}$ east of north. These values agree with previous interferometric modeling \citep{Vural14, Lazareff16}, VLT/NACO imaging \citep{Garufi17}, and ALMA cycle 2 observations \citep{Rubinstein18} of the object. This viewing geometry was adopted in all our TORUS radiative transfer modeling.

\item The TORUS radiative transfer models we explore which invoke vertical hydrostatic equilibrium in the circumstellar disk are unable to reproduce the SED of \object{HD\,142666}. This is consistent with previous results from radiation hydrodynamic and radiation hydrostatic modeling by \citet{Mulders12} and \citet{Flock16a} for other Herbig~Ae/Be stars. Using a series of TORUS models without vertical hydrostatic equilibrium invoked, we further investigated whether the inflation of the inner disk to greater scale heights, induced by turbulence arising from MRI \citep{Turner14, Flock16}, for example, could reproduce the observed SED. Among the models we explored, we found that those in which small grains ($0.1\,\mu$m in size) are the largest grains in the inner disk rim and thus determine the rim location and shape are unable to simultaneously fit the NIR, MIR and FIR portions of the SED. Instead, we found that our models required the presence of silicate dust of at least micrometer size to be present in the disk rim to be able to reproduce the SED across the full optical-to-millimeter wavelength range. 

\item TORUS models invoking the existence of micron grains were also found to provide an improved fit to the NIR visibilities compared to the models including only small grains. This is consistent with the original study of the grain size-dependence of the silicate dust destruction radius around Herbig Ae/Be stars by \citet{Monnier02}. Furthermore, we found that models in which rim curvature arises due to the dependence of the dust sublimation temperature on the local gas density \citep{Isella05} provide improved fits to the visibilities ($\chi^{2}_{\rm{r,vis}}=14.4$) compared to those in which the rim curvature arises from grain growth induced settling (the THM07 models; \citealt{Tannirkulam07}, $\chi^{2}_{\rm{r,vis}}=22.1$). In particular, the model providing the best fit to the SED and visibilities is the S:large model with disk scale height, $h_{\rm{0,gas}}=7\,$au and flaring parameter, $\beta=1.09$. A slight improvement to the fit for the S:large model is found using a viewing geometry of $i=58^{\circ}$ and PA$_{\rm{major}}=155^{\circ}$ ($\chi^{2}_{\rm{r}}=13.94$ assuming $337$ degrees of freedom).

\item The visibility and closure phase residuals in the best-fit S:large TORUS model point to the presence of additional complexity to the NIR emitting region which is unaccounted for in our TORUS models. The model closure phase signals are overestimated, indicating the emission is more centro-symmetric while the model visibilities are under-resolved along position angles tracing the apparent disk minor axis. In addition, we argue that the inclination we infer for HD~142666 is inconsistent with a disk-based origin for its UX~Ori-type variability. This is further indication for the requirement of additional model complexity. Additional optically thick material present in dusty disk winds \citep{Alexander07, Bans12} and/or the gaseous disk material interior to the dust sublimation rim \citep{Tannirkulam08a, Tannirkulam08b}, for example, appear to be required. 

\end{itemize}

\acknowledgments
C.L.D., S.K., A.K. and A.L. acknowledge support from the ERC Starting Grant ``ImagePlanetFormDiscs'' (Grant Agreement No. 639889), STFC Rutherford fellowship/grant (ST/J004030/1, ST/K003445/1) and Philip Leverhulme Prize (PLP-2013-110). 
J.D.M., F.B., and B.K. acknowledge support from NSF grants AST-1210972 and AST-1506540.
We would like to thank Bernard Lazareff, Jean-Baptiste Le Bouquin and Rachel Akeson for their assistance in acquiring archival data for \object{HD\,142666}. 
This work is based upon observations obtained with the Georgia State University Center for High Angular Resolution Astronomy Array at Mount Wilson Observatory. The CHARA Array is supported by the National Science Foundation under Grant No.\ AST-1211929. Institutional support has been provided from the GSU College of Arts and Sciences and the GSU Office of the Vice President for Research and Economic Development.
The calculations for this paper were performed on the University of Exeter Supercomputer, a DiRAC Facility jointly funded by STFC, the Large Facilities Capital Fund of BIS, and the University of Exeter.
This research has made use of: 
the NASA/IPAC Infrared Science Archive, which is operated by the Jet Propulsion Laboratory, California Institute of Technology, under contract with the National Aeronautics and Space Administration; 
the Keck Observatory Archive (KOA), which is operated by the W. M. Keck Observatory and the NASA Exoplanet Science Institute (NExScI), under contract with the National Aeronautics and Space Administration; the Jean-Marie Mariotti Center \texttt{OiDB} service\footnote{Available at http://oidb.jmmc.fr }; 
NASA's Astrophysics Data System;
the VizieR catalogue access tool, CDS, Strasbourg, France; 
the SIMBAD database, operated at CDS, Strasbourg, France; 
NumPy \citep{van2011numpy};
matplotlib, a Python library for publication quality graphics \citep{Hunter07};
Astropy, a community-developed core Python package for Astronomy \citep{astropy};
the Jean-Marie Mariotti Center \texttt{SearchCal} service\footnote{Available at http://www.jmmc.fr/searchcal}
co-developped by LAGRANGE and IPAG, and of CDS Astronomical Databases SIMBAD and VIZIER \footnote{Available at http://cdsweb.u-strasbg.fr/}.
This work has made use of services produced by the NASA Exoplanet Science Institute at the California Institute of Technology. 




\facility{VLTI, CHARA, Keck}.
\software{TORUS \citep{Harries00,Harries04,Kurosawa06,Tannirkulam07}, pysynphot \citep{pysynphot}, NumPy \citep{van2011numpy}, matplotlib \citep{Hunter07}, Astropy \citep{astropy}}



\appendix

\section{Multi-band photometry}\label{apen:phot}
The photometry used to build the SED for \object{HD\,142666} are listed in Table~\ref{tab:phot:HD}. These have been flux-converted, where necessary, using central wavelengths and zero-point magnitudes from \citet{Mann15}, \citet{Cutri03}, and \citet{Cutri12}.

\begin{deluxetable}{ccl}
\tabletypesize{\scriptsize}
\tablecolumns{3}
\tablecaption{Photometry for HD142666 \label{tab:phot:HD} }
\tablehead{
\colhead{$\lambda$} & \colhead{Flux}   & \colhead{Reference} \\
\colhead{($\mu$m)}     & \colhead{(Jy)}     & \colhead{} }
\startdata
$0.422$ & $0.65\pm0.012$ & \citet{Hog00}\\
$0.44$ & $0.53\pm0.024$ & \citet{Tannirkulam08b}\\
$0.535$ & $1.07\pm0.02$ & \citet{Hog00}\\
$0.55$ & $0.86\pm0.03$ & \citet{Tannirkulam08b}\\
$0.71$ & $0.96\pm0.05$ & \citet{Tannirkulam08b}\\
$0.7625$ & $1.46\pm1.34$ & \citet{Zacharias13}\\
$0.79$ & $1.14\pm0.04$ & \citet{Tannirkulam08b}\\
$1.235$ & $1.83\pm0.04$ & \citet{Ita10}\\
$1.662$ & $2.06\pm0.05$ & \citet{Ita10}\\
$2.159$ & $2.47\pm0.04$ & \citet{Ita10}\\
$3.368$ & $3.04\pm0.20$ & \citet{Cutri12cat}\\
$4.618$ & $3.78\pm0.17$ & \citet{Cutri12cat}\\
$9.0$ & $5.15\pm0.36$ & \citet{Ishihara10}\\
$12.0$ & $8.57\pm4.00$ & \citet{Helou88}\\
$12.082$ & $7.20\pm0.07$ & \citet{Cutri12cat}\\
$18.0$ & $6.58\pm0.01$ & \citet{Ishihara10}\\
$25.0$ & $11.20\pm6.00$ & \citet{Helou88}\\
$60.0$ & $7.47\pm5.00$ & \citet{Moshir90}\\
$65.0$ & $5.26\pm0.37$ & \citet{Yamamura10}\\
$70.0$ & $6.56\pm0.33$ & \citet{Pascual16}\\
$90.0$ & $5.73\pm0.32$ & \citet{Yamamura10}\\
$100.0$ & $5.91\pm0.30$ & \citet{Pascual16}\\
$140.0$ & $5.97\pm0.83$ & \citet{Yamamura10}\\
$160.0$ & $4.33\pm0.22$ & \citet{Pascual16}\\
$450.0$ & $1.09\pm0.06$ & \citet{Sylvester96}\\
$800.0$ & $0.35\pm0.02$ & \citet{Sylvester96}\\
$850.0$ & $0.26\pm0.08$ & \citet{diFrancesco08}\\
$1100.0$ & $0.18\pm0.01$ & \citet{Sylvester96}\\
$1200.0$ & $0.079\pm0.004$ & \citet{Natta04}\\
$1300.0$ & $0.127\pm0.009$ & \citet{Sylvester96}\\
$3100.0$ & $0.013\pm0.001$ & \citet{Natta04}\\
$3300.0$ & $0.011\pm0.001$ & \citet{Natta04}
\enddata
\end{deluxetable}

\vspace{1cm}

\bibliography{hd142666}

\begin{thebibliography}{}
\expandafter\ifx\csname natexlab\endcsname\relax\def\natexlab#1{#1}\fi
\providecommand{\url}[1]{\href{#1}{#1}}
\providecommand{\dodoi}[1]{doi:~\href{http://doi.org/#1}{\nolinkurl{#1}}}
\providecommand{\doeprint}[1]{\href{http://ascl.net/#1}{\nolinkurl{http://ascl.net/#1}}}
\providecommand{\doarXiv}[1]{\href{https://arxiv.org/abs/#1}{\nolinkurl{https://arxiv.org/abs/#1}}}

\bibitem[{{Akeson} {et~al.}(2000){Akeson}, {Ciardi}, {van Belle},
  {Creech-Eakman}, \& {Lada}}]{Akeson00}
{Akeson}, R.~L., {Ciardi}, D.~R., {van Belle}, G.~T., {Creech-Eakman}, M.~J.,
  \& {Lada}, E.~A. 2000, \apj, 543, 313, \dodoi{10.1086/317111}

\bibitem[{{Alexander} \& {Armitage}(2007)}]{Alexander07}
{Alexander}, R.~D., \& {Armitage}, P.~J. 2007, \mnras, 375, 500,
  \dodoi{10.1111/j.1365-2966.2006.11341.x}

\bibitem[{{Andrews} {et~al.}(2013){Andrews}, {Rosenfeld}, {Kraus}, \&
  {Wilner}}]{Andrews13}
{Andrews}, S.~M., {Rosenfeld}, K.~A., {Kraus}, A.~L., \& {Wilner}, D.~J. 2013,
  \apj, 771, 129, \dodoi{10.1088/0004-637X/771/2/129}

\bibitem[{{Astropy Collaboration} {et~al.}(2013){Astropy Collaboration},
  {Robitaille}, {Tollerud}, {Greenfield}, {Droettboom}, {Bray}, {Aldcroft},
  {Davis}, {Ginsburg}, {Price-Whelan}, {Kerzendorf}, {Conley}, {Crighton},
  {Barbary}, {Muna}, {Ferguson}, {Grollier}, {Parikh}, {Nair}, {Unther},
  {Deil}, {Woillez}, {Conseil}, {Kramer}, {Turner}, {Singer}, {Fox}, {Weaver},
  {Zabalza}, {Edwards}, {Azalee Bostroem}, {Burke}, {Casey}, {Crawford},
  {Dencheva}, {Ely}, {Jenness}, {Labrie}, {Lim}, {Pierfederici}, {Pontzen},
  {Ptak}, {Refsdal}, {Servillat}, \& {Streicher}}]{astropy}
{Astropy Collaboration}, {Robitaille}, T.~P., {Tollerud}, E.~J., {et~al.} 2013,
  \aap, 558, A33, \dodoi{10.1051/0004-6361/201322068}

\bibitem[{{Bailer-Jones} {et~al.}(2018){Bailer-Jones}, {Rybizki}, {Fouesneau},
  {Mantelet}, \& {Andrae}}]{Bailer18}
{Bailer-Jones}, C.~A.~L., {Rybizki}, J., {Fouesneau}, M., {Mantelet}, G., \&
  {Andrae}, R. 2018, \aj, 156, 58, \dodoi{10.3847/1538-3881/aacb21}

\bibitem[{{Bans} \& {K{\"o}nigl}(2012)}]{Bans12}
{Bans}, A., \& {K{\"o}nigl}, A. 2012, \apj, 758, 100,
  \dodoi{10.1088/0004-637X/758/2/100}

\bibitem[{{Beckwith} {et~al.}(1990){Beckwith}, {Sargent}, {Chini}, \&
  {Guesten}}]{Beckwith90}
{Beckwith}, S.~V.~W., {Sargent}, A.~I., {Chini}, R.~S., \& {Guesten}, R. 1990,
  \aj, 99, 924, \dodoi{10.1086/115385}

\bibitem[{{Bonneau} {et~al.}(2011){Bonneau}, {Delfosse}, {Mourard}, {Lafrasse},
  {Mella}, {Cetre}, {Clausse}, \& {Zins}}]{Bonneau11}
{Bonneau}, D., {Delfosse}, X., {Mourard}, D., {et~al.} 2011, \aap, 535, A53,
  \dodoi{10.1051/0004-6361/201015124}

\bibitem[{{Bonneau} {et~al.}(2006){Bonneau}, {Clausse}, {Delfosse}, {Mourard},
  {Cetre}, {Chelli}, {Cruzal{\`e}bes}, {Duvert}, \& {Zins}}]{Bonneau06}
{Bonneau}, D., {Clausse}, J.-M., {Delfosse}, X., {et~al.} 2006, \aap, 456, 789,
  \dodoi{10.1051/0004-6361:20054469}

\bibitem[{{Cardelli} {et~al.}(1989){Cardelli}, {Clayton}, \&
  {Mathis}}]{Cardelli89}
{Cardelli}, J.~A., {Clayton}, G.~C., \& {Mathis}, J.~S. 1989, \apj, 345, 245,
  \dodoi{10.1086/167900}

\bibitem[{{Colavita} {et~al.}(2013){Colavita}, {Wizinowich}, {Akeson},
  {Ragland}, {Woillez}, {Millan-Gabet}, {Serabyn}, {Abajian}, {Acton},
  {Appleby}, {Beletic}, {Beichman}, {Bell}, {Berkey}, {Berlin}, {Boden},
  {Booth}, {Boutell}, {Chaffee}, {Chan}, {Chin}, {Chock}, {Cohen}, {Cooper},
  {Crawford}, {Creech-Eakman}, {Dahl}, {Eychaner}, {Fanson}, {Felizardo},
  {Garcia-Gathright}, {Gathright}, {Hardy}, {Henderson}, {Herstein}, {Hess},
  {Hovland}, {Hrynevych}, {Johansson}, {Johnson}, {Kelley}, {Kendrick},
  {Koresko}, {Kurpis}, {Le Mignant}, {Lewis}, {Ligon}, {Lupton}, {McBride},
  {Medeiros}, {Mennesson}, {Moore}, {Morrison}, {Nance}, {Neyman}, {Niessner},
  {Paine}, {Palmer}, {Panteleeva}, {Papin}, {Parvin}, {Reder}, {Rudeen},
  {Saloga}, {Sargent}, {Shao}, {Smith}, {Smythe}, {Stomski}, {Summers},
  {Swain}, {Swanson}, {Thompson}, {Tsubota}, {Tumminello}, {Tyau}, {van Belle},
  {Vasisht}, {Vause}, {Vescelus}, {Walker}, {Wallace}, {Wehmeier}, \&
  {Wetherell}}]{Colavita13}
{Colavita}, M.~M., {Wizinowich}, P.~L., {Akeson}, R.~L., {et~al.} 2013, \pasp,
  125, 1226, \dodoi{10.1086/673475}

\bibitem[{{Cutri} {et~al.}(2003){Cutri}, {Skrutskie}, {van Dyk}, {Beichman},
  {Carpenter}, {Chester}, {Cambresy}, {Evans}, {Fowler}, {Gizis}, {Howard},
  {Huchra}, {Jarrett}, {Kopan}, {Kirkpatrick}, {Light}, {Marsh}, {McCallon},
  {Schneider}, {Stiening}, {Sykes}, {Weinberg}, {Wheaton}, {Wheelock}, \&
  {Zacarias}}]{Cutri03}
{Cutri}, R.~M., {Skrutskie}, M.~F., {van Dyk}, S., {et~al.} 2003, VizieR Online
  Data Catalog, 2246

\bibitem[{{Cutri} {et~al.}(2012{\natexlab{a}}){Cutri}, {Wright}, {Conrow},
  {Bauer}, {Benford}, {Brandenburg}, {Dailey}, {Eisenhardt}, {Evans},
  {Fajardo-Acosta}, {Fowler}, {Gelino}, {Grillmair}, {Harbut}, {Hoffman},
  {Jarrett}, {Kirkpatrick}, {Leisawitz}, {Liu}, {Mainzer}, {Marsh}, {Masci},
  {McCallon}, {Padgett}, {Ressler}, {Royer}, {Skrutskie}, {Stanford}, {Wyatt},
  {Tholen}, {Tsai}, {Wachter}, {Wheelock}, {Yan}, {Alles}, {Beck}, {Grav},
  {Masiero}, {McCollum}, {McGehee}, {Papin}, \& {Wittman}}]{Cutri12}
{Cutri}, R.~M., {Wright}, E.~L., {Conrow}, T., {et~al.} 2012{\natexlab{a}},
  {Explanatory Supplement to the WISE All-Sky Data Release Products}, Tech.
  rep.

\bibitem[{{Cutri} {et~al.}(2012{\natexlab{b}}){Cutri}, {Wright}, {Conrow},
  {Bauer}, {Benford}, {Brandenburg}, {Dailey}, {Eisenhardt}, {Evans},
  {Fajardo-Acosta}, {Fowler}, {Gelino}, {Grillmair}, {Harbut}, {Hoffman},
  {Jarrett}, {Kirkpatrick}, {Leisawitz}, {Liu}, {Mainzer}, {Marsh}, {Masci},
  {McCallon}, {Padgett}, {Ressler}, {Royer}, {Skrutskie}, {Stanford}, {Wyatt},
  {Tholen}, {Tsai}, {Wachter}, {Wheelock}, {Yan}, {Alles}, {Beck}, {Grav},
  {Masiero}, {McCollum}, {McGehee}, {Papin}, \& {Wittman}}]{Cutri12cat}
---. 2012{\natexlab{b}}, VizieR Online Data Catalog, 2311

\bibitem[{{Dent} {et~al.}(2005){Dent}, {Greaves}, \& {Coulson}}]{Dent05}
{Dent}, W.~R.~F., {Greaves}, J.~S., \& {Coulson}, I.~M. 2005, \mnras, 359, 663,
  \dodoi{10.1111/j.1365-2966.2005.08938.x}

\bibitem[{{Di Francesco} {et~al.}(2008){Di Francesco}, {Johnstone}, {Kirk},
  {MacKenzie}, \& {Ledwosinska}}]{diFrancesco08}
{Di Francesco}, J., {Johnstone}, D., {Kirk}, H., {MacKenzie}, T., \&
  {Ledwosinska}, E. 2008, \apjs, 175, 277, \dodoi{10.1086/523645}

\bibitem[{{Dominik} {et~al.}(2003){Dominik}, {Dullemond}, {Waters}, \&
  {Walch}}]{Dominik03}
{Dominik}, C., {Dullemond}, C.~P., {Waters}, L.~B.~F.~M., \& {Walch}, S. 2003,
  \aap, 398, 607, \dodoi{10.1051/0004-6361:20021629}

\bibitem[{{Draine}(2003)}]{Draine03}
{Draine}, B.~T. 2003, \apj, 598, 1026, \dodoi{10.1086/379123}

\bibitem[{{Draine} \& {Lee}(1984)}]{Draine84}
{Draine}, B.~T., \& {Lee}, H.~M. 1984, \apj, 285, 89, \dodoi{10.1086/162480}

\bibitem[{{Dullemond} {et~al.}(2001){Dullemond}, {Dominik}, \&
  {Natta}}]{Dullemond01}
{Dullemond}, C.~P., {Dominik}, C., \& {Natta}, A. 2001, \apj, 560, 957,
  \dodoi{10.1086/323057}

\bibitem[{{Duvert} {et~al.}(2017){Duvert}, {Young}, \& {Hummel}}]{Duvert17}
{Duvert}, G., {Young}, J., \& {Hummel}, C.~A. 2017, \aap, 597, A8,
  \dodoi{10.1051/0004-6361/201526405}

\bibitem[{{Eisner} {et~al.}(2009){Eisner}, {Graham}, {Akeson}, \&
  {Najita}}]{Eisner09}
{Eisner}, J.~A., {Graham}, J.~R., {Akeson}, R.~L., \& {Najita}, J. 2009, \apj,
  692, 309, \dodoi{10.1088/0004-637X/692/1/309}

\bibitem[{{Fischer} {et~al.}(2011){Fischer}, {Edwards}, {Hillenbrand}, \&
  {Kwan}}]{Fischer11}
{Fischer}, W., {Edwards}, S., {Hillenbrand}, L., \& {Kwan}, J. 2011, \apj, 730,
  73, \dodoi{10.1088/0004-637X/730/2/73}

\bibitem[{{Flock} {et~al.}(2016){Flock}, {Fromang}, {Turner}, \&
  {Benisty}}]{Flock16a}
{Flock}, M., {Fromang}, S., {Turner}, N.~J., \& {Benisty}, M. 2016, \apj, 827,
  144, \dodoi{10.3847/0004-637X/827/2/144}

\bibitem[{{Flock} {et~al.}(2017){Flock}, {Fromang}, {Turner}, \&
  {Benisty}}]{Flock16}
---. 2017, \apj, 835, 230, \dodoi{10.3847/1538-4357/835/2/230}

\bibitem[{{Gaia Collaboration} {et~al.}(2018){Gaia Collaboration}, {Brown},
  {Vallenari}, {Prusti}, {de Bruijne}, {Babusiaux}, \& {Bailer-Jones}}]{Gaia18}
{Gaia Collaboration}, {Brown}, A.~G.~A., {Vallenari}, A., {et~al.} 2018, ArXiv
  e-prints.
\newblock \doarXiv{1804.09365}

\bibitem[{{Gaia Collaboration} {et~al.}(2016){Gaia Collaboration}, {Prusti},
  {de Bruijne}, {Brown}, {Vallenari}, {Babusiaux}, {Bailer-Jones}, {Bastian},
  {Biermann}, {Evans}, \& et~al.}]{Gaia16}
{Gaia Collaboration}, {Prusti}, T., {de Bruijne}, J.~H.~J., {et~al.} 2016,
  \aap, 595, A1, \dodoi{10.1051/0004-6361/201629272}

\bibitem[{{Garufi} {et~al.}(2017){Garufi}, {Meeus}, {Benisty}, {Quanz},
  {Banzatti}, {Kama}, {Canovas}, {Eiroa}, {Schmid}, {Stolker}, {Pohl},
  {Rigliaco}, {M{\'e}nard}, {Meyer}, {van Boekel}, \& {Dominik}}]{Garufi17}
{Garufi}, A., {Meeus}, G., {Benisty}, M., {et~al.} 2017, \aap, 603, A21,
  \dodoi{10.1051/0004-6361/201630320}

\bibitem[{{Grinin} {et~al.}(1991){Grinin}, {Kiselev}, {Minikulov}, {Chernova},
  \& {Voshchinnikov}}]{Grinin91}
{Grinin}, V.~P., {Kiselev}, N.~N., {Minikulov}, N.~K., {Chernova}, G.~P., \&
  {Voshchinnikov}, N.~V. 1991, \apss, 186, 283, \dodoi{10.1007/BF02111202}

\bibitem[{{Guimar{\~a}es} {et~al.}(2006){Guimar{\~a}es}, {Alencar}, {Corradi},
  \& {Vieira}}]{Guimaraes06}
{Guimar{\~a}es}, M.~M., {Alencar}, S.~H.~P., {Corradi}, W.~J.~B., \& {Vieira},
  S.~L.~A. 2006, \aap, 457, 581, \dodoi{10.1051/0004-6361:20065005}

\bibitem[{{Harries}(2000)}]{Harries00}
{Harries}, T.~J. 2000, \mnras, 315, 722,
  \dodoi{10.1046/j.1365-8711.2000.03505.x}

\bibitem[{{Harries} {et~al.}(2004){Harries}, {Monnier}, {Symington}, \&
  {Kurosawa}}]{Harries04}
{Harries}, T.~J., {Monnier}, J.~D., {Symington}, N.~H., \& {Kurosawa}, R. 2004,
  \mnras, 350, 565, \dodoi{10.1111/j.1365-2966.2004.07668.x}

\bibitem[{{Haubois} {et~al.}(2014){Haubois}, {Bernaud}, {Mella}, {Duvert},
  {Benisty}, {B{\'e}rio}, {Bourges}, {Chelli}, {Chesneau}, {Lacour},
  {Lafrasse}, {Le Bouquin}, {Mourard}, {Nardetto}, \& {Olofsson}}]{Haubois14}
{Haubois}, X., {Bernaud}, P., {Mella}, G., {et~al.} 2014, in \procspie, Vol.
  9146, Optical and Infrared Interferometry IV, 91460O

\bibitem[{{Helou} \& {Walker}(1988)}]{Helou88}
{Helou}, G., \& {Walker}, D.~W., eds. 1988, {Infrared astronomical satellite
  (IRAS) catalogs and atlases. Volume 7: The small scale structure catalog},
  Vol.~7, 1--265

\bibitem[{{Herbig}(1960)}]{Herbig60}
{Herbig}, G.~H. 1960, \apjs, 4, 337, \dodoi{10.1086/190050}

\bibitem[{{Herbst} \& {Shevchenko}(1999)}]{Herbst99}
{Herbst}, W., \& {Shevchenko}, V.~S. 1999, \aj, 118, 1043,
  \dodoi{10.1086/300966}

\bibitem[{{Hern{\'a}ndez} {et~al.}(2004){Hern{\'a}ndez}, {Calvet},
  {Brice{\~n}o}, {Hartmann}, \& {Berlind}}]{Hernandez04}
{Hern{\'a}ndez}, J., {Calvet}, N., {Brice{\~n}o}, C., {Hartmann}, L., \&
  {Berlind}, P. 2004, \aj, 127, 1682, \dodoi{10.1086/381908}

\bibitem[{{Hildebrand}(1983)}]{Hildebrand83}
{Hildebrand}, R.~H. 1983, \qjras, 24, 267

\bibitem[{{Hillenbrand} {et~al.}(1992){Hillenbrand}, {Strom}, {Vrba}, \&
  {Keene}}]{Hillenbrand92}
{Hillenbrand}, L.~A., {Strom}, S.~E., {Vrba}, F.~J., \& {Keene}, J. 1992, \apj,
  397, 613, \dodoi{10.1086/171819}

\bibitem[{{H{\o}g} {et~al.}(2000){H{\o}g}, {Fabricius}, {Makarov}, {Urban},
  {Corbin}, {Wycoff}, {Bastian}, {Schwekendiek}, \& {Wicenec}}]{Hog00}
{H{\o}g}, E., {Fabricius}, C., {Makarov}, V.~V., {et~al.} 2000, \aap, 355, L27

\bibitem[{{Houck} {et~al.}(2004){Houck}, {Roellig}, {van Cleve}, {Forrest},
  {Herter}, {Lawrence}, {Matthews}, {Reitsema}, {Soifer}, {Watson}, {Weedman},
  {Huisjen}, {Troeltzsch}, {Barry}, {Bernard-Salas}, {Blacken}, {Brandl},
  {Charmandaris}, {Devost}, {Gull}, {Hall}, {Henderson}, {Higdon}, {Pirger},
  {Schoenwald}, {Sloan}, {Uchida}, {Appleton}, {Armus}, {Burgdorf},
  {Fajardo-Acosta}, {Grillmair}, {Ingalls}, {Morris}, \& {Teplitz}}]{Houck04}
{Houck}, J.~R., {Roellig}, T.~L., {van Cleve}, J., {et~al.} 2004, \apjs, 154,
  18, \dodoi{10.1086/423134}

\bibitem[{Hunter(2007)}]{Hunter07}
Hunter, J.~D. 2007, Computing In Science \& Engineering, 9, 90

\bibitem[{{Ilee} {et~al.}(2016){Ilee}, {Cyganowski}, {Nazari}, {Hunter},
  {Brogan}, {Forgan}, \& {Zhang}}]{Ilee16}
{Ilee}, J.~D., {Cyganowski}, C.~J., {Nazari}, P., {et~al.} 2016, \mnras, 462,
  4386, \dodoi{10.1093/mnras/stw1912}

\bibitem[{{Ilee} {et~al.}(2014){Ilee}, {Fairlamb}, {Oudmaijer},
  {Mendigut{\'{\i}}a}, {van den Ancker}, {Kraus}, \& {Wheelwright}}]{Ilee14}
{Ilee}, J.~D., {Fairlamb}, J., {Oudmaijer}, R.~D., {et~al.} 2014, \mnras, 445,
  3723, \dodoi{10.1093/mnras/stu1942}

\bibitem[{{Isella} \& {Natta}(2005)}]{Isella05}
{Isella}, A., \& {Natta}, A. 2005, \aap, 438, 899,
  \dodoi{10.1051/0004-6361:20052773}

\bibitem[{{Ishihara} {et~al.}(2010){Ishihara}, {Onaka}, {Kataza}, {Salama},
  {Alfageme}, {Cassatella}, {Cox}, {Garc{\'{\i}}a-Lario}, {Stephenson},
  {Cohen}, {Fujishiro}, {Fujiwara}, {Hasegawa}, {Ita}, {Kim}, {Matsuhara},
  {Murakami}, {M{\"u}ller}, {Nakagawa}, {Ohyama}, {Oyabu}, {Pyo}, {Sakon},
  {Shibai}, {Takita}, {Tanab{\'e}}, {Uemizu}, {Ueno}, {Usui}, {Wada},
  {Watarai}, {Yamamura}, \& {Yamauchi}}]{Ishihara10}
{Ishihara}, D., {Onaka}, T., {Kataza}, H., {et~al.} 2010, \aap, 514, A1,
  \dodoi{10.1051/0004-6361/200913811}

\bibitem[{{Ita} {et~al.}(2010){Ita}, {Matsuura}, {Ishihara}, {Oyabu}, {Takita},
  {Kataza}, {Yamamura}, {Matsunaga}, {Tanab{\'e}}, {Nakada}, {Fujiwara},
  {Wada}, {Onaka}, \& {Matsuhara}}]{Ita10}
{Ita}, Y., {Matsuura}, M., {Ishihara}, D., {et~al.} 2010, \aap, 514, A2,
  \dodoi{10.1051/0004-6361/200913695}

\bibitem[{{Kama} {et~al.}(2009){Kama}, {Min}, \& {Dominik}}]{Kama09}
{Kama}, M., {Min}, M., \& {Dominik}, C. 2009, \aap, 506, 1199,
  \dodoi{10.1051/0004-6361/200912068}

\bibitem[{{Kamp} \& {Dullemond}(2004)}]{Kamp04}
{Kamp}, I., \& {Dullemond}, C.~P. 2004, \apj, 615, 991, \dodoi{10.1086/424703}

\bibitem[{{Keller} {et~al.}(2008){Keller}, {Sloan}, {Forrest}, {Ayala},
  {D'Alessio}, {Shah}, {Calvet}, {Najita}, {Li}, {Hartmann}, {Sargent},
  {Watson}, \& {Chen}}]{Keller08}
{Keller}, L.~D., {Sloan}, G.~C., {Forrest}, W.~J., {et~al.} 2008, \apj, 684,
  411, \dodoi{10.1086/589818}

\bibitem[{{Kim} {et~al.}(1994){Kim}, {Martin}, \& {Hendry}}]{Kim94}
{Kim}, S.-H., {Martin}, P.~G., \& {Hendry}, P.~D. 1994, \apj, 422, 164,
  \dodoi{10.1086/173714}

\bibitem[{{Kraus} {et~al.}(2009){Kraus}, {Hofmann}, {Malbet}, {Meilland},
  {Natta}, {Schertl}, {Stee}, \& {Weigelt}}]{Kraus09}
{Kraus}, S., {Hofmann}, K.-H., {Malbet}, F., {et~al.} 2009, \aap, 508, 787,
  \dodoi{10.1051/0004-6361/200912990}

\bibitem[{{Kraus} {et~al.}(2008){Kraus}, {Preibisch}, \& {Ohnaka}}]{Kraus08}
{Kraus}, S., {Preibisch}, T., \& {Ohnaka}, K. 2008, \apj, 676, 490,
  \dodoi{10.1086/527427}

\bibitem[{{Kraus} {et~al.}(2014){Kraus}, {Monnier}, {Harries}, {Dong}, {Bate},
  {Whitney}, {Zhu}, {Buscher}, {Berger}, {Haniff}, {Ireland}, {Labadie},
  {Lacour}, {Petrov}, {Ridgway}, {Surdej}, {ten Brummelaar}, {Tuthill}, \& {van
  Belle}}]{Kraus14}
{Kraus}, S., {Monnier}, J., {Harries}, T., {et~al.} 2014, in \procspie, Vol.
  9146, Optical and Infrared Interferometry IV, 914611

\bibitem[{{Kraus} {et~al.}(2017){Kraus}, {Kluska}, {Kreplin}, {Bate},
  {Harries}, {Hofmann}, {Hone}, {Monnier}, {Weigelt}, {Anugu}, {de Wit}, \&
  {Wittkowski}}]{Kraus17}
{Kraus}, S., {Kluska}, J., {Kreplin}, A., {et~al.} 2017, \apjl, 835, L5,
  \dodoi{10.3847/2041-8213/835/1/L5}

\bibitem[{{Kreplin} {et~al.}(2016){Kreplin}, {Madlener}, {Chen}, {Weigelt},
  {Kraus}, {Grinin}, {Tambovtseva}, \& {Kishimoto}}]{Kreplin16}
{Kreplin}, A., {Madlener}, D., {Chen}, L., {et~al.} 2016, \aap, 590, A96,
  \dodoi{10.1051/0004-6361/201628281}

\bibitem[{{Kreplin} {et~al.}(2013){Kreplin}, {Weigelt}, {Kraus}, {Grinin},
  {Hofmann}, {Kishimoto}, {Schertl}, {Tambovtseva}, {Clausse}, {Massi},
  {Perraut}, \& {Stee}}]{Kreplin13}
{Kreplin}, A., {Weigelt}, G., {Kraus}, S., {et~al.} 2013, \aap, 551, A21,
  \dodoi{10.1051/0004-6361/201220806}

\bibitem[{{Kurosawa} {et~al.}(2006){Kurosawa}, {Harries}, \&
  {Symington}}]{Kurosawa06}
{Kurosawa}, R., {Harries}, T.~J., \& {Symington}, N.~H. 2006, \mnras, 370, 580,
  \dodoi{10.1111/j.1365-2966.2006.10527.x}

\bibitem[{{Kurucz}(1979)}]{Kurucz79}
{Kurucz}, R.~L. 1979, \apjs, 40, 1, \dodoi{10.1086/190589}

\bibitem[{{Lazareff} {et~al.}(2017){Lazareff}, {Berger}, {Kluska}, {Le
  Bouquin}, {Benisty}, {Malbet}, {Koen}, {Pinte}, {Thi}, {Absil}, {Baron},
  {Delboulb{\'e}}, {Duvert}, {Isella}, {Jocou}, {Juhasz}, {Kraus}, {Lachaume},
  {M{\'e}nard}, {Millan-Gabet}, {Monnier}, {Moulin}, {Perraut}, {Rochat},
  {Soulez}, {Tallon}, {Thi{\'e}baut}, {Traub}, \& {Zins}}]{Lazareff16}
{Lazareff}, B., {Berger}, J.-P., {Kluska}, J., {et~al.} 2017, \aap, 599, A85,
  \dodoi{10.1051/0004-6361/201629305}

\bibitem[{{Le Bouquin} {et~al.}(2011){Le Bouquin}, {Berger}, {Lazareff},
  {Zins}, {Haguenauer}, {Jocou}, {Kern}, {Millan-Gabet}, {Traub}, {Absil},
  {Augereau}, {Benisty}, {Blind}, {Bonfils}, {Bourget}, {Delboulbe},
  {Feautrier}, {Germain}, {Gitton}, {Gillier}, {Kiekebusch}, {Kluska},
  {Knudstrup}, {Labeye}, {Lizon}, {Monin}, {Magnard}, {Malbet}, {Maurel},
  {M{\'e}nard}, {Micallef}, {Michaud}, {Montagnier}, {Morel}, {Moulin},
  {Perraut}, {Popovic}, {Rabou}, {Rochat}, {Rojas}, {Roussel}, {Roux},
  {Stadler}, {Stefl}, {Tatulli}, \& {Ventura}}]{leBouquin11}
{Le Bouquin}, J.-B., {Berger}, J.-P., {Lazareff}, B., {et~al.} 2011, \aap, 535,
  A67, \dodoi{10.1051/0004-6361/201117586}

\bibitem[{{Lindegren} {et~al.}(2016){Lindegren}, {Lammers}, {Bastian},
  {Hern{\'a}ndez}, {Klioner}, {Hobbs}, {Bombrun}, {Michalik}, {Ramos-Lerate},
  {Butkevich}, {Comoretto}, {Joliet}, {Holl}, {Hutton}, {Parsons},
  {Steidelm{\"u}ller}, {Abbas}, {Altmann}, {Andrei}, {Anton}, {Bach},
  {Barache}, {Becciani}, {Berthier}, {Bianchi}, {Biermann}, {Bouquillon},
  {Bourda}, {Br{\"u}semeister}, {Bucciarelli}, {Busonero}, {Carlucci},
  {Casta{\~n}eda}, {Charlot}, {Clotet}, {Crosta}, {Davidson}, {de Felice},
  {Drimmel}, {Fabricius}, {Fienga}, {Figueras}, {Fraile}, {Gai}, {Garralda},
  {Geyer}, {Gonz{\'a}lez-Vidal}, {Guerra}, {Hambly}, {Hauser}, {Jordan},
  {Lattanzi}, {Lenhardt}, {Liao}, {L{\"o}ffler}, {McMillan}, {Mignard}, {Mora},
  {Morbidelli}, {Portell}, {Riva}, {Sarasso}, {Serraller}, {Siddiqui}, {Smart},
  {Spagna}, {Stampa}, {Steele}, {Taris}, {Torra}, {van Reeven}, {Vecchiato},
  {Zschocke}, {de Bruijne}, {Gracia}, {Raison}, {Lister}, {Marchant},
  {Messineo}, {Soffel}, {Osorio}, {de Torres}, \& {O'Mullane}}]{Lindegren16}
{Lindegren}, L., {Lammers}, U., {Bastian}, U., {et~al.} 2016, \aap, 595, A4,
  \dodoi{10.1051/0004-6361/201628714}

\bibitem[{{Lucy}(1999)}]{Lucy99}
{Lucy}, L.~B. 1999, \aap, 344, 282

\bibitem[{{Makarov} {et~al.}(1994){Makarov}, {Bastian}, {Hoeg}, {Grossmann}, \&
  {Wicenec}}]{Makarov94}
{Makarov}, V., {Bastian}, U., {Hoeg}, E., {Grossmann}, V., \& {Wicenec}, A.
  1994, Information Bulletin on Variable Stars, 4118

\bibitem[{{Mamajek}(2012)}]{Mamajek12}
{Mamajek}, E.~E. 2012, \apjl, 754, L20, \dodoi{10.1088/2041-8205/754/2/L20}

\bibitem[{{Mann} \& {von Braun}(2015)}]{Mann15}
{Mann}, A.~W., \& {von Braun}, K. 2015, \pasp, 127, 102, \dodoi{10.1086/680012}

\bibitem[{{Manoj} {et~al.}(2006){Manoj}, {Bhatt}, {Maheswar}, \&
  {Muneer}}]{Manoj06}
{Manoj}, P., {Bhatt}, H.~C., {Maheswar}, G., \& {Muneer}, S. 2006, \apj, 653,
  657, \dodoi{10.1086/508764}

\bibitem[{{McClure} {et~al.}(2013){McClure}, {D'Alessio}, {Calvet},
  {Espaillat}, {Hartmann}, {Sargent}, {Watson}, {Ingleby}, \&
  {Hern{\'a}ndez}}]{McClure13}
{McClure}, M.~K., {D'Alessio}, P., {Calvet}, N., {et~al.} 2013, \apj, 775, 114,
  \dodoi{10.1088/0004-637X/775/2/114}

\bibitem[{{McDonald} {et~al.}(2017){McDonald}, {Zijlstra}, \&
  {Watson}}]{McDonald17}
{McDonald}, I., {Zijlstra}, A.~A., \& {Watson}, R.~A. 2017, \mnras, 471, 770,
  \dodoi{10.1093/mnras/stx1433}

\bibitem[{{Meeus} {et~al.}(1998){Meeus}, {Waelkens}, \& {Malfait}}]{Meeus98}
{Meeus}, G., {Waelkens}, C., \& {Malfait}, K. 1998, \aap, 329, 131

\bibitem[{{Meeus} {et~al.}(2001){Meeus}, {Waters}, {Bouwman}, {van den Ancker},
  {Waelkens}, \& {Malfait}}]{Meeus01}
{Meeus}, G., {Waters}, L.~B.~F.~M., {Bouwman}, J., {et~al.} 2001, \aap, 365,
  476, \dodoi{10.1051/0004-6361:20000144}

\bibitem[{{Millan-Gabet} {et~al.}(1999){Millan-Gabet}, {Schloerb}, {Traub},
  {Malbet}, {Berger}, \& {Bregman}}]{Millan99}
{Millan-Gabet}, R., {Schloerb}, F.~P., {Traub}, W.~A., {et~al.} 1999, \apjl,
  513, L131, \dodoi{10.1086/311926}

\bibitem[{{Monnier} \& {Millan-Gabet}(2002)}]{Monnier02}
{Monnier}, J.~D., \& {Millan-Gabet}, R. 2002, \apj, 579, 694,
  \dodoi{10.1086/342917}

\bibitem[{{Monnier} {et~al.}(2005){Monnier}, {Millan-Gabet}, {Billmeier},
  {Akeson}, {Wallace}, {Berger}, {Calvet}, {D'Alessio}, {Danchi}, {Hartmann},
  {Hillenbrand}, {Kuchner}, {Rajagopal}, {Traub}, {Tuthill}, {Boden}, {Booth},
  {Colavita}, {Gathright}, {Hrynevych}, {Le Mignant}, {Ligon}, {Neyman},
  {Swain}, {Thompson}, {Vasisht}, {Wizinowich}, {Beichman}, {Beletic},
  {Creech-Eakman}, {Koresko}, {Sargent}, {Shao}, \& {van Belle}}]{Monnier05}
{Monnier}, J.~D., {Millan-Gabet}, R., {Billmeier}, R., {et~al.} 2005, \apj,
  624, 832, \dodoi{10.1086/429266}

\bibitem[{{Monnier} {et~al.}(2006){Monnier}, {Berger}, {Millan-Gabet}, {Traub},
  {Schloerb}, {Pedretti}, {Benisty}, {Carleton}, {Haguenauer}, {Kern},
  {Labeye}, {Lacasse}, {Malbet}, {Perraut}, {Pearlman}, \& {Zhao}}]{Monnier06}
{Monnier}, J.~D., {Berger}, J.-P., {Millan-Gabet}, R., {et~al.} 2006, \apj,
  647, 444, \dodoi{10.1086/505340}

\bibitem[{{Mora} {et~al.}(2001){Mora}, {Mer{\'{\i}}n}, {Solano}, {Montesinos},
  {de Winter}, {Eiroa}, {Ferlet}, {Grady}, {Davies}, {Miranda}, {Oudmaijer},
  {Palacios}, {Quirrenbach}, {Harris}, {Rauer}, {Collier Cameron}, {Deeg},
  {Garz{\'o}n}, {Penny}, {Schneider}, {Tsapras}, \& {Wesselius}}]{Mora01}
{Mora}, A., {Mer{\'{\i}}n}, B., {Solano}, E., {et~al.} 2001, \aap, 378, 116,
  \dodoi{10.1051/0004-6361:20011098}

\bibitem[{{Moshir} {et~al.}(1990){Moshir}, {Copan}, {Conrow}, {McCallon},
  {Hacking}, {Gregorich}, {Rohrbach}, {Melnyk}, {Rice}, {Fullmer}, \&
  {Chester}}]{Moshir90}
{Moshir}, M., {Copan}, G., {Conrow}, T., {et~al.} 1990, in IRAS Faint Source
  Catalogue, version 2.0 (1990)

\bibitem[{{Mulders} \& {Dominik}(2012)}]{Mulders12}
{Mulders}, G.~D., \& {Dominik}, C. 2012, \aap, 539, A9,
  \dodoi{10.1051/0004-6361/201118127}

\bibitem[{{Natta} {et~al.}(1997){Natta}, {Grinin}, {Mannings}, \&
  {Ungerechts}}]{Natta97}
{Natta}, A., {Grinin}, V.~P., {Mannings}, V., \& {Ungerechts}, H. 1997, \apj,
  491, 885, \dodoi{10.1086/305006}

\bibitem[{{Natta} {et~al.}(2001){Natta}, {Prusti}, {Neri}, {Wooden}, {Grinin},
  \& {Mannings}}]{Natta01}
{Natta}, A., {Prusti}, T., {Neri}, R., {et~al.} 2001, \aap, 371, 186,
  \dodoi{10.1051/0004-6361:20010334}

\bibitem[{{Natta} {et~al.}(2004){Natta}, {Testi}, {Neri}, {Shepherd}, \&
  {Wilner}}]{Natta04}
{Natta}, A., {Testi}, L., {Neri}, R., {Shepherd}, D.~S., \& {Wilner}, D.~J.
  2004, \aap, 416, 179, \dodoi{10.1051/0004-6361:20035620}

\bibitem[{{Newville} {et~al.}(2014){Newville}, {Stensitzki}, {Allen}, \&
  {Ingargiola}}]{Newville14}
{Newville}, M., {Stensitzki}, T., {Allen}, D.~B., \& {Ingargiola}, A. 2014,
  {LMFIT: Non-Linear Least-Square Minimization and Curve-Fitting for Python¶},
  \dodoi{10.5281/zenodo.11813}

\bibitem[{{Pascual} {et~al.}(2016){Pascual}, {Montesinos}, {Meeus}, {Marshall},
  {Mendigut{\'{\i}}a}, \& {Sandell}}]{Pascual16}
{Pascual}, N., {Montesinos}, B., {Meeus}, G., {et~al.} 2016, \aap, 586, A6,
  \dodoi{10.1051/0004-6361/201526605}

\bibitem[{{Pauls} {et~al.}(2005){Pauls}, {Young}, {Cotton}, \&
  {Monnier}}]{Pauls05}
{Pauls}, T.~A., {Young}, J.~S., {Cotton}, W.~D., \& {Monnier}, J.~D. 2005,
  \pasp, 117, 1255, \dodoi{10.1086/444523}

\bibitem[{{Pecaut} \& {Mamajek}(2013)}]{Pecaut13}
{Pecaut}, M.~J., \& {Mamajek}, E.~E. 2013, \apjs, 208, 9,
  \dodoi{10.1088/0067-0049/208/1/9}

\bibitem[{{Pollack} {et~al.}(1994){Pollack}, {Hollenbach}, {Beckwith},
  {Simonelli}, {Roush}, \& {Fong}}]{Pollack94}
{Pollack}, J.~B., {Hollenbach}, D., {Beckwith}, S., {et~al.} 1994, \apj, 421,
  615, \dodoi{10.1086/173677}

\bibitem[{{Pontoppidan} {et~al.}(2007){Pontoppidan}, {Dullemond}, {Blake},
  {Boogert}, {van Dishoeck}, {Evans}, {Kessler-Silacci}, \&
  {Lahuis}}]{Pontoppidan07}
{Pontoppidan}, K.~M., {Dullemond}, C.~P., {Blake}, G.~A., {et~al.} 2007, \apj,
  656, 980, \dodoi{10.1086/510570}

\bibitem[{{Ricci} {et~al.}(2014){Ricci}, {Testi}, {Natta}, {Scholz}, {de
  Gregorio-Monsalvo}, \& {Isella}}]{Ricci14}
{Ricci}, L., {Testi}, L., {Natta}, A., {et~al.} 2014, \apj, 791, 20,
  \dodoi{10.1088/0004-637X/791/1/20}

\bibitem[{{Rubinstein} {et~al.}(2018){Rubinstein}, {Macias}, {Espaillat},
  {Zhang}, {Calvet}, \& {Robinson}}]{Rubinstein18}
{Rubinstein}, A.~E., {Macias}, E., {Espaillat}, C.~C., {et~al.} 2018, ArXiv
  e-prints.
\newblock \doarXiv{1804.07343}

\bibitem[{{Schegerer} {et~al.}(2013){Schegerer}, {Ratzka}, {Schuller}, {Wolf},
  {Mosoni}, \& {Leinert}}]{Schegerer13}
{Schegerer}, A.~A., {Ratzka}, T., {Schuller}, P.~A., {et~al.} 2013, \aap, 555,
  A103, \dodoi{10.1051/0004-6361/201220190}

\bibitem[{{Shakura} \& {Sunyaev}(1973)}]{Shakura73}
{Shakura}, N.~I., \& {Sunyaev}, R.~A. 1973, \aap, 24, 337

\bibitem[{{Siess} {et~al.}(2000){Siess}, {Dufour}, \& {Forestini}}]{Siess00}
{Siess}, L., {Dufour}, E., \& {Forestini}, M. 2000, \aap, 358, 593

\bibitem[{{Strom} {et~al.}(1972){Strom}, {Strom}, {Yost}, {Carrasco}, \&
  {Grasdalen}}]{Strom72}
{Strom}, S.~E., {Strom}, K.~M., {Yost}, J., {Carrasco}, L., \& {Grasdalen}, G.
  1972, \apj, 173, 353, \dodoi{10.1086/151425}

\bibitem[{{STScI Development Team}(2013)}]{pysynphot}
{STScI Development Team}. 2013, {pysynphot: Synthetic photometry software
  package}, Astrophysics Source Code Library.
\newblock \doeprint{1303.023}

\bibitem[{{Sylvester} {et~al.}(1996){Sylvester}, {Skinner}, {Barlow}, \&
  {Mannings}}]{Sylvester96}
{Sylvester}, R.~J., {Skinner}, C.~J., {Barlow}, M.~J., \& {Mannings}, V. 1996,
  \mnras, 279, 915, \dodoi{10.1093/mnras/279.3.915}

\bibitem[{{Tambovtseva} \& {Grinin}(2008)}]{Tambovtseva08}
{Tambovtseva}, L.~V., \& {Grinin}, V.~P. 2008, Astronomy Letters, 34, 231,
  \dodoi{10.1134/S1063773708040026}

\bibitem[{{Tannirkulam} {et~al.}(2007){Tannirkulam}, {Harries}, \&
  {Monnier}}]{Tannirkulam07}
{Tannirkulam}, A., {Harries}, T.~J., \& {Monnier}, J.~D. 2007, \apj, 661, 374,
  \dodoi{10.1086/513265}

\bibitem[{{Tannirkulam} {et~al.}(2008{\natexlab{a}}){Tannirkulam}, {Monnier},
  {Millan-Gabet}, {Harries}, {Pedretti}, {ten Brummelaar}, {McAlister},
  {Turner}, {Sturmann}, \& {Sturmann}}]{Tannirkulam08a}
{Tannirkulam}, A., {Monnier}, J.~D., {Millan-Gabet}, R., {et~al.}
  2008{\natexlab{a}}, \apjl, 677, L51, \dodoi{10.1086/587873}

\bibitem[{{Tannirkulam} {et~al.}(2008{\natexlab{b}}){Tannirkulam}, {Monnier},
  {Harries}, {Millan-Gabet}, {Zhu}, {Pedretti}, {Ireland}, {Tuthill}, {ten
  Brummelaar}, {McAlister}, {Farrington}, {Goldfinger}, {Sturmann}, {Sturmann},
  \& {Turner}}]{Tannirkulam08b}
{Tannirkulam}, A., {Monnier}, J.~D., {Harries}, T.~J., {et~al.}
  2008{\natexlab{b}}, \apj, 689, 513, \dodoi{10.1086/592346}

\bibitem[{{ten Brummelaar} {et~al.}(2005){ten Brummelaar}, {McAlister},
  {Ridgway}, {Bagnuolo}, {Turner}, {Sturmann}, {Sturmann}, {Berger}, {Ogden},
  {Cadman}, {Hartkopf}, {Hopper}, \& {Shure}}]{Brummelaar05}
{ten Brummelaar}, T.~A., {McAlister}, H.~A., {Ridgway}, S.~T., {et~al.} 2005,
  \apj, 628, 453, \dodoi{10.1086/430729}

\bibitem[{{ten Brummelaar} {et~al.}(2012){ten Brummelaar}, {Sturmann},
  {McAlister}, {Sturmann}, {Turner}, {Farrington}, {Schaefer}, {Goldfinger}, \&
  {Kloppenborg}}]{Brummelaar12}
{ten Brummelaar}, T.~A., {Sturmann}, J., {McAlister}, H.~A., {et~al.} 2012, in
  \procspie, Vol. 8445, Optical and Infrared Interferometry III, 84453C

\bibitem[{{ten Brummelaar} {et~al.}(2013){ten Brummelaar}, {Sturmann},
  {Ridgway}, {Sturmann}, {Turner}, {McAlister}, {Farrington}, {Beckmann},
  {Weigelt}, \& {Shure}}]{Brummelaar13}
{ten Brummelaar}, T.~A., {Sturmann}, J., {Ridgway}, S.~T., {et~al.} 2013,
  Journal of Astronomical Instrumentation, 2, 1340004,
  \dodoi{10.1142/S2251171713400047}

\bibitem[{{Testi} {et~al.}(2014){Testi}, {Birnstiel}, {Ricci}, {Andrews},
  {Blum}, {Carpenter}, {Dominik}, {Isella}, {Natta}, {Williams}, \&
  {Wilner}}]{Testi14}
{Testi}, L., {Birnstiel}, T., {Ricci}, L., {et~al.} 2014, Protostars and
  Planets VI, 339, \dodoi{10.2458/azu_uapress_9780816531240-ch015}

\bibitem[{{Turner} {et~al.}(2014){Turner}, {Benisty}, {Dullemond}, \&
  {Hirose}}]{Turner14}
{Turner}, N.~J., {Benisty}, M., {Dullemond}, C.~P., \& {Hirose}, S. 2014, \apj,
  780, 42, \dodoi{10.1088/0004-637X/780/1/42}

\bibitem[{{van Boekel} {et~al.}(2003){van Boekel}, {Waters}, {Dominik},
  {Bouwman}, {de Koter}, {Dullemond}, \& {Paresce}}]{vanBoekel03}
{van Boekel}, R., {Waters}, L.~B.~F.~M., {Dominik}, C., {et~al.} 2003, \aap,
  400, L21, \dodoi{10.1051/0004-6361:20030141}

\bibitem[{Van Der~Walt {et~al.}(2011)Van Der~Walt, Colbert, \&
  Varoquaux}]{van2011numpy}
Van Der~Walt, S., Colbert, S.~C., \& Varoquaux, G. 2011, Computing in Science
  \& Engineering, 13, 22

\bibitem[{{Vinkovi{\'c}} \& {Jurki{\'c}}(2007)}]{Vinkovic07}
{Vinkovi{\'c}}, D., \& {Jurki{\'c}}, T. 2007, \apj, 658, 462,
  \dodoi{10.1086/511327}

\bibitem[{{Vural} {et~al.}(2014){Vural}, {Kreplin}, {Kishimoto}, {Weigelt},
  {Hofmann}, {Kraus}, {Schertl}, {Dugu{\'e}}, {Duvert}, {Lagarde}, \&
  {Massi}}]{Vural14}
{Vural}, J., {Kreplin}, A., {Kishimoto}, M., {et~al.} 2014, \aap, 564, A118,
  \dodoi{10.1051/0004-6361/201322997}

\bibitem[{{Walker} {et~al.}(2004){Walker}, {Wood}, {Lada}, {Robitaille},
  {Bjorkman}, \& {Whitney}}]{Walker04}
{Walker}, C., {Wood}, K., {Lada}, C.~J., {et~al.} 2004, \mnras, 351, 607,
  \dodoi{10.1111/j.1365-2966.2004.07807.x}

\bibitem[{{Walker} \& {Wolstencroft}(1988)}]{Walker88}
{Walker}, H.~J., \& {Wolstencroft}, R.~D. 1988, \pasp, 100, 1509,
  \dodoi{10.1086/132357}

\bibitem[{{Whitney} {et~al.}(2004){Whitney}, {Indebetouw}, {Bjorkman}, \&
  {Wood}}]{Whitney04}
{Whitney}, B.~A., {Indebetouw}, R., {Bjorkman}, J.~E., \& {Wood}, K. 2004,
  \apj, 617, 1177, \dodoi{10.1086/425608}

\bibitem[{{Wood} {et~al.}(2002){Wood}, {Wolff}, {Bjorkman}, \&
  {Whitney}}]{Wood02}
{Wood}, K., {Wolff}, M.~J., {Bjorkman}, J.~E., \& {Whitney}, B. 2002, \apj,
  564, 887, \dodoi{10.1086/324285}

\bibitem[{{Yamamura} {et~al.}(2010){Yamamura}, {Makiuti}, {Ikeda}, {Fukuda},
  {Oyabu}, {Koga}, \& {White}}]{Yamamura10}
{Yamamura}, I., {Makiuti}, S., {Ikeda}, N., {et~al.} 2010, VizieR Online Data
  Catalog, 2298

\bibitem[{{Zacharias} {et~al.}(2013){Zacharias}, {Finch}, {Girard}, {Henden},
  {Bartlett}, {Monet}, \& {Zacharias}}]{Zacharias13}
{Zacharias}, N., {Finch}, C.~T., {Girard}, T.~M., {et~al.} 2013, \aj, 145, 44,
  \dodoi{10.1088/0004-6256/145/2/44}

\bibitem[{{Zwintz} {et~al.}(2009){Zwintz}, {Kallinger}, {Guenther},
  {Gruberbauer}, {Huber}, {Rowe}, {Kuschnig}, {Weiss}, {Matthews}, {Moffat},
  {Rucinski}, {Sasselov}, {Walker}, \& {Casey}}]{Zwintz09}
{Zwintz}, K., {Kallinger}, T., {Guenther}, D.~B., {et~al.} 2009, \aap, 494,
  1031, \dodoi{10.1051/0004-6361:200811116}

\end{thebibliography}

\end{document}